\documentclass[prd,12pt,letterpaper,showpacs,groupedaddress,superscriptaddress,nofootinbib,floatfix,preprintnumbers,tightenlines]{revtex4-2}
\usepackage{hyperref,amssymb,amsmath,graphicx,xcolor,cancel,mathrsfs,multirow,bm,mathtools}
\usepackage[caption=false]{subfig}
\usepackage{bm}

\newcommand{\Mod}[1]{\ (\mathrm{mod}\ #1)}

\begin{document}
\begin{figure}[!t]
	
	\vskip -1.5cm
	\leftline{\includegraphics[width=0.25\textwidth]{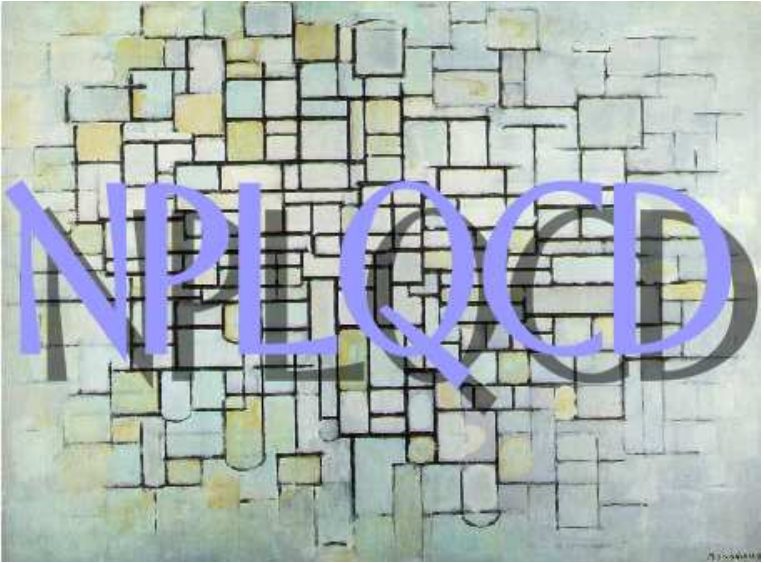}}
\end{figure}

\title{Sparsening Algorithm for Multi-Hadron Lattice QCD Correlation Functions}

\author{W.~Detmold}
\affiliation{Center for Theoretical Physics, Massachusetts Institute of Technology, Boston, MA 02139, USA}
\author{D.J.~Murphy}
\affiliation{Center for Theoretical Physics, Massachusetts Institute of Technology, Boston, MA 02139, USA}
\author{A.V.~Pochinsky}
\affiliation{Center for Theoretical Physics, Massachusetts Institute of Technology, Boston, MA 02139, USA}
\author{M.J.~Savage}
\affiliation{Institute for Nuclear Theory, University of Washington, Seattle, WA 98195, USA}
\author{P.E.~Shanahan}
\affiliation{Center for Theoretical Physics, Massachusetts Institute of Technology, Boston, MA 02139, USA}
\author{M.L.~Wagman}
\affiliation{Center for Theoretical Physics, Massachusetts Institute of Technology, Boston, MA 02139, USA}

\date{\today}

\preprint{MIT-CTP/5127}
\preprint{INT-PUB-19-026}



\begin{abstract}
Modern advances in algorithms for lattice QCD calculations have steadily driven down the resources required to generate gauge field ensembles and calculate quark propagators, such that, in cases relevant to nuclear physics, performing quark contractions to assemble correlation functions from propagators has become the dominant cost. This work explores a propagator sparsening algorithm for forming correlation functions describing multi-hadron systems, such as light nuclei, with reduced computational cost. The algorithm constructs correlation functions from sparsened propagators defined on a coarsened lattice geometry, where the sparsened propagators are obtained from propagators computed on the full lattice. This algorithm is used to study the low-energy QCD ground-state spectrum using a single Wilson-clover lattice ensemble with \( m_{\pi} \approx 800 \) MeV. It is found that the extracted ground state masses and binding energies, as well as their statistical uncertainties, are consistent when determined from correlation functions constructed from sparsened and full propagators. In addition, while evidence of modified couplings to excited states is observed in sparsened correlation functions, it is demonstrated that these effects can be removed, if desired, with an inexpensive modification to the sparsened estimator. 

\end{abstract}

\maketitle

\section{Introduction}
\label{sec:intro}
Lattice Quantum Chromodynamics (QCD) provides an \textit{ab-initio} method for predicting
the low-lying spectrum, structure, and reactions of hadrons and nuclei from the dynamics of their constituent quarks and gluons. In practice, this is a computationally demanding task, and requires the use of state-of-the-art supercomputers, as well as the development of increasingly sophisticated numerical algorithms \cite{Joo:2019byq}. Continued progress, especially toward understanding the properties of increasingly heavy nuclei, will require further advances in both hardware and algorithms \cite{Aoki:2012tk,Beane:2006mx,Beane:2009py,Beane:2009gs,Beane:2010hg,Beane:2010em,Beane:2011iw,Beane:2012vq,Beane:2012ey,Beane:2013br,Beane:2014ora,Berkowitz:2015eaa,Beane:2015yha,Berkowitz:2017smo,Berkowitz:2019yrf,Chang:2015qxa,Chang:2017eiq,Doi:2012xd,Francis:2013lva,Francis:2018qch,HALQCD:2012aa,Hanlon:2018yfv,Iritani:2018zbt,Iritani:2018sra,Ishii:2006ec,Nemura:2017bbw,Orginos:2015aya,Savage:2016kon,Shanahan:2017bgi,Tiburzi:2017iux,Wagman:2017tmp,Winter:2017bfs,Yamazaki:2009ua,Yamazaki:2011nd,Yamazaki:2012hi,Yamazaki:2015asa,Yamazaki:2015vjn,Yamazaki:2017jfh}. 

A prototypical lattice QCD calculation proceeds in three stages. First, Monte Carlo importance sampling techniques are applied to the QCD path integral, generating a Markov chain of representative configurations of the gauge field. This ensemble generation is an expensive task that is often undertaken as a community effort \cite{Joo:2019byq}, with the same gauge field configurations shared between many physics calculations. In the second phase, the lattice Dirac operator is repeatedly inverted for each gauge field configuration to determine quark propagators nonperturbatively. Finally, these quark propagators are contracted together to form correlation functions describing the physics of particular states of interest. The effects of the lattice regularization on physical observables, as well as the effects of potential mistunings in the bare input quark masses, can be systematically removed by repeating this procedure to generate a series of simulations with different lattice spacings, simulation volumes, and masses. One can then perform controlled interpolations and extrapolations to the infinite volume, continuum, physical quark mass limit, to provide QCD predictions which can be directly compared to experimental results where they exist, or to make predictions for quantities that cannot be accessed experimentally.

Generating gauge field ensembles and quark propagators is common to lattice calculations of many different physical quantities, and has historically dominated the cost of these calculations. As a result, improving the efficiency of algorithms used for gauge field generation, such as hybrid Monte Carlo (HMC) \cite{Duane:1987de}, as well as the sparse matrix inverters needed to compute quark propagators, has been a major focus of algorithmic research and software development. Multigrid algorithms \cite{Babich:2010qb,Osborn:2010mb,Boyle:2014rwa,PhysRevD.92.114516,Detmold:2016rnh,Clark:2016rdz,Yamaguchi:2016kop,Bacchio:2017pcp,Brower:2018ymy,Richtmann:2019eyj}, which exploit the local coherence of QCD by inverting a cheaper approximation to the Dirac operator defined on a coarsened lattice, have been particularly successful in accelerating gauge field generation and propagator inversions in recent years, leading to \( \mathcal{O}(10-100) \) fold improvements in the efficiencies of these tasks. Similar ideas have also found success as a technique for reducing the memory footprint of eigenvectors of the lattice Dirac operator \cite{Clark:2017wom}. As a result of these advances, the cost of the contraction stage of lattice QCD calculations targeting nuclei has become relatively more expensive, and has, in some cases, become the dominant cost of the entire calculation. New efforts to address this situation and improve the efficiency of contractions are needed.

This work investigates the feasibility of an algorithm exploiting local coherence to reduce the numerical cost of computing correlation functions of single hadrons and light nuclei, based on sparsening. Section \ref{sec:methods} begins by discussing a simple prescription for sparsening, and details the construction of multi-hadron correlation functions such as those that describe the properties of light nuclei. Section \ref{subsec:hadron_sparsening} examines the impact of sparsening on hadronic correlation functions, demonstrating that the ground-state energies extracted from these correlation functions are unaltered within the statistical resolution of this calculation, and Section \ref{subsec:excited_states} introduces an improved estimator for controlling modifications of the couplings to excited states introduced by sparsening. Finally, Section \ref{subsec:nuclei} examines the impact of sparsening on the ground states of light nuclei.

\section{Methodology}
\label{sec:methods}
\label{sec:methods}

Naively, the quark contractions required to form correlation functions describing many-body systems require prohibitively large computational resources in general, since the number of quark contractions grows exponentially with the number of quark fields. A number of lattice QCD collaborations have instead used more efficient ``baryon block'' algorithms \cite{Basak:2005ir,Beane:2005rj,Beane:2006mx,Beane:2008dv,Doi:2012xd,Detmold:2012eu}, which have enabled first-principles calculations of the spectra and matrix elements of light nuclei, for example in Refs.~\cite{Aoki:2012tk,Beane:2006mx,Beane:2009py,Beane:2009gs,Beane:2010hg,Beane:2010em,Beane:2011iw,Beane:2012vq,Beane:2012ey,Beane:2013br,Beane:2014ora,Berkowitz:2015eaa,Beane:2015yha,Berkowitz:2017smo,Berkowitz:2019yrf,Chang:2015qxa,Chang:2017eiq,Doi:2012xd,Francis:2013lva,Francis:2018qch,HALQCD:2012aa,Hanlon:2018yfv,Iritani:2018zbt,Iritani:2018sra,Ishii:2006ec,Nemura:2017bbw,Orginos:2015aya,Savage:2016kon,Shanahan:2017bgi,Tiburzi:2017iux,Wagman:2017tmp,Winter:2017bfs,Yamazaki:2009ua,Yamazaki:2011nd,Yamazaki:2012hi,Yamazaki:2015asa,Yamazaki:2015vjn,Yamazaki:2017jfh}. These algorithms work by first constructing partially-contracted ``blocks'' from quark propagators \(S\):
\begin{equation}
\label{eqn:baryon_block}
  \mathcal{B}_{b}^{a_{1},a_{2},a_{3}} \left( \vec{p},t; x_{0} \right) = \sum_{\vec{x}} e^{i \vec{p} \cdot \vec{x}} \sum_{k=1}^{N_{B(b)}} \widetilde{w}_{b}^{(c_{1},c_{2},c_{3}),k} \sum_{i_{1},i_{2},i_{3}} \epsilon^{i_{1},i_{2},i_{3}} S_{c_{i_{1}}}^{a_{1}} \left( x; x_{0} \right) S_{c_{i_{2}}}^{a_{2}} \left( x; x_{0} \right) S_{c_{i_{3}}}^{a_{3}} \left( x; x_{0} \right),
\end{equation}
describing the propagation from \( x_{0} = (\vec{x}_{0},t_{0}) \) to \( x = (\vec{x},t) \) of a baryon with quantum numbers \(b\) and momentum \(\vec{p}\). Here \(a_{i}\) and \(c_{i}\) are combined spin-color-flavor indices, and, in many cases, particular choices of the weights \( \widetilde{w}_{b}^{(a_{1},a_{2},a_{3}),k} \) corresponding to interpolating operators with the correct transformation properties to project onto the wavefunctions of many-body states of interest are known \cite{Basak:2005ir}. Expressing the contractions for multi-hadron nuclear correlation functions in terms of nucleon-level blocks can often significantly reduce the computational cost by improving the projection onto the state of interest, especially if the blocks are stored and re-used between different calculations \cite{Beane:2009gs,Beane:2009py}. 

The dominant cost of assembling the baryon blocks defined by Eq.~\eqref{eqn:baryon_block} is associated with the Fourier transforms (FTs) used to project onto states with definite momenta. These FTs are a natural target of a multigrid-type algorithm, since spatially blocking the lattice by a factor of \(N\) reduces the number of modes by a factor of \(N^{3}\). The local coherence of QCD implies that blocked and unblocked calculations should result in the same values of low-energy hadronic observables, up to uncertainties from statistical sampling and discretization effects, provided that \( N a \lesssim m_{\pi}^{-1} \), where \( a \) is the lattice spacing and \(m_{\pi}\) is the mass of the lightest hadronic state (pion). In this work, a particularly simple spatial blocking procedure for quark propagators is explored: the lattice is blocked uniformly in all spatial directions, and the value of the propagator evaluated on the first site of each block is used to define sparsened propagators on the coarsened lattice. While in principle one could imagine exploring more sophisticted blocking procedures --- such as a renormalization-group-based block average, or a projection onto the coarsened lattice defined by blocked low-mode eigenvectors of the Dirac operator --- such a study is left for future work.

In the following sections, we distinguish between full correlation functions constructed from quark propagators defined on the full lattice
\begin{equation}
\label{eqn:full_corr_def}
\begin{split}
  C_{\rm full} \left( \vec{p} , t ; \vec{x}_0, t_0 \right) &= \Big\langle 0 \Big\vert \sum_{\vec{x}\in\Lambda_{3}} e^{i \vec{p} \cdot \vec{x}} \mathscr{O} \left ( \vec{x},t \right) \mathscr{O}^{\dagger} \left( \vec{x}_0, t_0 \right) \Big\vert 0 \Big\rangle \\
  \Lambda_3 &= \big\{ \left( n_{1} , n_{2} , n_{3} \right) \big\vert 0 \leq n_{i} < L \big\} \\
\end{split}
\end{equation}
and sparsened correlation functions
\begin{equation}
\label{eqn:sparse_corr_def}
\begin{split}
  C_{\rm sparse} \left( \vec{p} , t ; \vec{x}_0, t_0 \right) &= \Big\langle 0 \Big\vert \sum_{\vec{x}\in\widetilde{\Lambda}_{3}(N)} e^{i \vec{p} \cdot \vec{x}} \mathscr{O} \left ( \vec{x},t \right) \mathscr{O}^{\dagger} \left( \vec{x}_0, t_0 \right) \Big\vert 0 \Big\rangle \\
  \widetilde{\Lambda}_3(N) &= \big\{ \left( \widetilde{n}_{1} , \widetilde{n}_{2} , \widetilde{n}_{3} \right) \big\vert 0 \leq \widetilde{n}_{i} < L; \widetilde{n}_{i} \equiv 0 \Mod{N} \big\} \\
\end{split}
\end{equation}
constructed from sparsened propagators defined on a coarsened sublattice \( \widetilde{\Lambda}_{3}(N) \subset \Lambda_{3} \), as described above, with \( \mathscr{O}^{\dagger} \) and \( \mathscr{O} \) appropriate creation and annihilation operators for the state of interest, and \( \vec{x}_{0} \in \widetilde{\Lambda}_{3}(N) \). Ultimately, the effect of sparsening, as it has been implemented in this work, is to modify the structure of the interpolating operator used at the sink. Since any choice of interpolating operator with the correct quantum numbers is equally valid for probing a given state in the lattice theory, this implementation of sparsening is guaranteed to preserve the values of physical observables, such as the finite volume energy spectrum, but can, however, modify the relative overlaps onto the ground and excited states in a particular channel. In the approach explored in this work, sparsening is expected to modify couplings to excited states at short Euclidean time separations, since Eq.~\eqref{eqn:sparse_corr_def} can be understood as an incomplete momentum projection over a subset of the allowed lattice modes. The degree to which the couplings to excited states are modified by sparsening, as well as the degree to which it impacts the statistical uncertainties of observables such as hadron energies, are empirical questions that are explored in the next section.

\section{Results}
Results are reported for the low-energy QCD spectrum computed on a single \( 32^{3} \times 48 \) lattice ensemble with the Wilson-clover fermion action \cite{Sheikholeslami:1985ij} and L\"{u}scher-Weisz gauge action \cite{Luscher:1984xn}. This ensemble was generated using three degenerate flavors of quarks with masses tuned to the strange quark mass, leading to \(m_{\pi} \approx 806 \) MeV, and a lattice spacing \( a \approx 0.145 \) fm determined by \(\Upsilon\) spectroscopy \cite{Beane:2012vq}. Throughout, lattice momenta \( \vec{p} \) are specified in terms of the dimensionless wavenumber \( \vec{n} \), where \(  \vec{p} = 2 \pi \vec{n} / L \) and \( L = 32 \) is the spatial extent of the lattice. The correlation functions described in this work are computed from Gaussian-smeared propagators constructed using 30 iterations of APE smearing \cite{Albanese:1987ds} at the source and sink with radius \( \rho = 4.35 \) in lattice units, and sparsened according to the procedure described in Section \ref{sec:methods}. Further details can be found in Ref.~\cite{Beane:2012vq}. The distribution of source positions throughout the spacetime volume is varied depending on the quantity being studied: in Section \ref{subsec:hadron_sparsening} all source locations with  \( \vert \vec{x}_{0} \vert / a \leq 12 \), where the components of \( \vec{x}_{0} \) are multiples of 4 and \( t_{0} / a = 12 \), are included, while in Section \ref{subsec:nuclei}, the source locations are randomly distributed throughout the four-dimensional spacetime volume. In all cases measurements were performed on 900 independent gauge field configurations.

Throughout this work the lattice has been blocked by a factor of \( N = 4 \) lattice units in the spatial directions to define sparsened propagators and correlation functions. This blocking is chosen to be consistent with the expected scale of spatial correlations in hadronic two-point functions, \( ( a m_{\pi} / 2 )^{-1} \approx 3.4 \), in the lattice units of the ensemble used for this study. In Ref.~\cite{Wagman:2016bam}, for example, it has been demonstrated that the nucleon correlation function approximately factorizes as 
\begin{equation}
  \left\langle C_{N}(t) \right\rangle \approx \left\langle e^{R_{N}(t)} \right\rangle \left\langle e^{i \theta_{N}}(t) \right\rangle \sim \left[ \left( e^{-m_{\pi} t / 2} \right) \left( e^{-(m_{N}/3 - m_{\pi}/2)t} \right) \right]^{3}, 
\end{equation}
where \( R_{N}(t) \) and \( \theta_{N}(t) \) denote the magnitude and phase of the nucleon two-point function, respectively. An analogous factorization is expected to hold for other hadronic states. In addition, in the context of this work, we have numerically studied the magnitude of correlations in hadronic two-point functions as the spatial locations of the quark propagators used to compute these two-point functions are varied, and find results that are consistent with \( ( a m_{\pi} / 2)^{-1} \) as the relevant scale; the interested reader is referred to Ref.~\cite{Murphy:2019lp} for additional detail.

\subsection{Sparsened Hadronic Correlation Functions}
\label{subsec:hadron_sparsening}

The viability of the proposed sparsening algorithm as a cost reduction technique for lattice QCD calculations depends primarily on the degree to which it preserves the precision with which matrix elements and the low-energy spectrum of QCD can be extracted, as well as the speedup of computing quark contractions that it enables. While the blocking described in the preceding paragraph was designed to preserve long-distance physics, the incomplete momentum projection implied by Eq.~\eqref{eqn:sparse_corr_def} effectively alters the lattice interpolating operator at the sink, and therefore alters the overlap onto the different hadronic states in the QCD spectrum. This section explores the practical ramifications of modifying the sink structure by comparing results extracted from the pion, \( \rho \) meson, nucleon, and \( \Delta \) baryon two-point functions computed using either full or sparsened propagators and for all lattice momenta with \( \vert \vec{n} \vert \leq \sqrt{5}\). While sparsening offers no significant calculational speedup for correlation functions describing single hadrons, these are the simplest and most statistically precise quantities available to study in lattice QCD, and thus a natural starting point to examine the impact of sparsening. Similar studies of more complicated matrix elements involving these states are deferred to future work.

\subsubsection{Consistency of Full and Sparsened Two-Point Correlation Functions}
\label{subsubsec:hadron_scatter}

\begin{figure}[!ht]
\centering
  \subfloat{\includegraphics[width=0.3\linewidth]{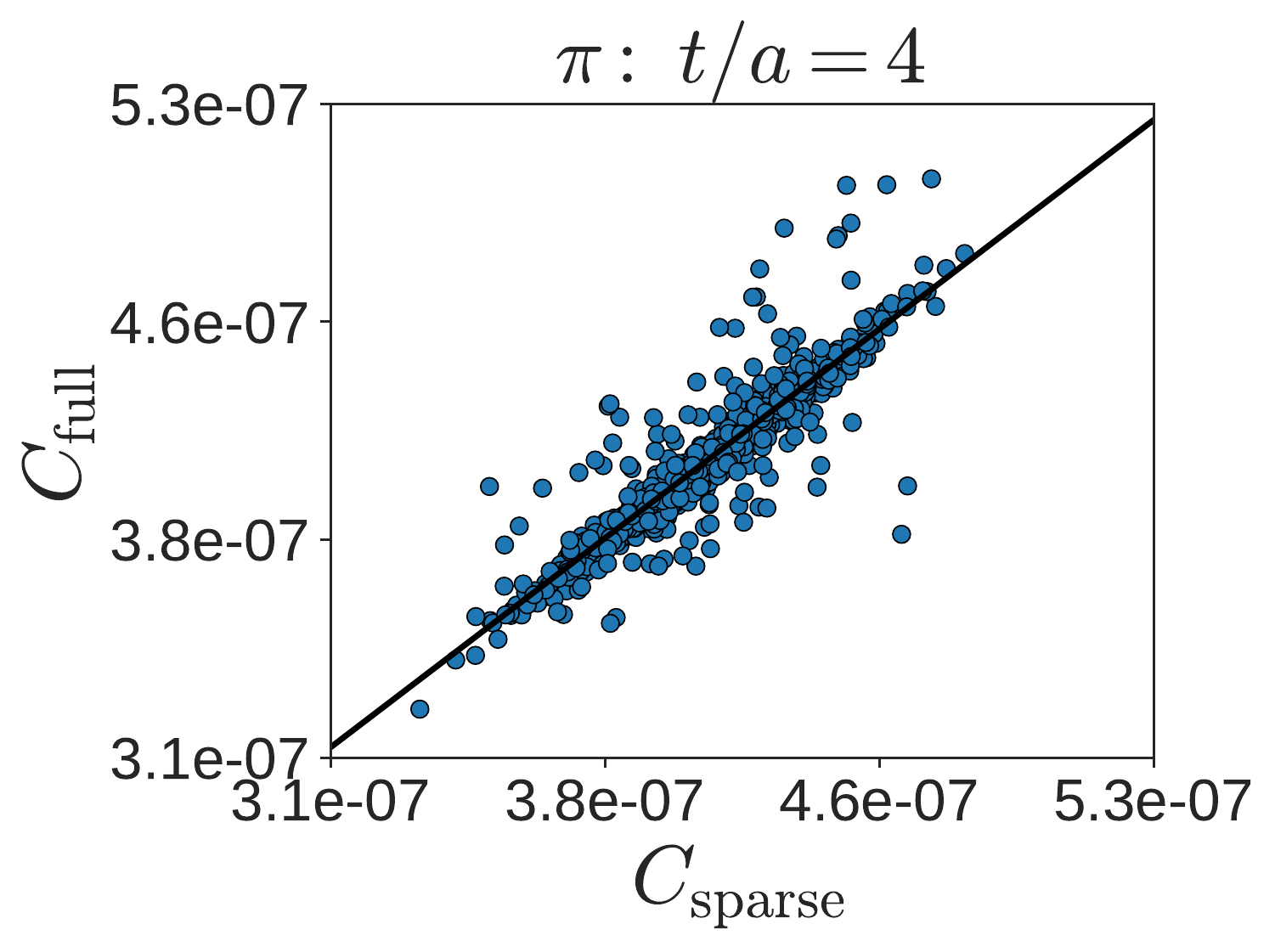}}
  \subfloat{\includegraphics[width=0.3\linewidth]{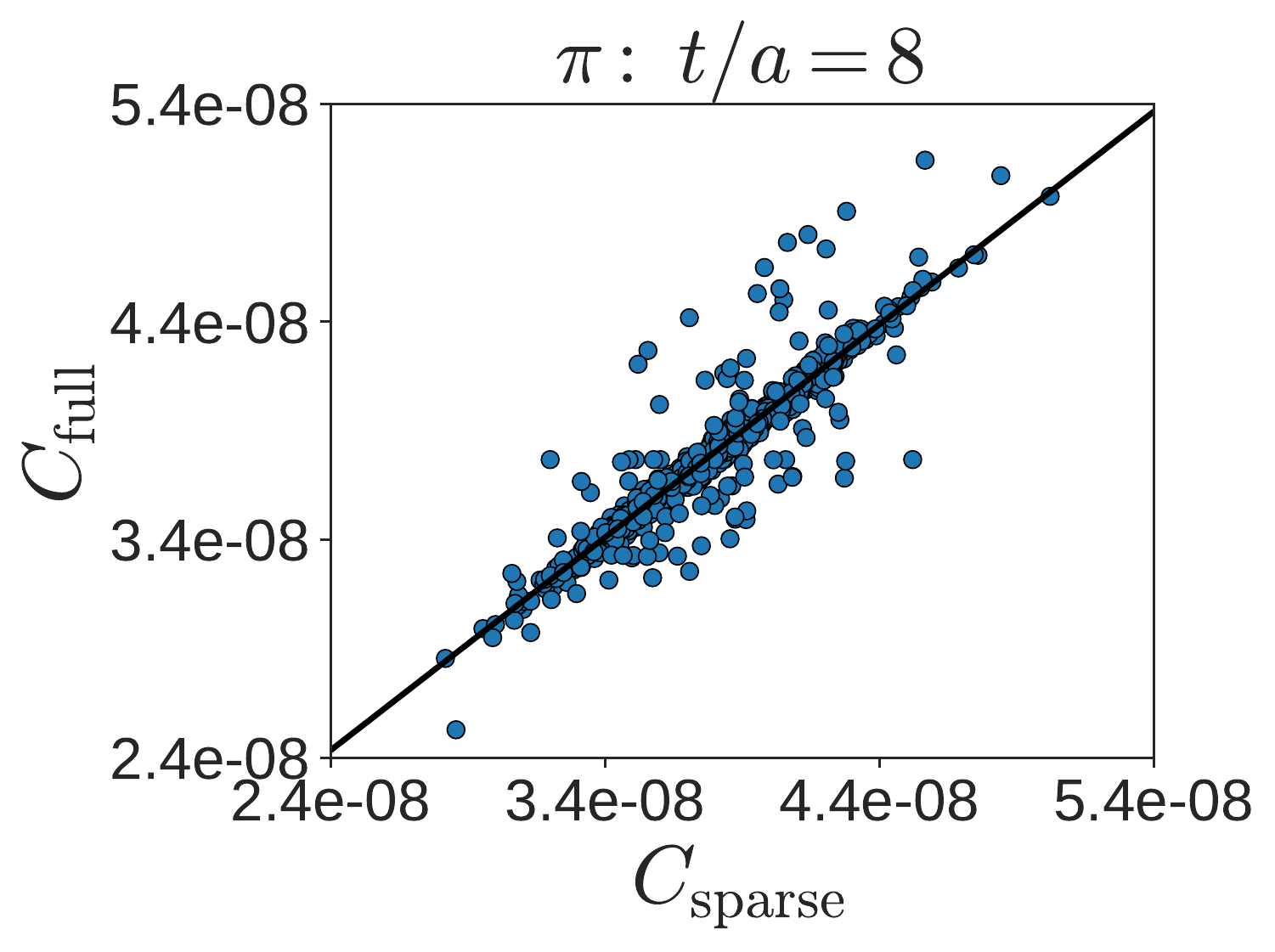}}
  \subfloat{\includegraphics[width=0.3\linewidth]{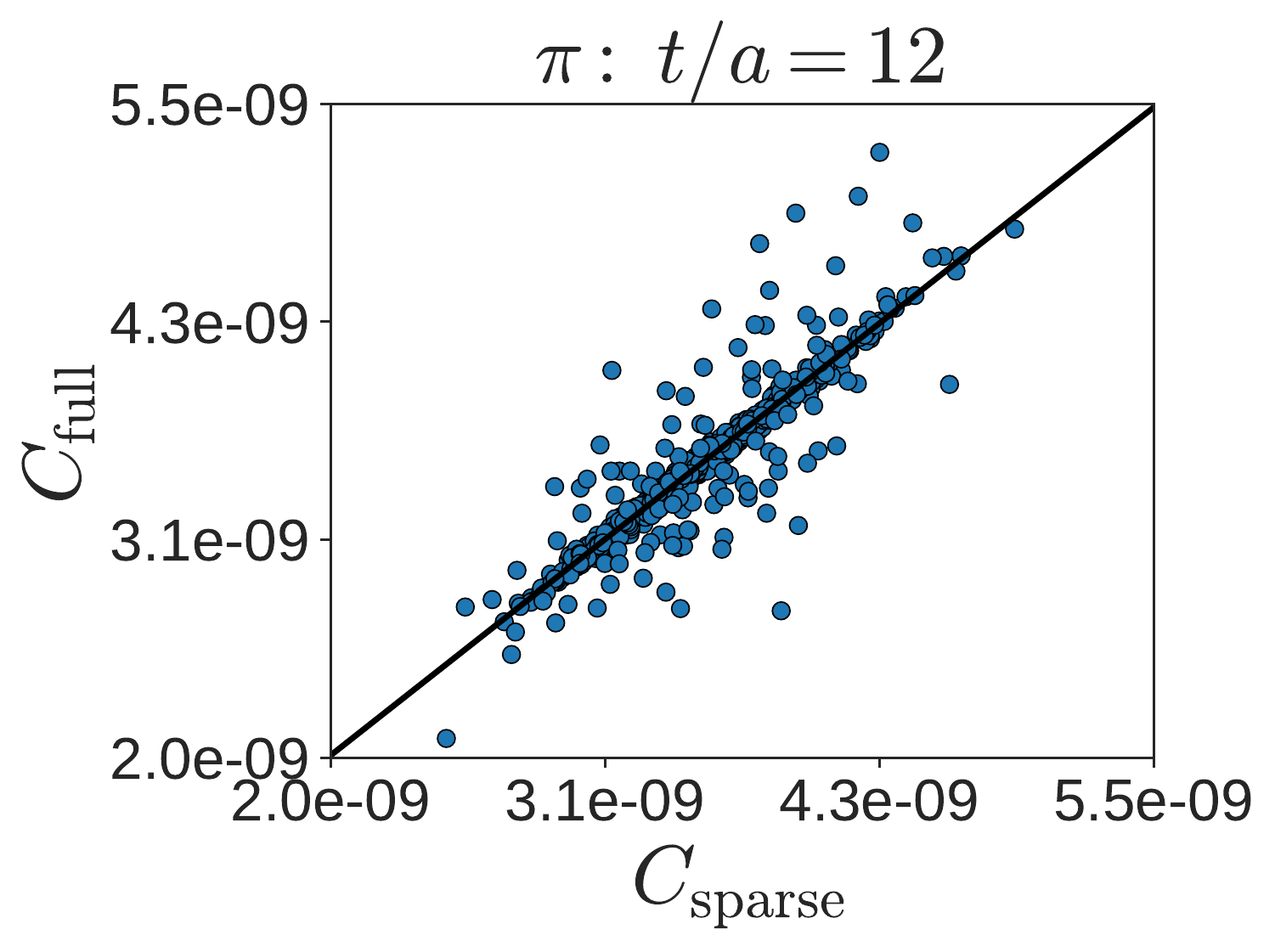}} \\
  \subfloat{\includegraphics[width=0.3\linewidth]{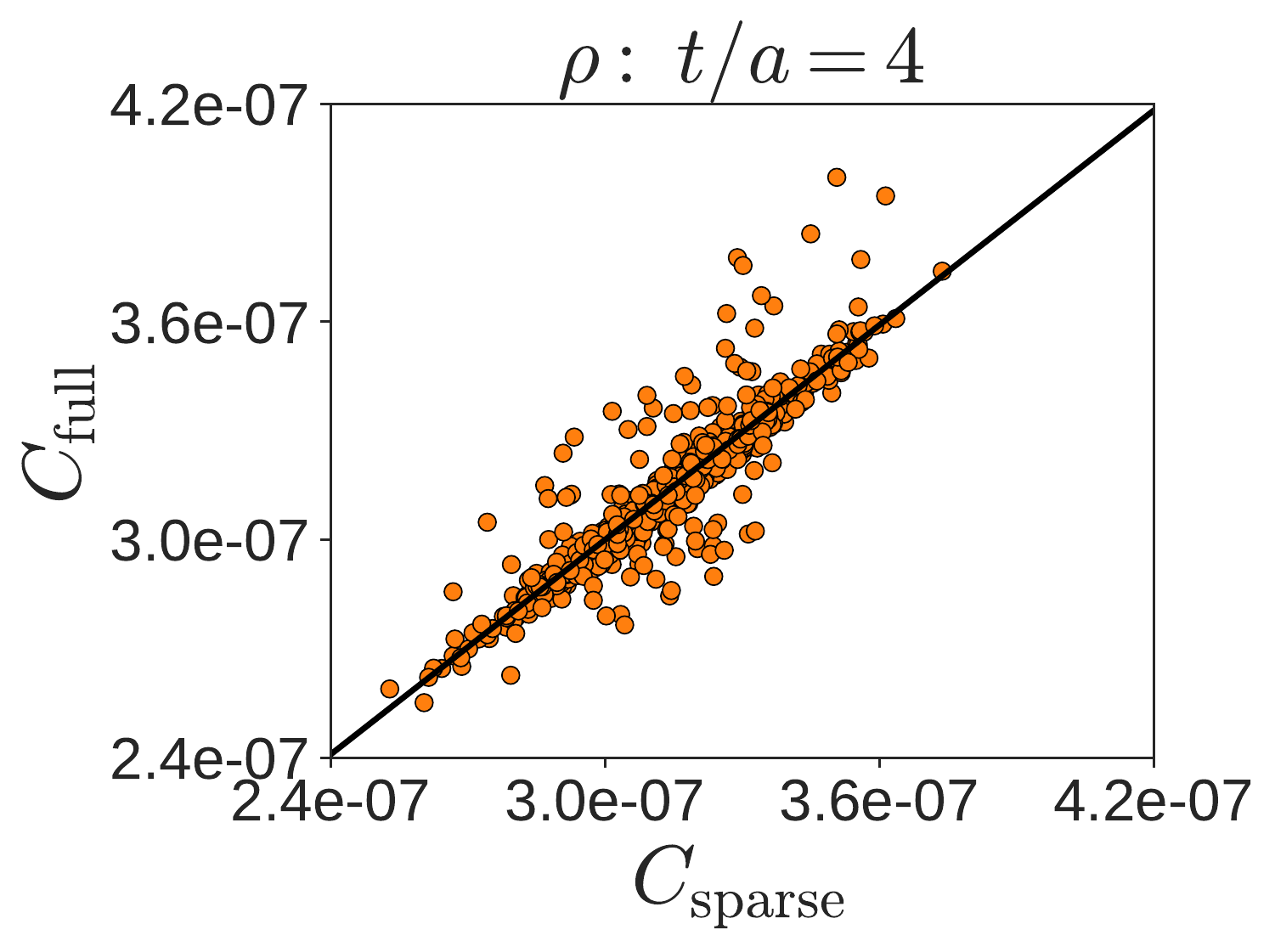}}
  \subfloat{\includegraphics[width=0.3\linewidth]{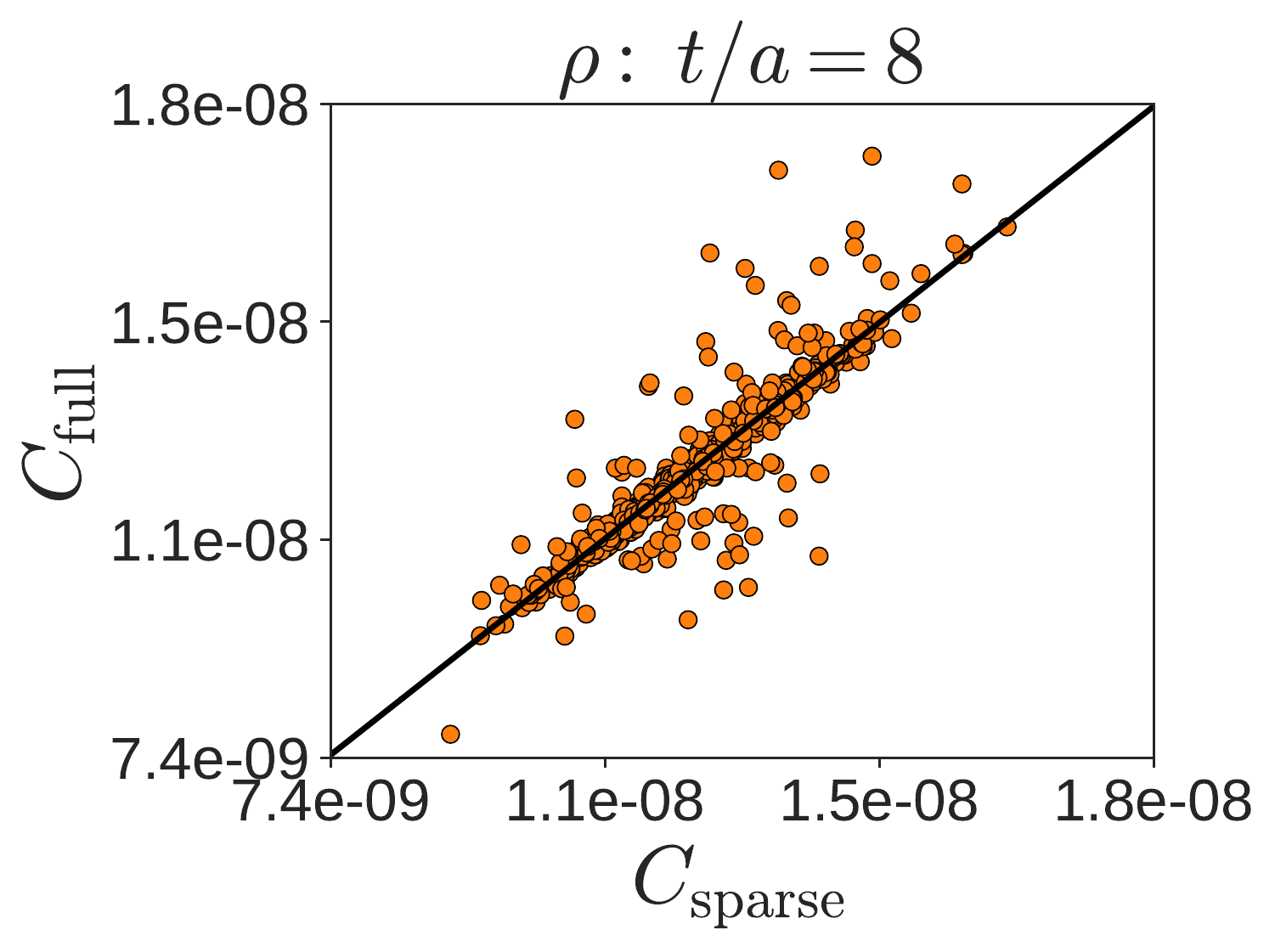}}
  \subfloat{\includegraphics[width=0.3\linewidth]{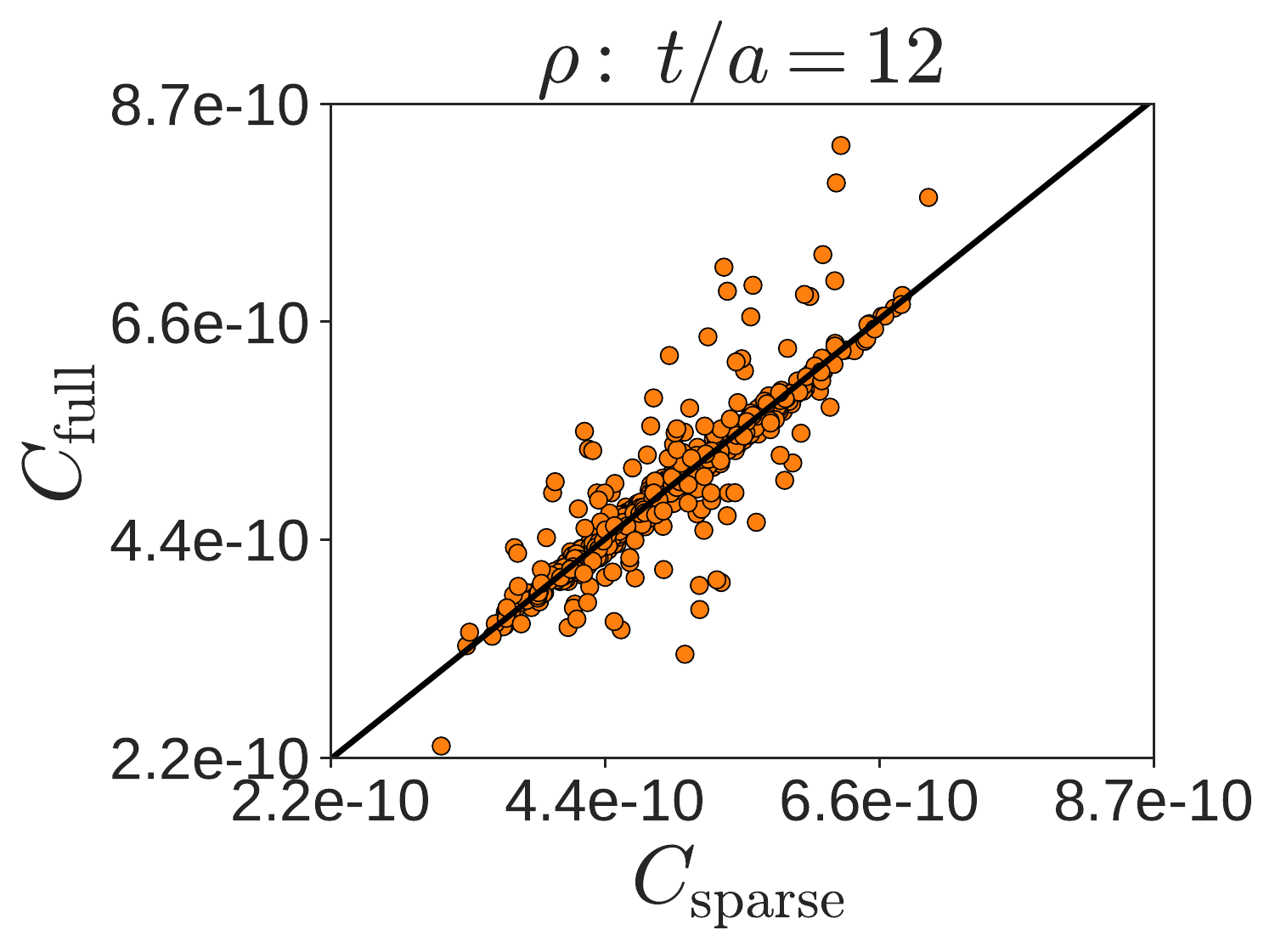}} \\
  \subfloat{\includegraphics[width=0.3\linewidth]{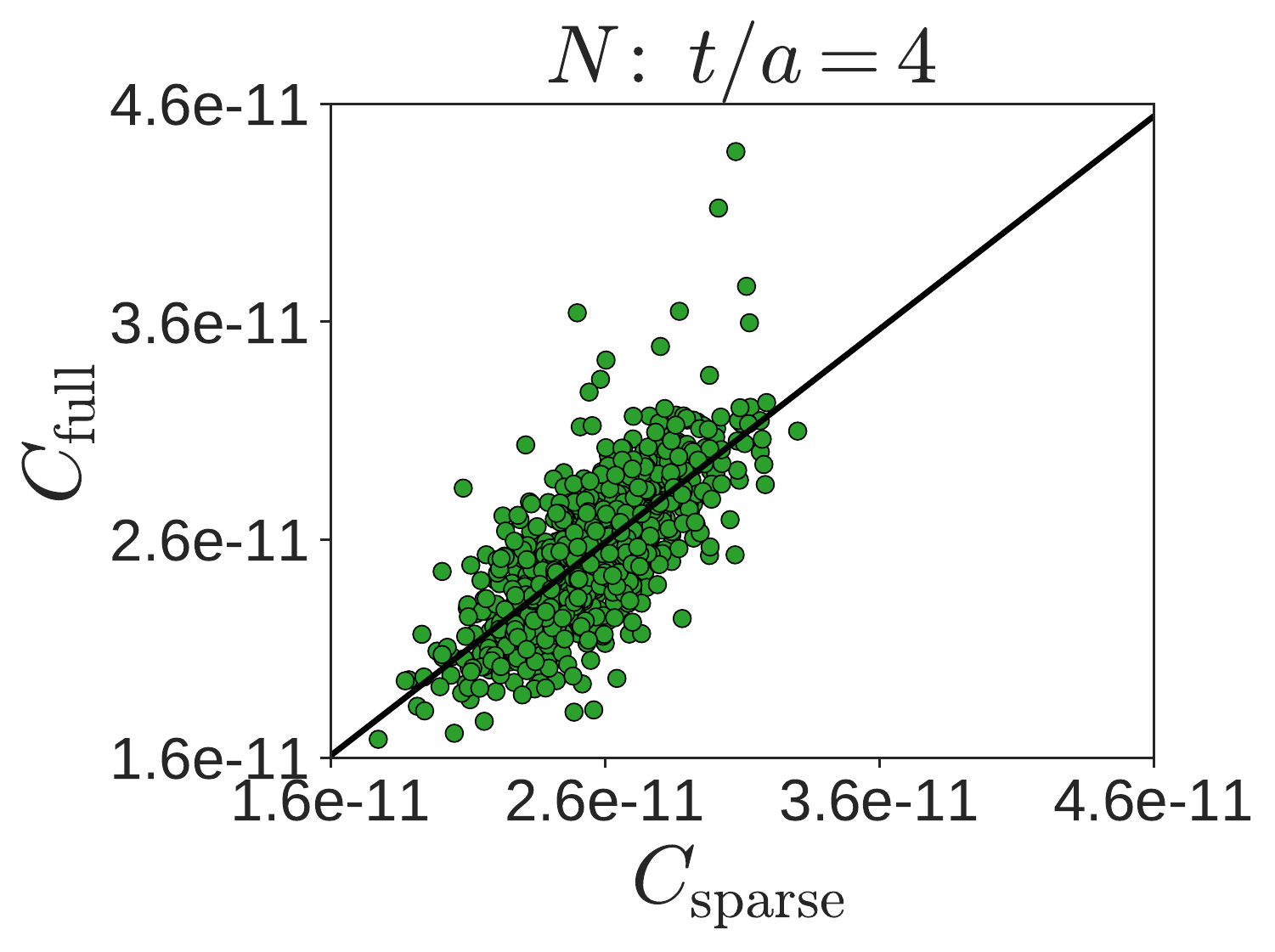}}
  \subfloat{\includegraphics[width=0.3\linewidth]{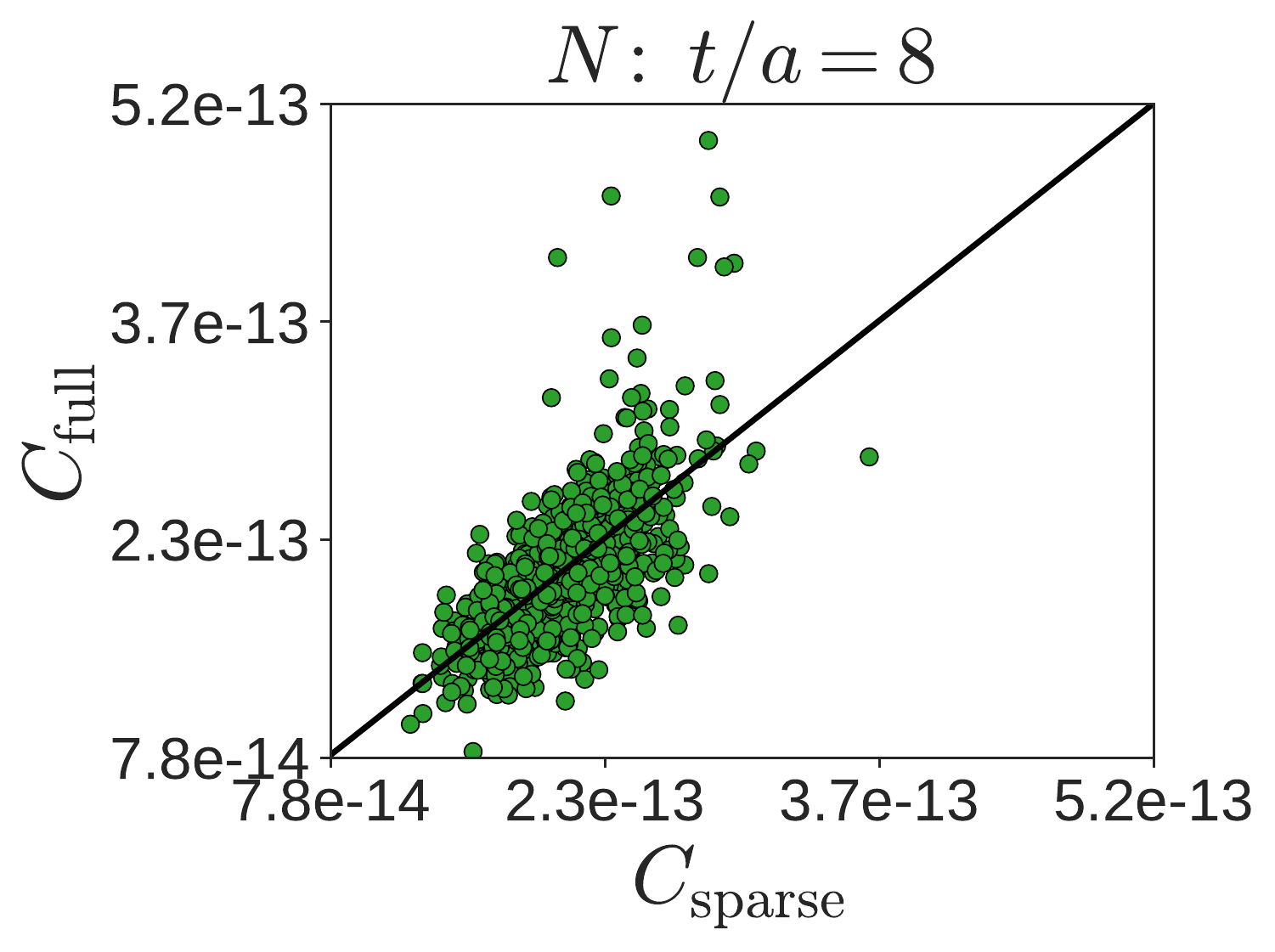}}
  \subfloat{\includegraphics[width=0.3\linewidth]{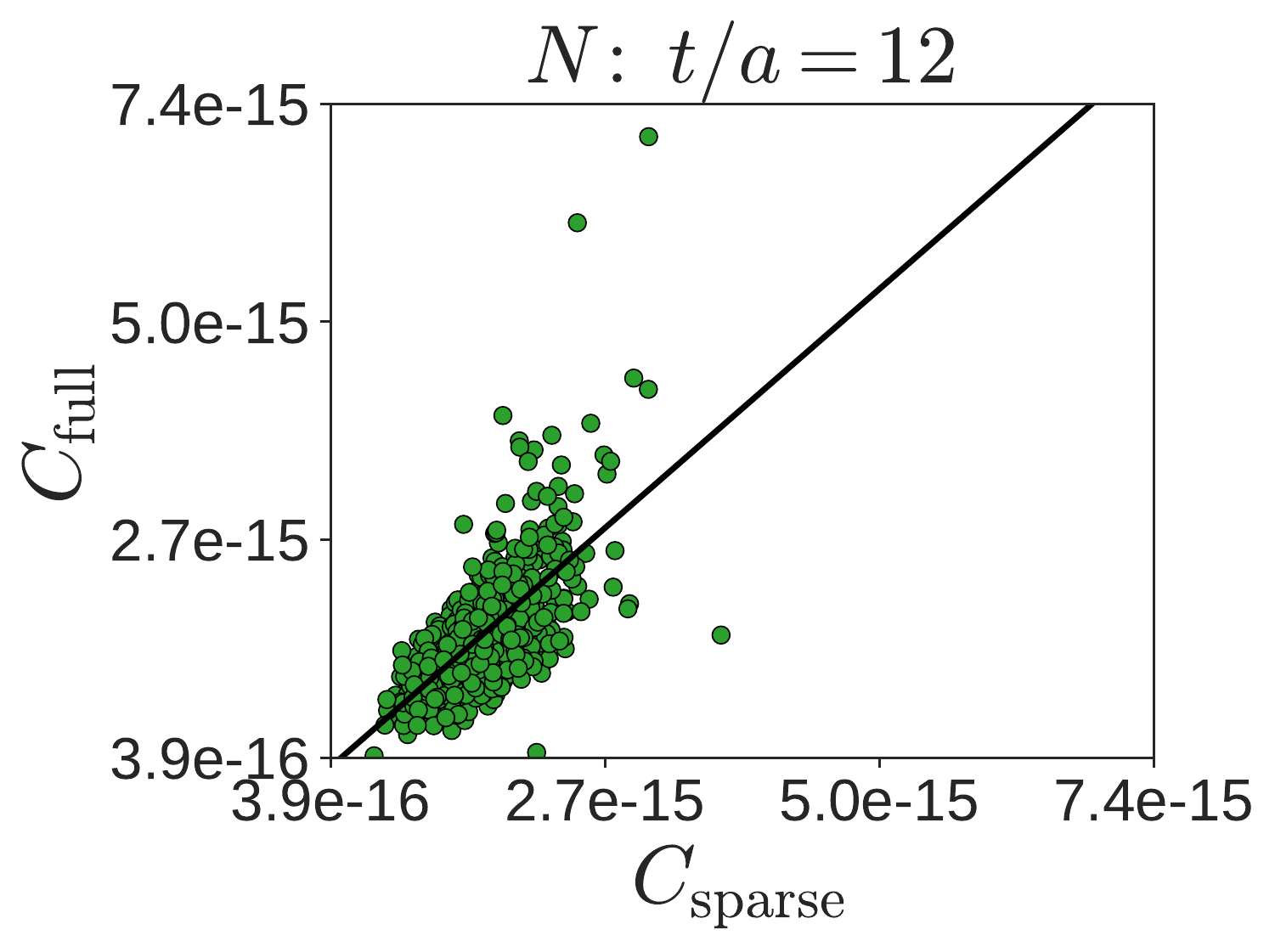}} \\
  \subfloat{\includegraphics[width=0.3\linewidth]{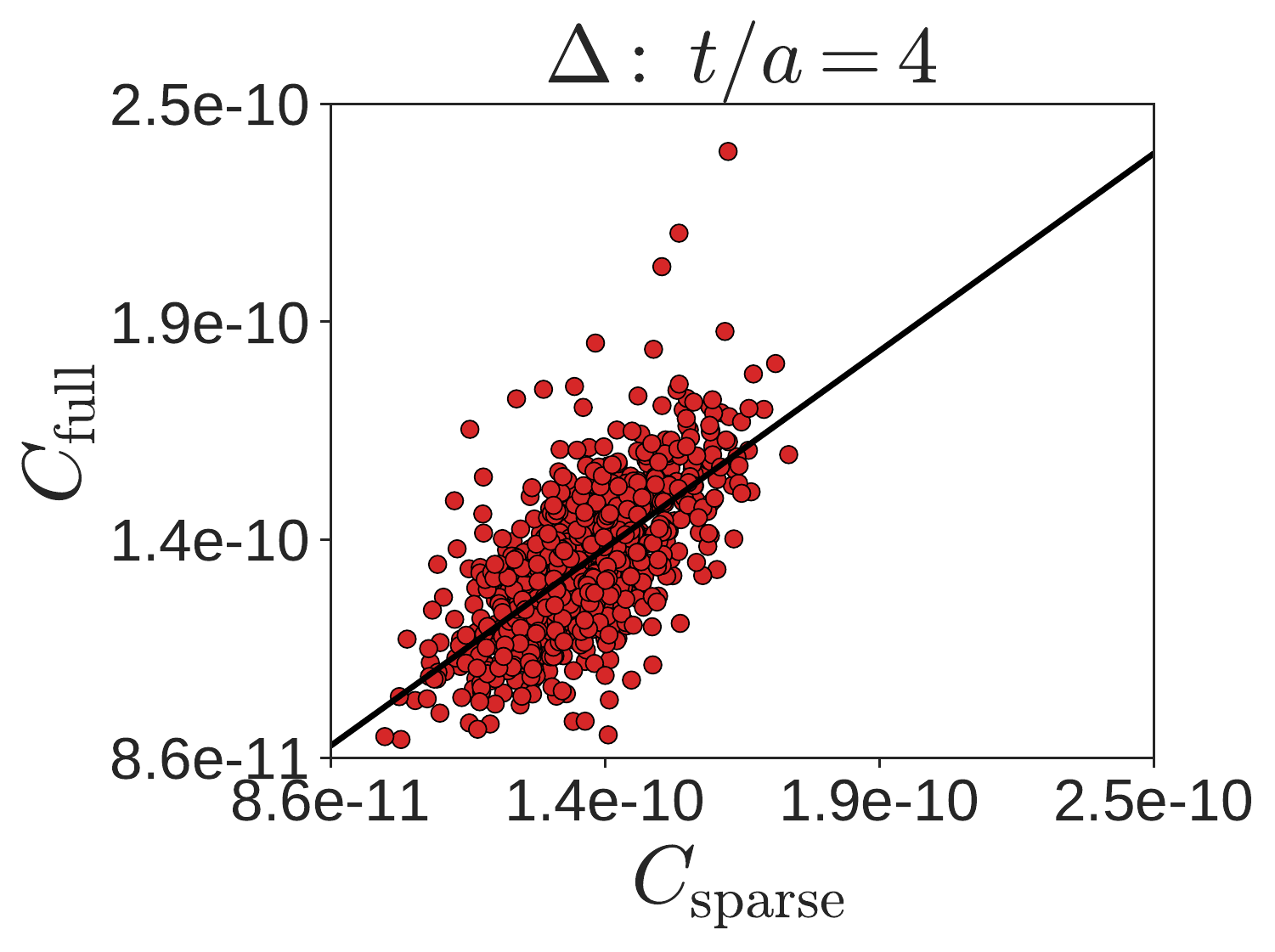}}
  \subfloat{\includegraphics[width=0.3\linewidth]{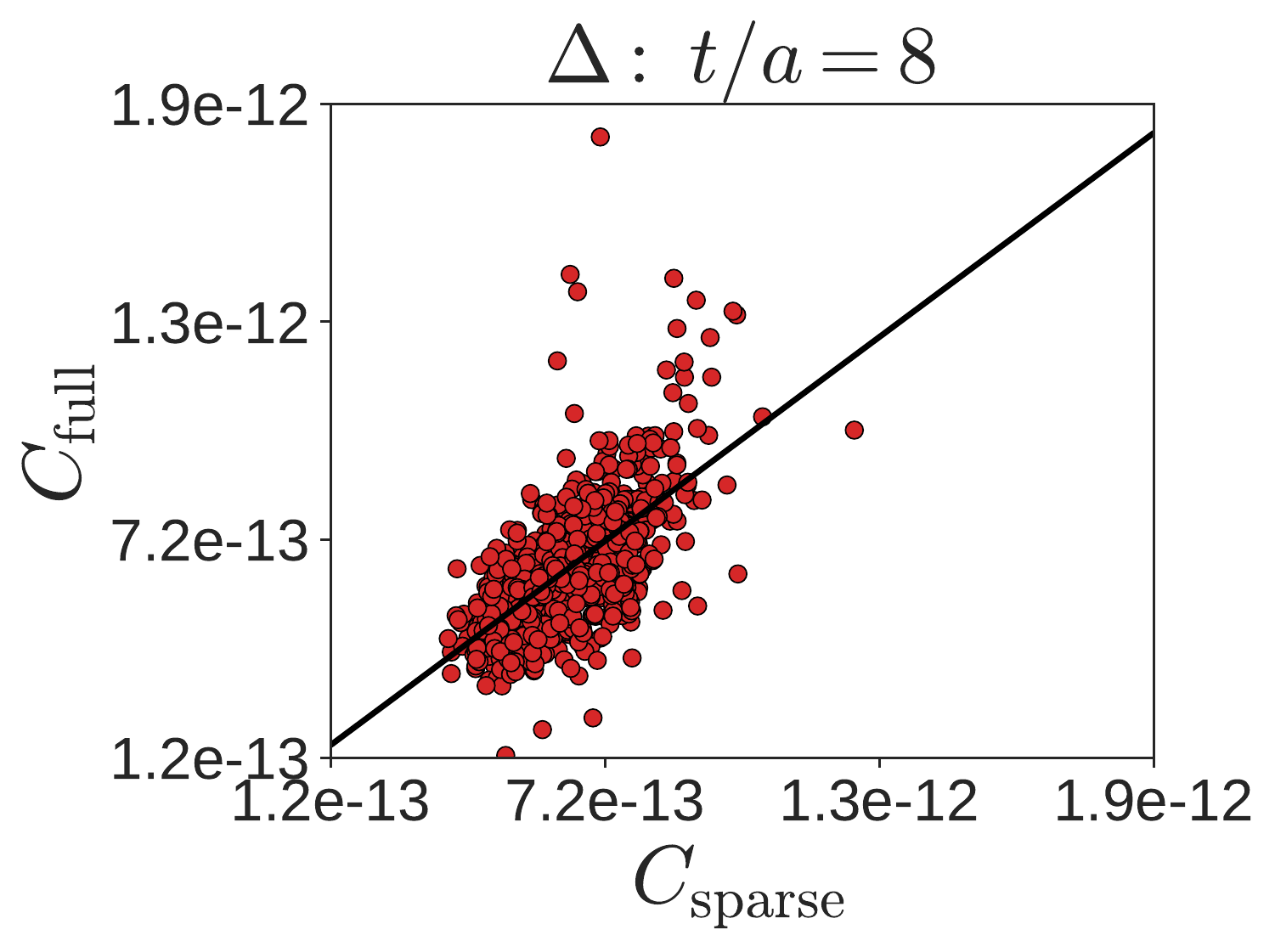}}
  \subfloat{\includegraphics[width=0.3\linewidth]{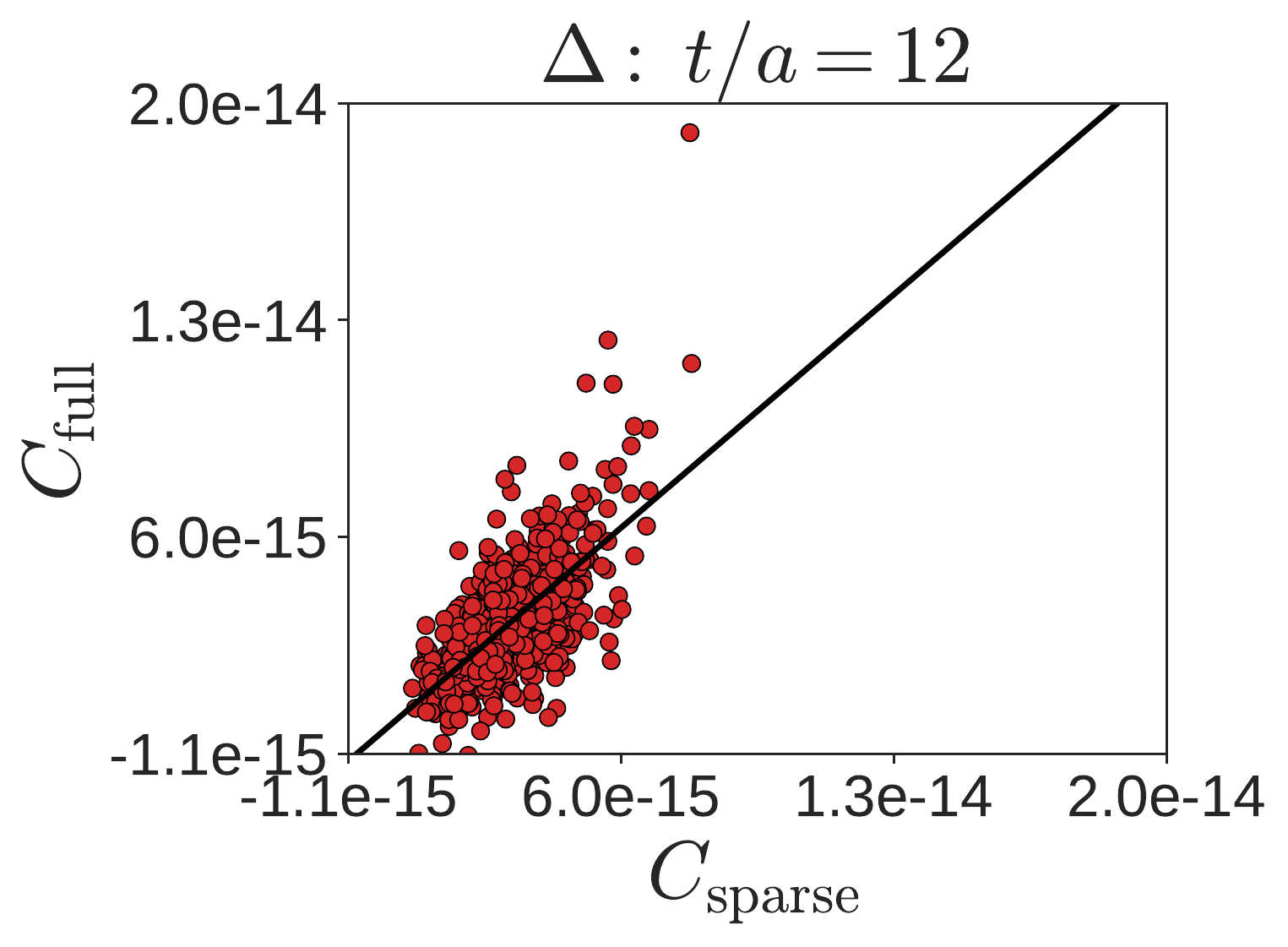}}
  \caption{Scatter plots of source location-averaged sparse two-point correlator data against source location-averaged full two-point correlator data, as well as the corresponding linear regression, for the pion (first row), \( \rho \) meson (second row), nucleon (third row), and \( \Delta \) baryon (fourth row), with fixed Euclidean time separations of 4 (left column), 8 (middle column), and 12 (right column) lattice units. The parameters of each regression are summarized in Table \ref{tab:hadron_regression}.}
\label{fig:hadron_scatter}
\end{figure}

To understand correlations between measurements, linear regressions of the sparsened two-point correlation functions, Eq.~\eqref{eqn:sparse_corr_def}, against the corresponding full two-point correlation functions, Eq.~\eqref{eqn:full_corr_def}, are computed and summarized in Table \ref{tab:hadron_regression} and Figure \ref{fig:hadron_scatter}. For each gauge field configuration used in the calculation, the source location-averaged sparse data is plotted against the source location-averaged full data for a fixed choice of the Euclidean time separation. This procedure is repeated for sink times \( t/a \in \{ 4,8,12 \} \), where the range is chosen to overlap with both the short-time, excited state-dominated regime as well as the late-time, ground state-dominated regime observed in the effective mass plots shown in Figure \ref{fig:hadron_eff_mass}. Results are shown for hadrons at rest; similar correlations are observed for hadrons with non-zero momenta, which were studied for all states with \( \vert \vec{n} \vert \leq \sqrt{5} \).

\begin{table}[!ht]
\setlength{\tabcolsep}{10pt}
\begin{tabular}{ccccc}
\hline
\hline
  \rule{0cm}{0.4cm}\textbf{State} & $\bm{t/a}$ & $\bm{R^{2}}$ & \textbf{Slope} & \textbf{Intercept} \\
\hline
\rule{0cm}{0.4cm}\multirow{3}{*}{$\pi$} & 4 & 0.82 & 0.95(2) & $2.2(0.8) \times 10^{-8}$ \\
 & 8 & 0.85 & 0.97(2) & $1.1(0.6) \times 10^{-9}$ \\
 & 12 & 0.86 & 0.98(2) & $5.8(6.5) \times 10^{-11}$ \\
\hline
\rule{0cm}{0.4cm}\multirow{3}{*}{$\rho$} & 4 & 0.85 & 0.98(2) & $6.0(4.9) \times 10^{-9}$ \\
 & 8 & 0.86 & 1.00(1) & $-0.2(1.8) \times 10^{-10}$ \\
 & 12 & 0.85 & 1.01(2) & $-3.2(8.7) \times 10^{-12}$ \\
\hline
\rule{0cm}{0.4cm}\multirow{3}{*}{$N$} & 4 & 0.52 & 0.97(3) & $6.3(8.6) \times 10^{-13}$ \\
 & 8 & 0.45 & 0.98(6) & $0.6(1.1) \times 10^{-14}$ \\
 & 12 & 0.41 & 1.09(8) & $-1.2(1.3) \times 10^{-16}$ \\
\hline
\rule{0cm}{0.4cm}\multirow{3}{*}{$\Delta$} & 4 & 0.45 & 0.92(4) & $9.6(4.9) \times 10^{-12}$ \\
 & 8 & 0.41 & 0.95(5) & $3.3(2.8) \times 10^{-14}$ \\
 & 12 & 0.44 & 1.06(7) & $-1.1(2.1) \times 10^{-16}$ \\
\hline
\hline
\end{tabular}
  \caption{Coefficients of determination (\(R^{2}\)), slopes, and intercepts for linear regressions of source location-averaged sparse two-point correlator data against source location-averaged full two-point correlator data. The corresponding data sets and fitted models are plotted in Figure \ref{fig:hadron_scatter}.}
\label{tab:hadron_regression}
\end{table}

The full and sparsened data sets are observed to be statistically consistent, as evidenced by regression intercepts which are consistent with zero and regression slopes which are consistent with unity. The degree to which the full data can be described by a linear function of the sparse data is summarized by the coefficient of determination
\begin{equation}
  R^{2} = 1 - \frac{\sum_{\alpha} \left( y_{\alpha} - f_{\alpha} \right)^{2}}{\sum_{\alpha} \left( y_{\alpha} - \overline{y} \right)^{2}},
\end{equation}
where \( y_{\alpha} \) and \( \overline{y} \) are the full correlation function computed on a gauge field configuration indexed by \( \alpha \) and the ensemble average, respectively, and \( f_{\alpha} \) is a linear function of the corresponding sparse correlation function. The extreme limits \( R^{2} = 0 \) and \( R^{2} = 1 \) correspond to no relationship and a perfect linear relationship, respectively. Somewhat larger \(R^{2}\) values are observed for the mesons than for the baryons --- this is expected since the baryon signals are contaminated with more statistical noise.

\subsubsection{Effective Energies}
\label{subsubsec:eff_energies}

\begin{figure}[!ht]
  \centering
  \subfloat{\includegraphics[width=0.48\linewidth]{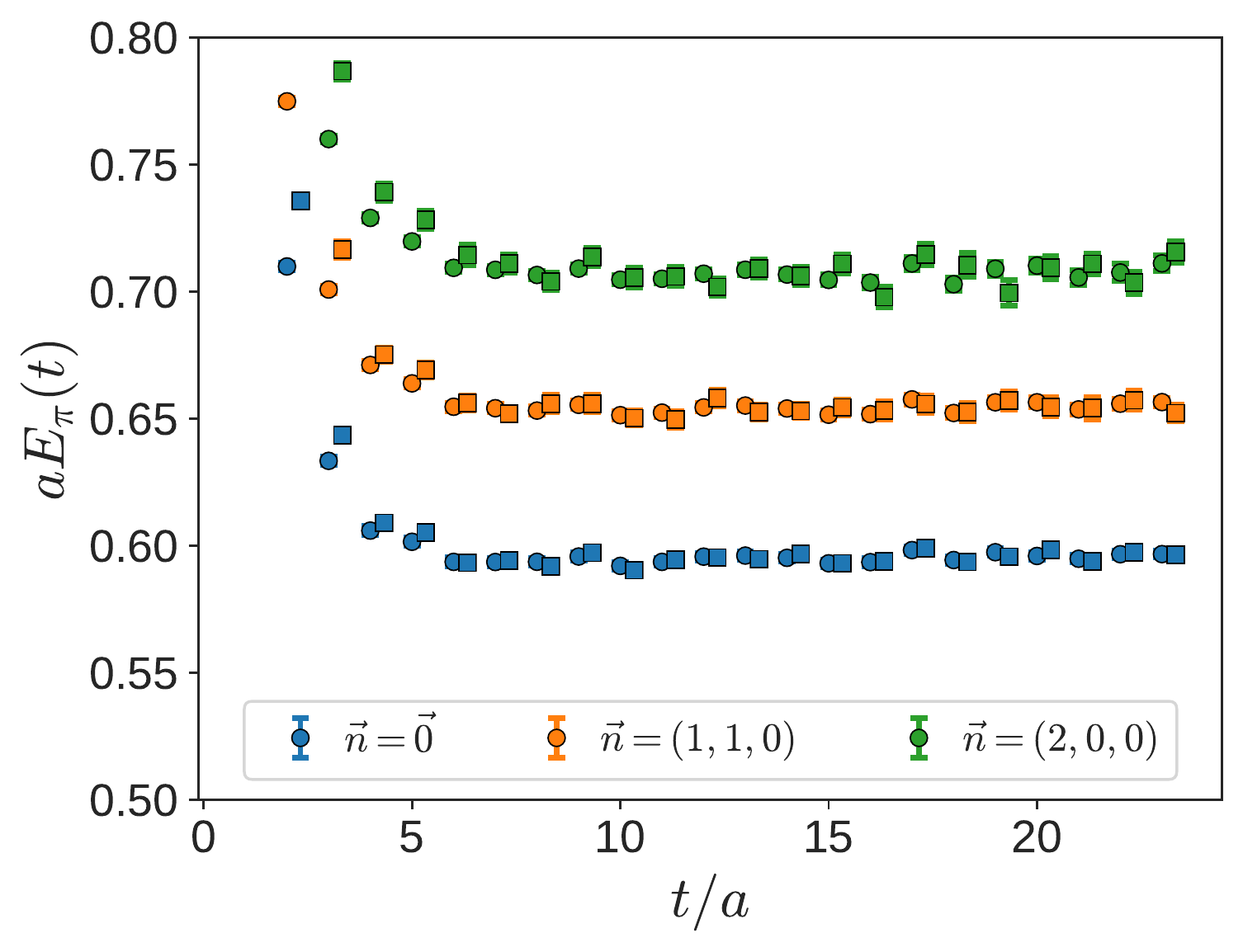}}
  \subfloat{\includegraphics[width=0.48\linewidth]{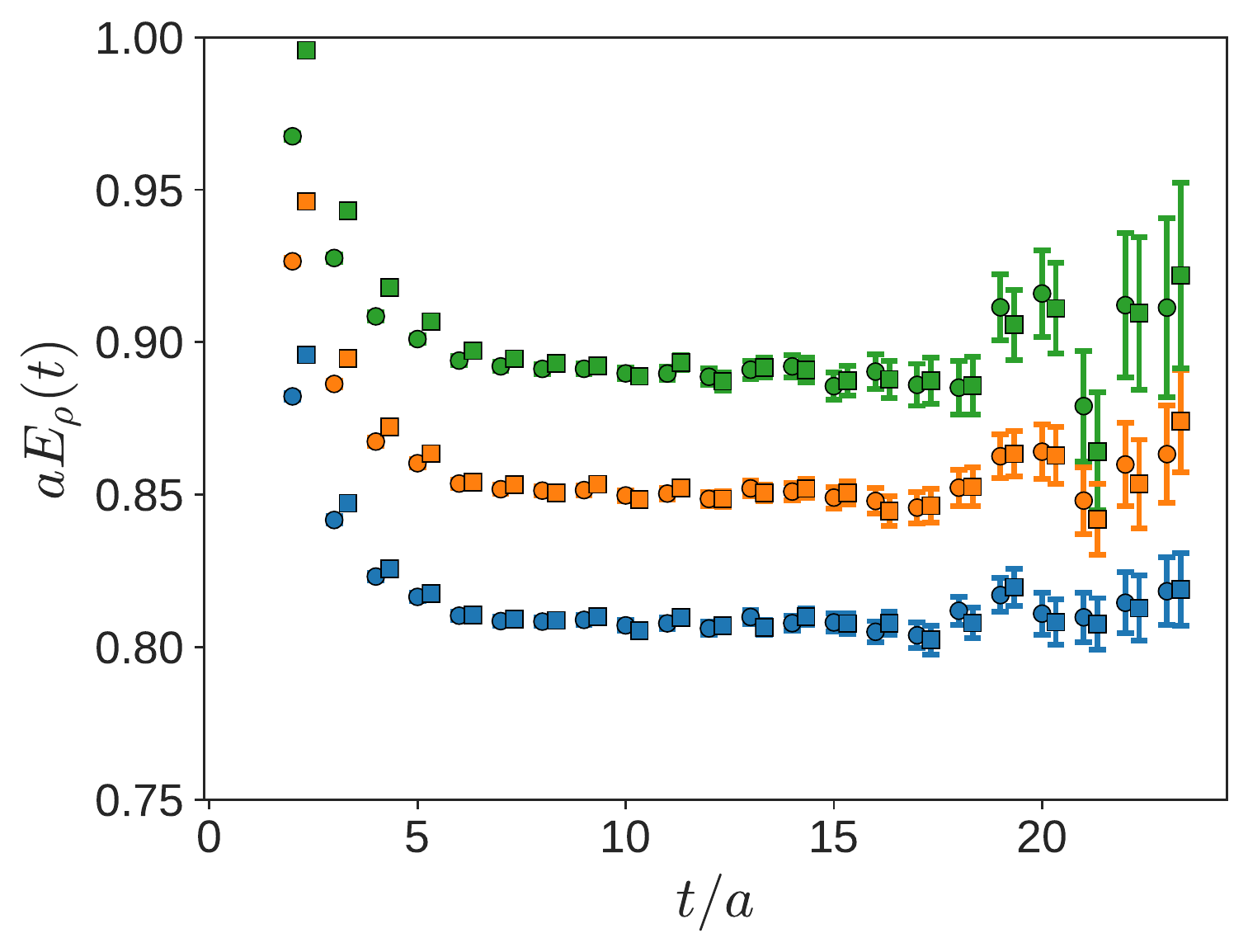}} \\
  \subfloat{\includegraphics[width=0.48\linewidth]{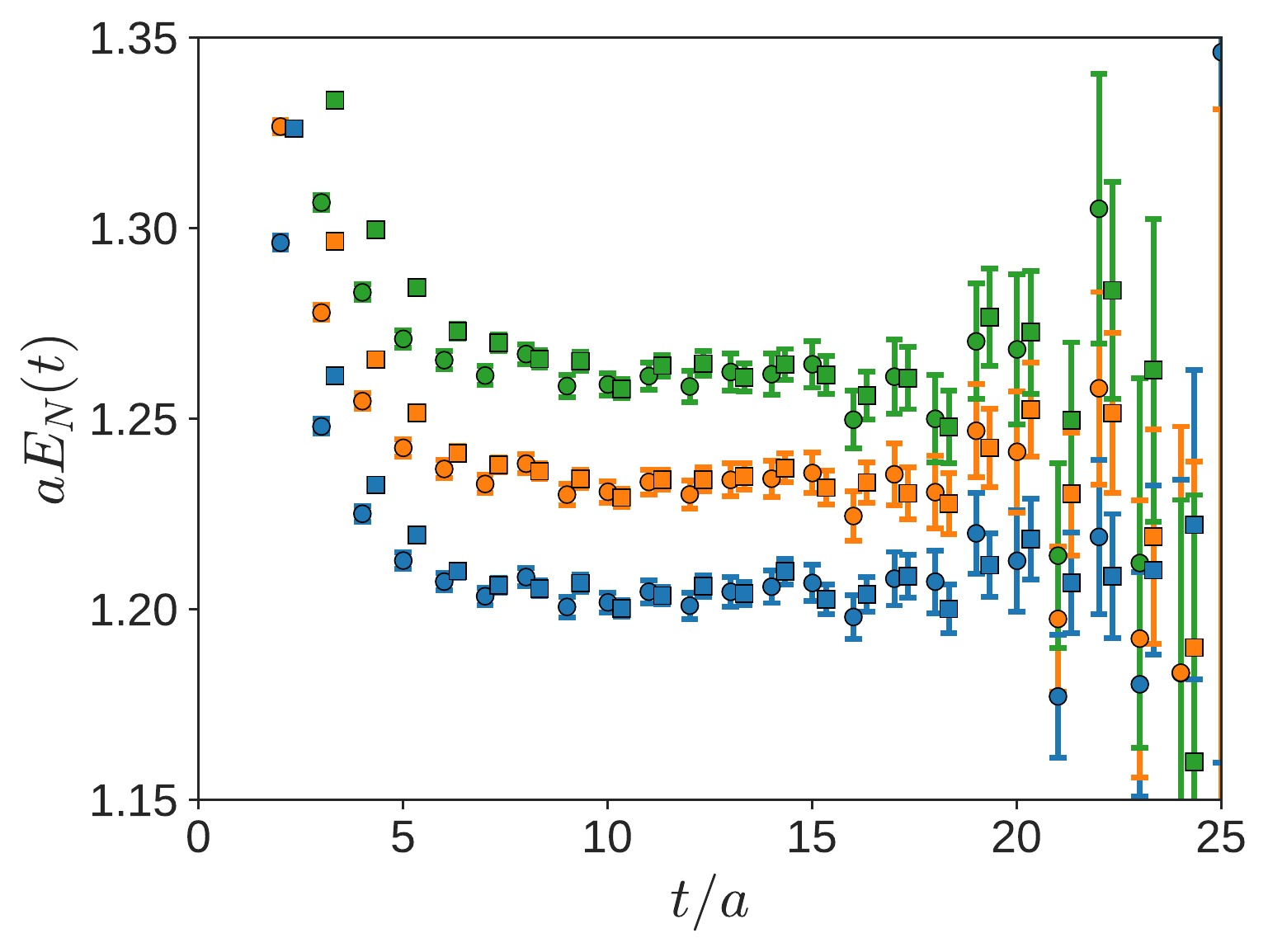}}
  \subfloat{\includegraphics[width=0.48\linewidth]{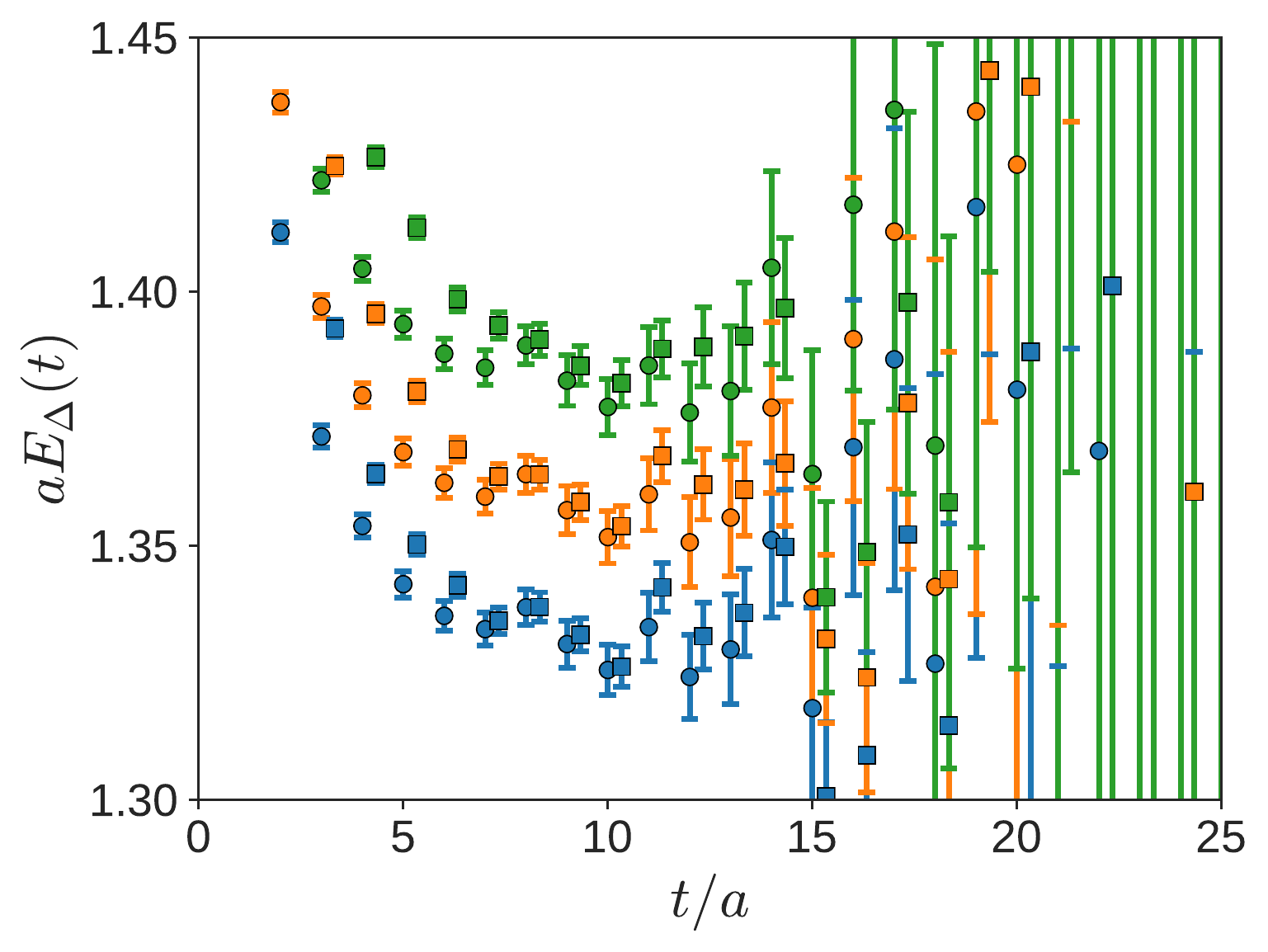}}
  \caption{Effective energies, Eq.~\eqref{eqn:eff_energy}, of the pion (upper left), \( \rho \) meson (upper right), nucleon (lower left), and \( \Delta \) baryon (lower right), with lattice momenta \( \vert \vec{n} \vert^{2} \in \{ 0, 2, 4 \} \). Circles denote data computed from full two-point correlation functions (Eq.~\eqref{eqn:full_corr_def}), whereas squares denote data computed from sparsened two-point correlation functions (Eq.~\eqref{eqn:sparse_corr_def}). The sparsened data has been slightly shifted along the time axis for clarity.}
  \label{fig:hadron_eff_mass}
\end{figure}

Figure \ref{fig:hadron_eff_mass} depicts the effective energy function
\begin{equation}
\label{eqn:eff_energy}
a E_{\rm eff}(t) = 
\begin{dcases}  
  \cosh^{-1} \left[ \frac{C(t-1) + C(t+1)}{2 C(t)} \right], \,\, {\rm mesons} \\
  \sinh^{-1} \left[ \frac{C(t-1) - C(t+1)}{2 C(t)} \right], \,\, {\rm baryons}
\end{dcases}
\end{equation}
of hadrons with lattice momenta \( \vert \vec{n} \vert \in \{0,\sqrt{2},2\} \). At large Euclidean time separations, \( t / a \gg 1\), the effective mass asymptotically approaches the energy of the ground state with the quantum numbers of the interpolating operator from which it is constructed. It is observed that the full and sparsened effective energies reach consistent plateaux for \( t/a \gtrsim 8 \) --- in particular, both the asymptotic value of the effective energy, and the range of Euclidean times over which the effective energy signal exhibits a stable plateau, are consistent --- suggesting that the proposed sparsening preserves the energy and signal quality arising from the ground state contribution.

\begin{figure}[!ht]
\centering
  \subfloat{\includegraphics[width=0.48\linewidth]{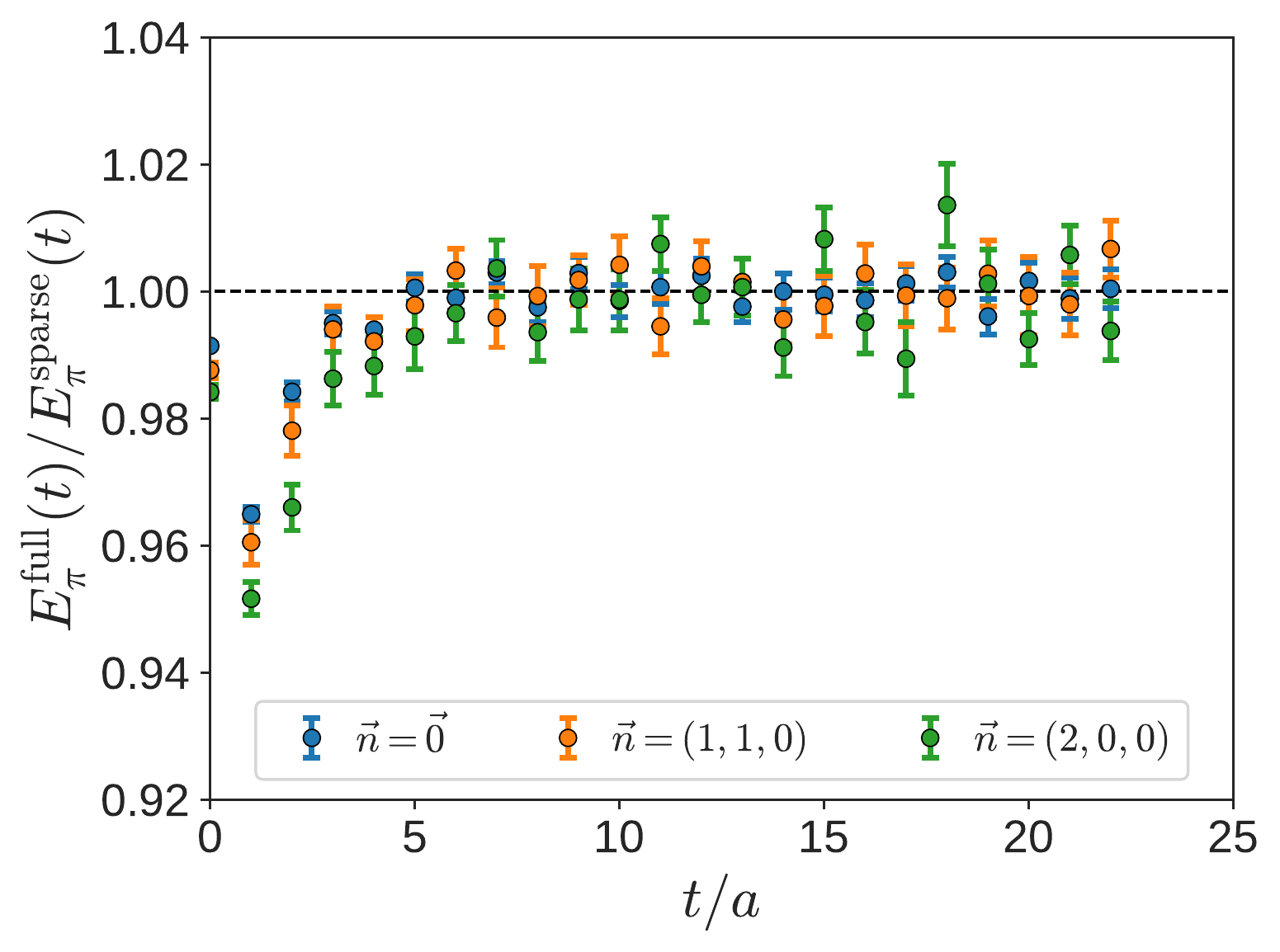}}
  \subfloat{\includegraphics[width=0.48\linewidth]{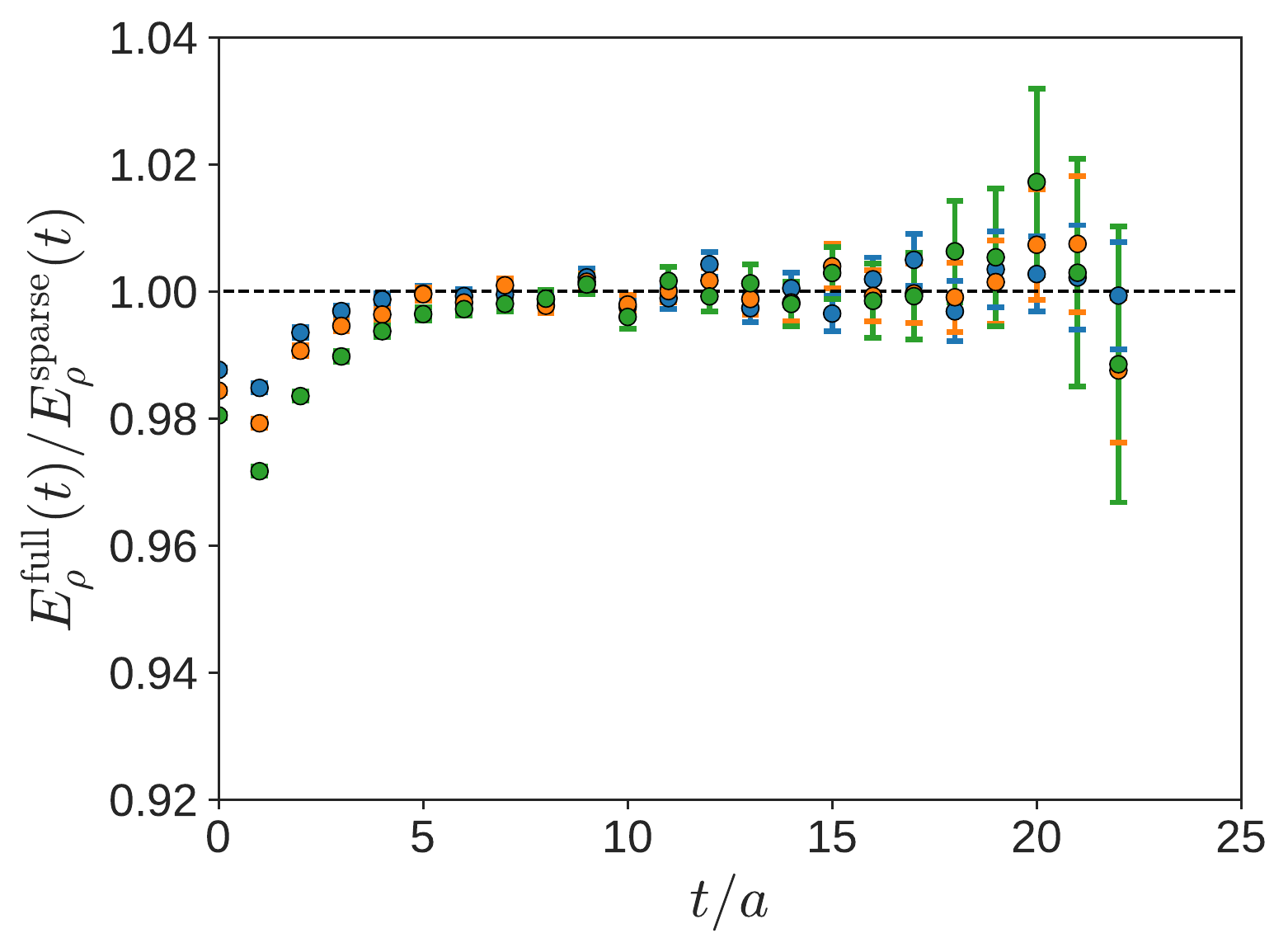}} \\
  \subfloat{\includegraphics[width=0.48\linewidth]{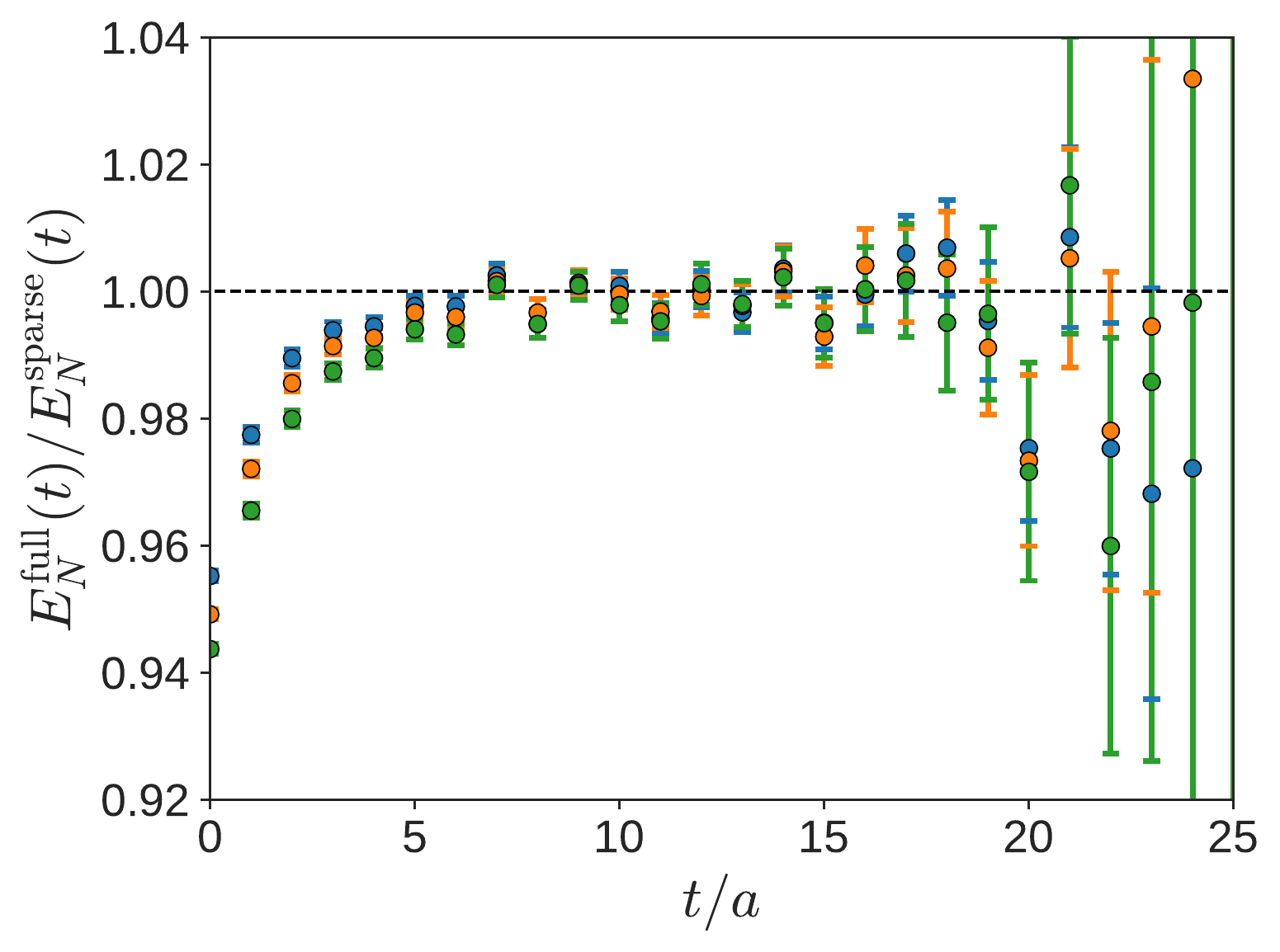}}
  \subfloat{\includegraphics[width=0.48\linewidth]{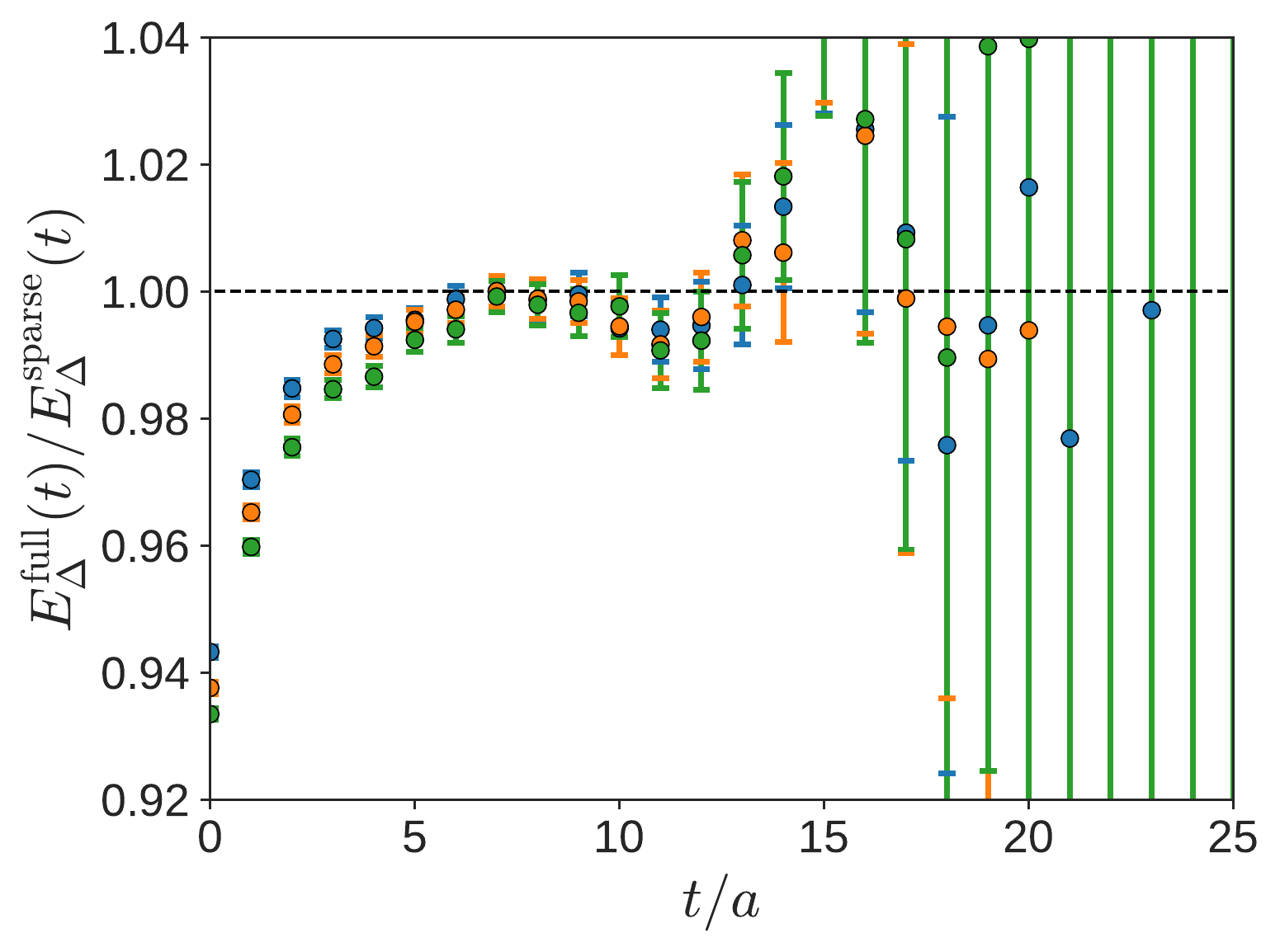}}
  \caption{Correlated ratios of the full and sparsened effective energy signals for the pion (upper left), \( \rho \) meson (upper right), nucleon (lower left), and \( \Delta \) baryon (lower right), with lattice momenta \( \vert \vec{n} \vert^{2} \in \{ 0, 2, 4 \} \).}
\label{fig:hadron_eff_mass_ratios}
\end{figure}

At small Euclidean time separations \( t/a \lesssim 6 \), the effective mass is contaminated by contributions from higher energy excited states, which are exponentially suppressed in \( t \). In this regime, there are statistically significant deviations between the full and sparsened results; this deviation is further emphasized in Figure \ref{fig:hadron_eff_mass_ratios}, which shows the correlated ratio of the full and sparsened effective mass signals. From Figures \ref{fig:hadron_eff_mass} and \ref{fig:hadron_eff_mass_ratios}, it is possible to infer a few effects of sparsening on the coupling to the excited state spectrum of QCD: first, that excited state contamination is more prominent in the sparsened signals, as evidenced by the larger deviations from the asymptotic ground state plateau at early times in Figure \ref{fig:hadron_eff_mass}, and second, that these deviations are also more prominent when projecting onto states with higher lattice momenta, as observed in Figure \ref{fig:hadron_eff_mass_ratios}.

\subsubsection{Ground State Energy Extraction}
\label{subsubsec:hadron_fits}

Ground state energies can be extracted from two-point correlation functions by determining the parameters \( \vec{\beta} \) minimizing
\begin{equation}
  \label{eqn:correlated_chi2pdof}
  \chi^{2} ( \vec{\beta} ) = \sum_{t_{i} \in T_{\rm fit}} \sum_{t_{j} \in T_{\rm fit}} \big( C ( \vec{p}, t_{i} ) - f ( t_{i}; \vec{\beta} ) \big) \left( \Sigma^{-1} \right)_{ij} \big( C ( \vec{p}, t_{j} ) - f ( t_{j}; \vec{\beta} ) \big),
\end{equation}
where
\begin{equation}
  \label{eqn:cov_matrix}
  \Sigma_{ij} = \big\langle \left( C_{\alpha}(\vec{p},t_{i}) - C(\vec{p},t_{i}) \right) \left( C_{\alpha}(\vec{p},t_{j}) - C(\vec{p},t_{j}) \right) \big\rangle_{\alpha}
\end{equation}
is the covariance matrix describing correlations between time slices, \( C_{\alpha}(\vec{p},t) \) is the full or sparsened two-point function computed on a fixed gauge field configuration indexed by \( \alpha \) and averaged over all source locations, and \( f(t;\vec{\beta}) \) is an appropriate fit ansatz. In this notation \( T_{\rm fit} \) denotes the range of time separations included in the fit, \( \langle \cdots \rangle_{\alpha} \) denotes an ensemble average across measurements on independent gauge field configurations, and \( C(\vec{p},t) = \left\langle C_{\alpha}(\vec{p},t) \right\rangle_{\alpha} \). For simplicity, \( T_{\rm fit} \) is chosen to lie inside the asymptotic plateau regions exhibited in Figure \ref{fig:hadron_eff_mass}, where the correlation function is saturated by the ground state contribution, and the ans\"{a}tze are simple, single-exponential forms, with
\begin{equation}
f(t; Z_{\rm src}, Z_{\rm snk}, E) = 
\begin{dcases}
  \frac{Z_{\rm src} Z_{\rm snk}^{*}}{2 E} \left( e^{- E t} + e^{-E \left( T - t \right)} \right), \,\, {\rm mesons} \\
  \frac{Z_{\rm src} Z_{\rm snk}^{*}}{2 E} e^{- E t}, \,\, {\rm baryons}
\end{dcases}
\end{equation}
where \( T/a = 48 \) is the temporal extent of the lattice. The results of these fits are summarized in Table \ref{tab:hadron_fit_results}: it is observed for all hadron species and momenta that the extracted ground state energies are consistent --- both in terms of the central values and statistical uncertainties --- whether the fits are performed to full or sparsened correlation functions.

\begin{table}[!h]
\setlength{\tabcolsep}{6pt}
  \resizebox{\linewidth}{!}{
\begin{tabular}{c|c|c|ccc|ccc}
\hline
\hline
  \rule{0cm}{0.4cm}& & & & \textbf{Full} & & & \textbf{Sparse} & \\
  \textbf{State} & $\bm{T_{\rm fit}}$ & $\bm{\vec{n}}$ & $\bm{a E}$ & $\bm{\chi^{2}/}$\textbf{dof} & $\bm{\kappa \left( \Sigma \right)}$ & $\bm{a E}$ & $\bm{\chi^{2}/}$\textbf{dof} & $\bm{\kappa \left( \Sigma \right)}$ \\
\hline
  \rule{0cm}{0.4cm}\multirow{6}{*}{$\pi$} & \multirow{6}{*}{$[7,18]$} & (0,0,0) & $0.59477(30)$ & $1.30(71)$ & $3.83 \times 10^{6}$ & $0.59471(30)$ & $1.39(74)$ & $2.56 \times 10^{6}$ \\
  & & (1,0,0) & $0.62505(36)$ & $1.31(72)$ & $3.50 \times 10^{8}$ & $0.62500(35)$ & $1.76(80)$ & $4.58 \times 10^{7}$ \\
  & & (1,1,0) & $0.65368(38)$ & $1.32(72)$ & $4.95 \times 10^{8}$ & $0.65368(36)$ & $1.90(84)$ & $8.83 \times 10^{7}$ \\
  & & (1,1,1) & $0.68102(41)$ & $1.38(74)$ & $6.60 \times 10^{8}$ & $0.68086(38)$ & $1.06(63)$ & $1.58 \times 10^{8}$ \\
  & & (2,0,0) & $0.70648(44)$ & $1.21(69)$ & $7.92 \times 10^{8}$ & $0.70651(42)$ & $1.43(72)$ & $1.70 \times 10^{8}$ \\
  & & (2,1,0) & $0.73169(48)$ & $1.36(74)$ & $9.83 \times 10^{8}$ & $0.73225(44)$ & $1.28(69)$ & $2.19 \times 10^{8}$ \\
\hline
  \rule{0cm}{0.4cm}\multirow{6}{*}{$\rho$} & \multirow{6}{*}{$[7,18]$} & (0,0,0) & $0.80795(55)$ & $0.51(45)$ & $5.37 \times 10^{7}$ & $0.80805(54)$ & $0.58(48)$ & $3.81 \times 10^{7}$ \\
  & & (1,0,0) & $0.82954(56)$ & $0.52(46)$ & $7.10 \times 10^{7}$ & $0.82972(55)$ & $0.57(48)$ & $5.28 \times 10^{7}$ \\
  & & (1,1,0) & $0.85050(59)$ & $0.46(43)$ & $8.41 \times 10^{7}$ & $0.85066(57)$ & $0.77(55)$ & $6.12 \times 10^{7}$ \\
  & & (1,1,1) & $0.87087(61)$ & $0.40(40)$ & $9.31 \times 10^{7}$ & $0.87110(59)$ & $0.48(44)$ & $7.28 \times 10^{7}$ \\
  & & (2,0,0) & $0.89025(63)$ & $0.39(40)$ & $1.00 \times 10^{8}$ & $0.89103(62)$ & $0.71(53)$ & $7.33 \times 10^{7}$ \\
  & & (2,1,0) & $0.90952(66)$ & $0.38(39)$ & $1.07 \times 10^{8}$ & $0.91029(65)$ & $0.50(45)$ & $7.89 \times 10^{7}$ \\
\hline
  \rule{0cm}{0.4cm}\multirow{6}{*}{$N$} & \multirow{6}{*}{$[10,17]$} & (0,0,0) & $1.2039(20)$ & $0.65(65)$ & $5.08 \times 10^{7}$ & $1.2054(13)$ & $0.68(67)$ & $3.18 \times 10^{7}$ \\
  & & (1,0,0) & $1.2183(20)$ & $0.64(65)$ & $5.49 \times 10^{7}$ & $1.2195(14)$ & $0.59(62)$ & $3.23 \times 10^{7}$ \\
  & & (1,1,0) & $1.2326(21)$ & $0.62(64)$ & $5.84 \times 10^{7}$ & $1.2342(14)$ & $0.19(36)$ & $3.46 \times 10^{7}$ \\
  & & (1,1,1) & $1.2467(22)$ & $0.57(61)$ & $6.11 \times 10^{7}$ & $1.2486(15)$ & $0.51(58)$ & $3.98 \times 10^{7}$ \\
  & & (2,0,0) & $1.2604(23)$ & $0.58(62)$ & $6.08 \times 10^{7}$ & $1.2629(16)$ & $0.36(49)$ & $3.77 \times 10^{7}$ \\
  & & (2,1,0) & $1.2742(24)$ & $0.50(58)$ & $6.22 \times 10^{7}$ & $1.2763(16)$ & $0.68(67)$ & $3.85 \times 10^{7}$ \\
\hline
  \rule{0cm}{0.4cm}\multirow{6}{*}{$\Delta$} & \multirow{6}{*}{$[7,13]$} & (0,0,0) & $1.3329(26)$ & $1.17(97)$ & $6.04 \times 10^{6}$ & $1.3342(17)$ & $1.5(1.1)$ & $4.35 \times 10^{6}$ \\
  & & (1,0,0) & $1.3461(26)$ & $1.10(94)$ & $6.55 \times 10^{6}$ & $1.3477(17)$ & $2.0(1.3)$ & $4.89 \times 10^{6}$ \\
  & & (1,1,0) & $1.3591(26)$ & $1.04(91)$ & $7.03 \times 10^{6}$ & $1.3609(17)$ & $1.16(96)$ & $5.36 \times 10^{6}$ \\
  & & (1,1,1) & $1.3720(27)$ & $0.98(89)$ & $7.46 \times 10^{6}$ & $1.3740(18)$ & $1.3(1.0)$ & $5.53 \times 10^{6}$ \\
  & & (2,0,0) & $1.3846(27)$ & $0.88(84)$ & $7.81 \times 10^{6}$ & $1.3874(18)$ & $0.66(73)$ & $5.76 \times 10^{6}$ \\
  & & (2,1,0) & $1.3972(28)$ & $0.83(82)$ & $8.15 \times 10^{6}$ & $1.4001(19)$ & $0.94(87)$ & $6.04 \times 10^{6}$ \\
\hline
\hline
\end{tabular}
  }
  \caption{Summary of fits to extract the ground state energies of the pion, \( \rho \) meson, nucleon, and \( \Delta \) baryon. \(T_{\rm fit}\) denotes the range of Euclidean times included in the fit in lattice units, \( \vert \vec{n} \vert \) is the wavenumber describing the total momentum carried by the hadron, \( a E \) is the extracted ground state energy, \( \chi^{2} / \)dof is obtained by minimizing Eq.~\eqref{eqn:correlated_chi2pdof}, and \( \kappa(\Sigma) \) denotes the condition number of the covariance matrix (Eq.~\eqref{eqn:cov_matrix}). The first set of fit results (middle three columns) are from fits to full correlation functions (Eq.~\eqref{eqn:full_corr_def}), while the second set of fit results (rightmost three columns) are from fits to sparsened correlation functions (Eq.~\eqref{eqn:sparse_corr_def}). The statistical uncertainties of fitted quantities are computed using the jackknife resampling technique.}
\label{tab:hadron_fit_results}
\end{table}

\subsection{Excited States in Sparsened Correlation Functions}
\label{subsec:excited_states}

In the preceding subsections, it was observed that while sparsening can modify the couplings to excited states in the early Euclidean time regime of lattice two-point correlators, this had no statistically significant effect on the extraction of ground-state hadron energies. It is conceivable, however, that in other calculations which are more sensitive to short-distance effects --- for example, spectroscopic calculations using a large, variational basis of interpolating operators to extract ground and excited state energies --- modified couplings to higher energy states may be more of a concern, especially if the couplings are enhanced or if couplings to additional excited states not present in the full correlation functions are induced. One way to systematically control this is to modify the form of the sparsened estimator for the two-point correlation function:
\begin{equation}
  \label{eqn:improved_sparse_corr}
  \widetilde{C}(\vec{p},t) = \frac{1}{N_{\rm sparse}} \sum_{x_{0} \in \Lambda_{\rm sparse}} C_{\rm sparse}(\vec{p},t;x_{0}) + \frac{1}{N_{\rm \Delta}} \sum_{x_{0}' \in \Lambda_{\rm \Delta}} \big( C_{\rm full}(\vec{p},t;x_{0}') - C_{\rm sparse}(\vec{p},t;x_{0}') \big).
\end{equation}
For this construction to be useful in practice, one must be able to determine the modified estimator precisely at low cost. The proposed idea, similar in spirit to the all-mode averaging technique introduced in Ref.~\cite{Shintani:2014vja}, is to compute the inexpensive, sparsened correlation functions by averaging over \( N_{\rm sparse} \) independent propagator source locations \( x_{0} \in \Lambda_{\rm sparse} \), as well as the full correlation functions by averaging over a smaller subset of \( N_{\rm \Delta} \) propagator source locations with \( \Lambda_{\rm \Delta} \subset \Lambda_{\rm sparse} \). One can then form the estimator of Eq.~\eqref{eqn:improved_sparse_corr}, where the second term interpolates between Eq.~\eqref{eqn:sparse_corr_def} (\(N_{\Delta} = 0\)) and Eq.~\eqref{eqn:full_corr_def} (\(N_{\Delta} = N_{\rm sparse}\)). This can reduce the additional excited state contamination while still leading to significant cost reductions in practice, provided the observed differences can be effectively removed when \( N_{\Delta} \ll N_{\rm sparse} \).

\begin{figure}[!ht]
\centering
  \subfloat{\includegraphics[width=0.48\linewidth]{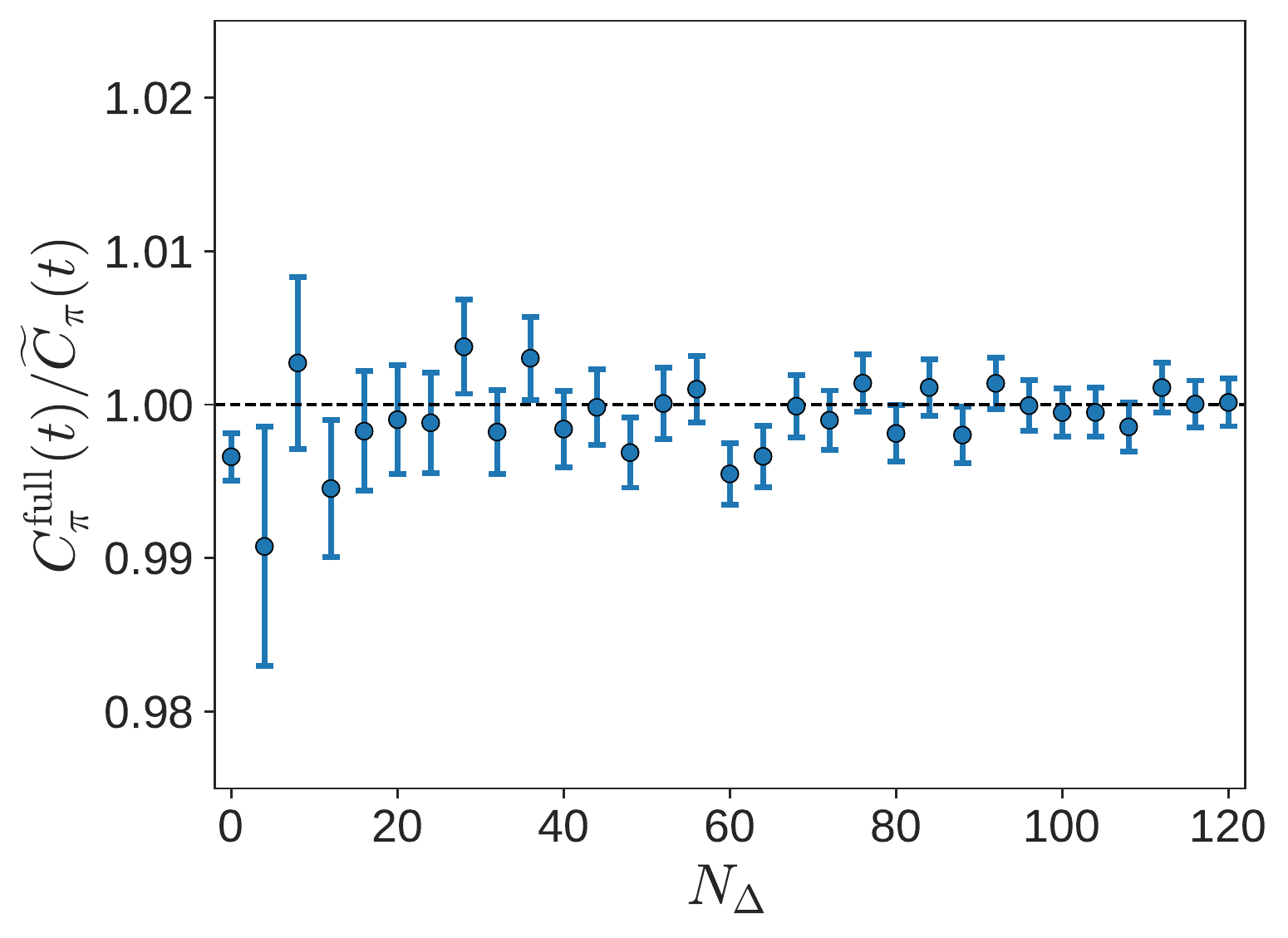}}
  \subfloat{\includegraphics[width=0.48\linewidth]{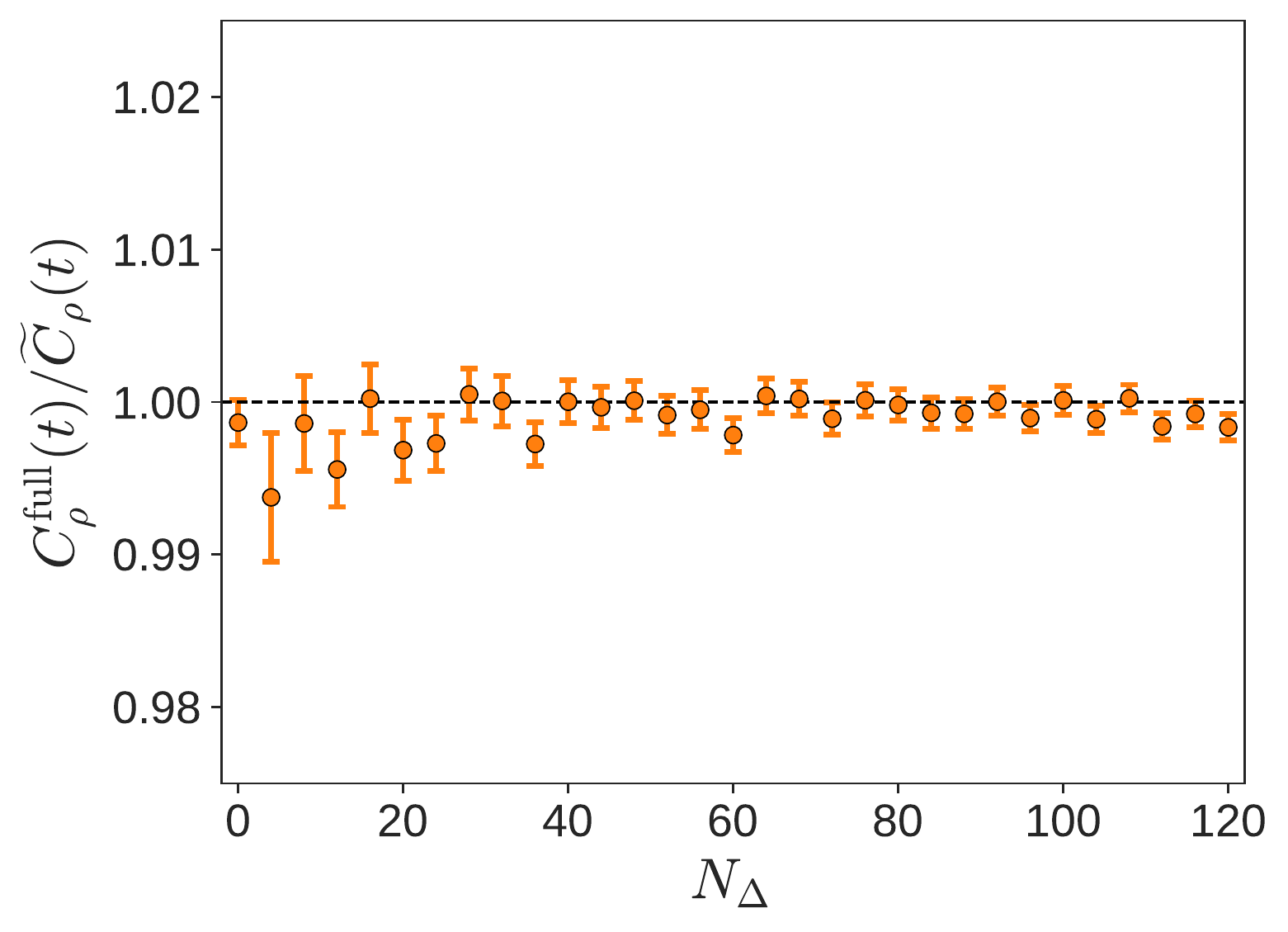}} \\
  \subfloat{\includegraphics[width=0.48\linewidth]{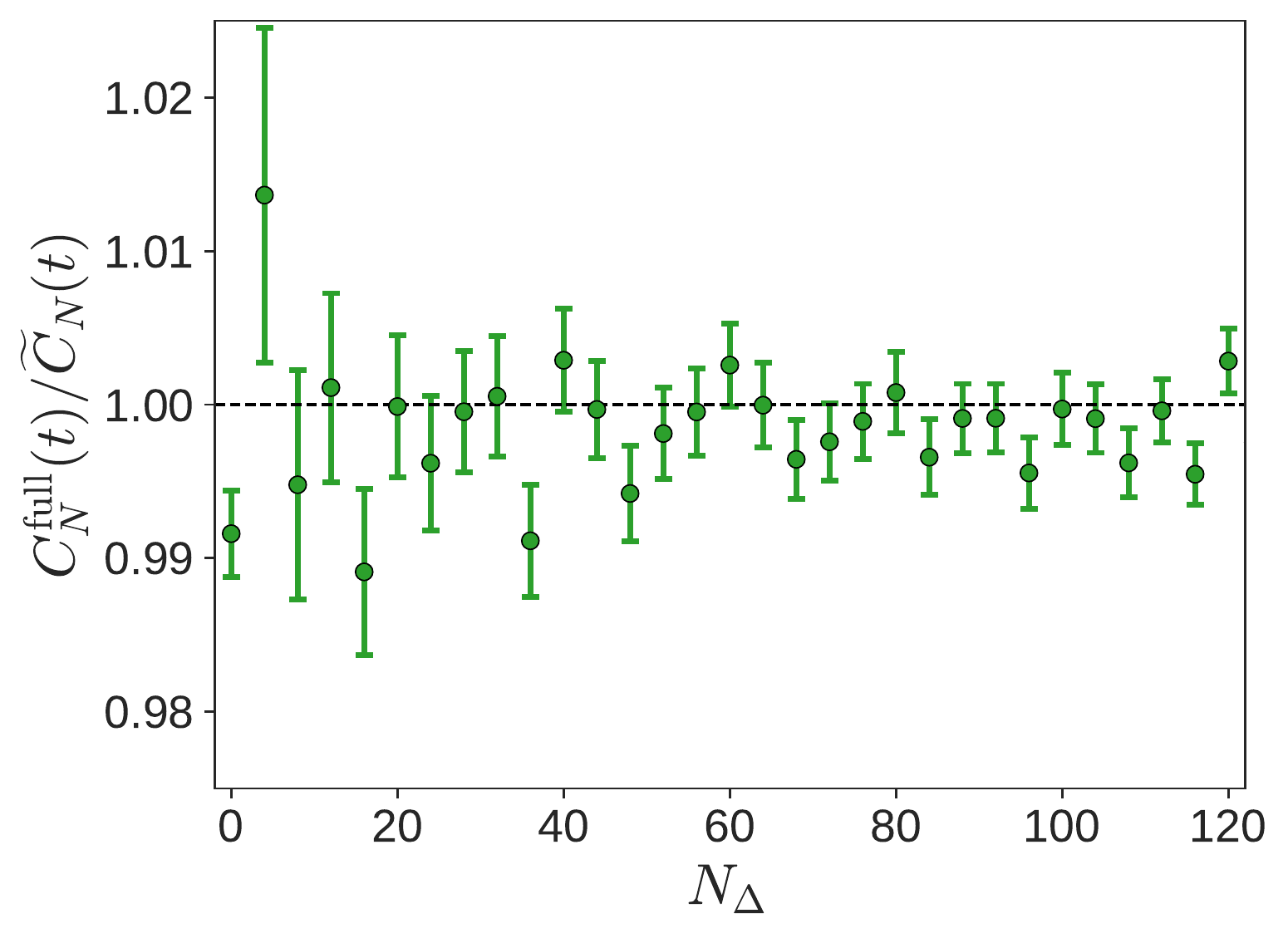}}
  \subfloat{\includegraphics[width=0.48\linewidth]{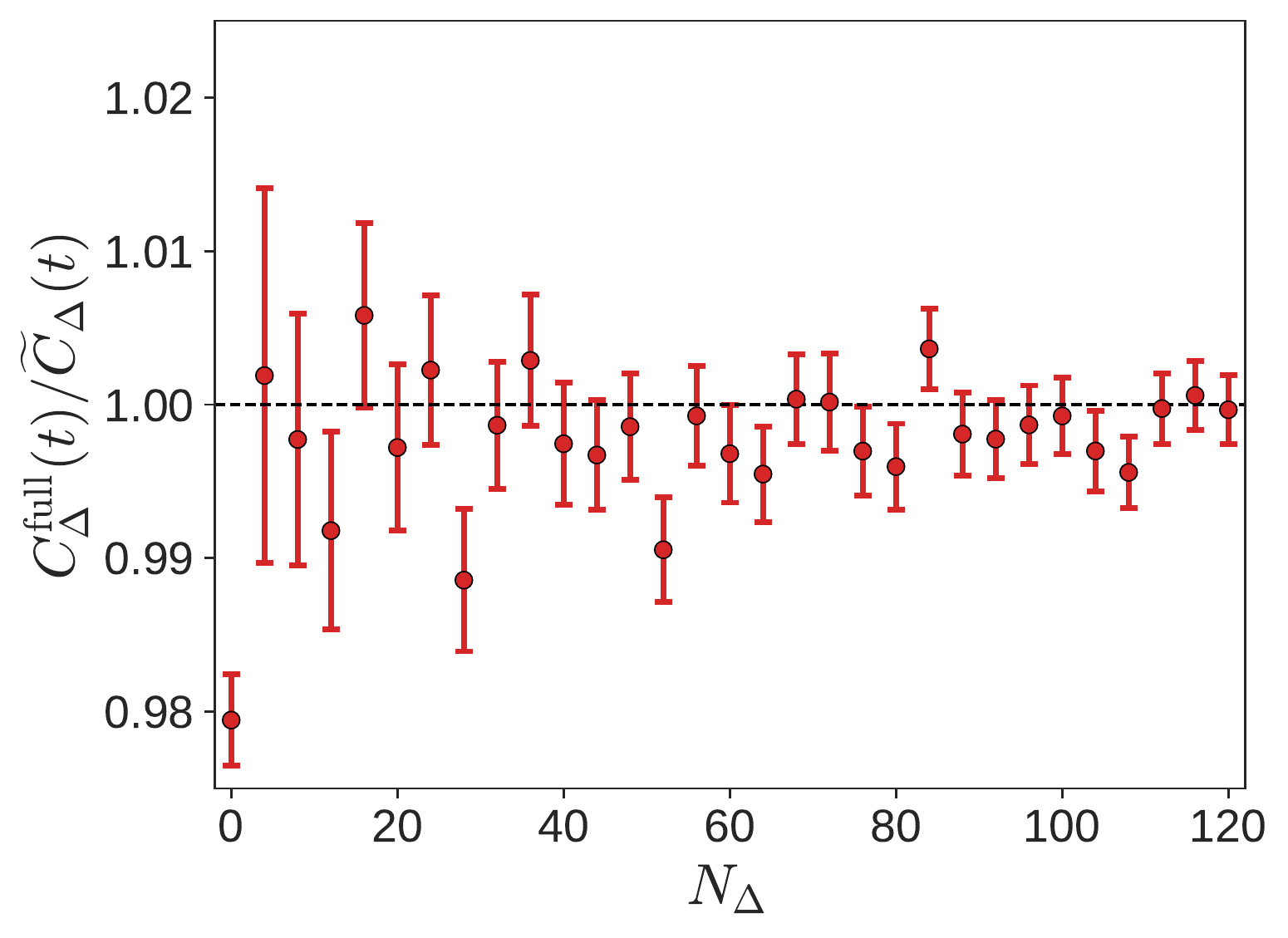}}
  \caption{Correlated ratios of the full (Eq.~\eqref{eqn:full_corr_def}) and modified sparse (Eq.~\eqref{eqn:improved_sparse_corr}) two-point correlators for the pion (upper left), \( \rho \) meson (upper right), nucleon (lower left), and \( \Delta \) baryon (lower right), as a function of \(N_{\Delta}\). The Euclidean time separation in the two-point correlators is held fixed at \(t/a=3\).}
\label{fig:hadron_excited_states}
\end{figure}

Figure \ref{fig:hadron_excited_states} shows the ratio of the full two-point correlator (Eq.~\eqref{eqn:full_corr_def}) to the modified sparse estimator \(\widetilde{C}\) (Eq.~\eqref{eqn:improved_sparse_corr}) as a function of \( N_{\rm \Delta} \), with \(t/a = 3\) held fixed. The full set of source locations is used for \(\Lambda_{\rm sparse}\) (\(N_{\rm sparse} = 123\)), and, for each value of \(N_{\Delta}\), a random subset \(\Lambda_{\rm \Delta}\) is drawn independently to compute the correction term.

While, in this study, a small choice of \(N_{\Delta}\) is sufficient to remove the additional excited state contamination observed in the sparsened correlation functions, this comes at the cost of inflating the statistical error. As \(N_{\Delta}\) is increased the correlated ratio asymptotically approaches a regime where it is consistent with unity and no inflation of the statistical uncertainty is observed. However, the reader should be cautioned against inferring too much from the dependence on \( N_{\Delta} \) in Figure \ref{fig:hadron_excited_states}. In this study \( \Lambda_{\Delta} \) has been drawn from a collection of closely spaced, and thus highly correlated, propagators, with all sources on a single time slice. If the propagator source locations were instead distributed randomly throughout the lattice, it is likely that the modified estimator would converge more quickly in \( N_{\Delta} \). Verifying this conjecture is left for future work.

\subsection{Sparsened Nuclear Correlation Functions}
\label{subsec:nuclei}

While the results for single hadron correlation functions described in the previous section are encouraging for the use of the sparsening technique proposed in this work, the necessary quark contractions are also inexpensive, and thus there is no clear scenario where this technique might be useful in practice. Computing correlation functions for nuclear systems composed of multiple hadrons, however, quickly becomes computationally challenging, and requires the use of more sophisticated techniques such as the baryon block algorithm \cite{Basak:2005ir,Beane:2005rj,Beane:2006mx,Beane:2008dv,Doi:2012xd,Detmold:2012eu} described in Section \ref{sec:methods}. A typical calculation involves constructing and combining many such blocks for different choices of source and sink smearings and locations, which is often the dominant cost in the entire workflow. This is further compounded in calculations which employ background field methods to compute matrix elements involving current insertions, since one must also compute blocks for multiple values of the background field strength \cite{Savage:2016kon}. By drastically reducing the cost of building baryon blocks, sparsening can either help to reduce the overall computational cost of such calculations, or else enable the use of a much larger basis of interpolating operators, smearings, and background fields at fixed computational cost.

This section investigates the effects of sparsening on the extraction of ground state energies of more complicated bound states consisting of multiple nucleons, in analogy to Section \ref{subsubsec:hadron_fits}. The states considered include the \(^{1}S_{0}\) and \(^{3}S_{1}\) \(NN\) bound states\footnote{At the heavy, \(SU(3)\)-symmetric quark mass point used in these calculations, the dinucleon state is observed to be bound \cite{Beane:2012vq}, unlike in nature.} (dinucleon and deuteron, respectively), and the \(^{3}{\rm He}\) and \(^{4}{\rm He}\) isotopes of Helium, and have been previously studied in Refs.~\cite{Beane:2012vq,Beane:2013br,Wagman:2017tmp}. Two classes of fits are performed. In the first, the energies of these states are extracted directly by fitting an exponential ansatz to the Euclidean time dependence of the two-point correlation function. In the second, the binding energies of each state are instead extracted from an exponential fit to a suitable correlated ratio: for a bound state of \( A \) nucleons the ratio used is
\begin{equation}
\label{eqn:binding_energy_ratio}
  R_{A}(t) = \frac{C_{A}(t)}{ \left[ C_{N}(t) \right]^{A} } \stackrel{t \gg 1}{\propto} \exp \left( - \Delta E t \right),
\end{equation}
where \(C_{A}(t)\) is the multi-hadron two-point correlation function, \(C_{N}(t)\) is the single nucleon two-point correlation function, and \( \Delta E \equiv E_{A} - A E_{N} \) is the binding energy. The advantage of the ratio \(R_{A}(t)\) is that it naturally accounts for the strong correlations in the statistical fluctuations of \(C_{A}(t)\) and \(C_{N}(t)\), which must be taken into account to properly determine the statistical uncertainty of \( \Delta E \). In addition, \( t \) must be chosen sufficiently large that both \( C_{A}(t) \) and \( C_{N}(t) \) are ground state dominated.

\begin{figure}[!ht]
\centering
  \subfloat{\includegraphics[width=0.3\linewidth]{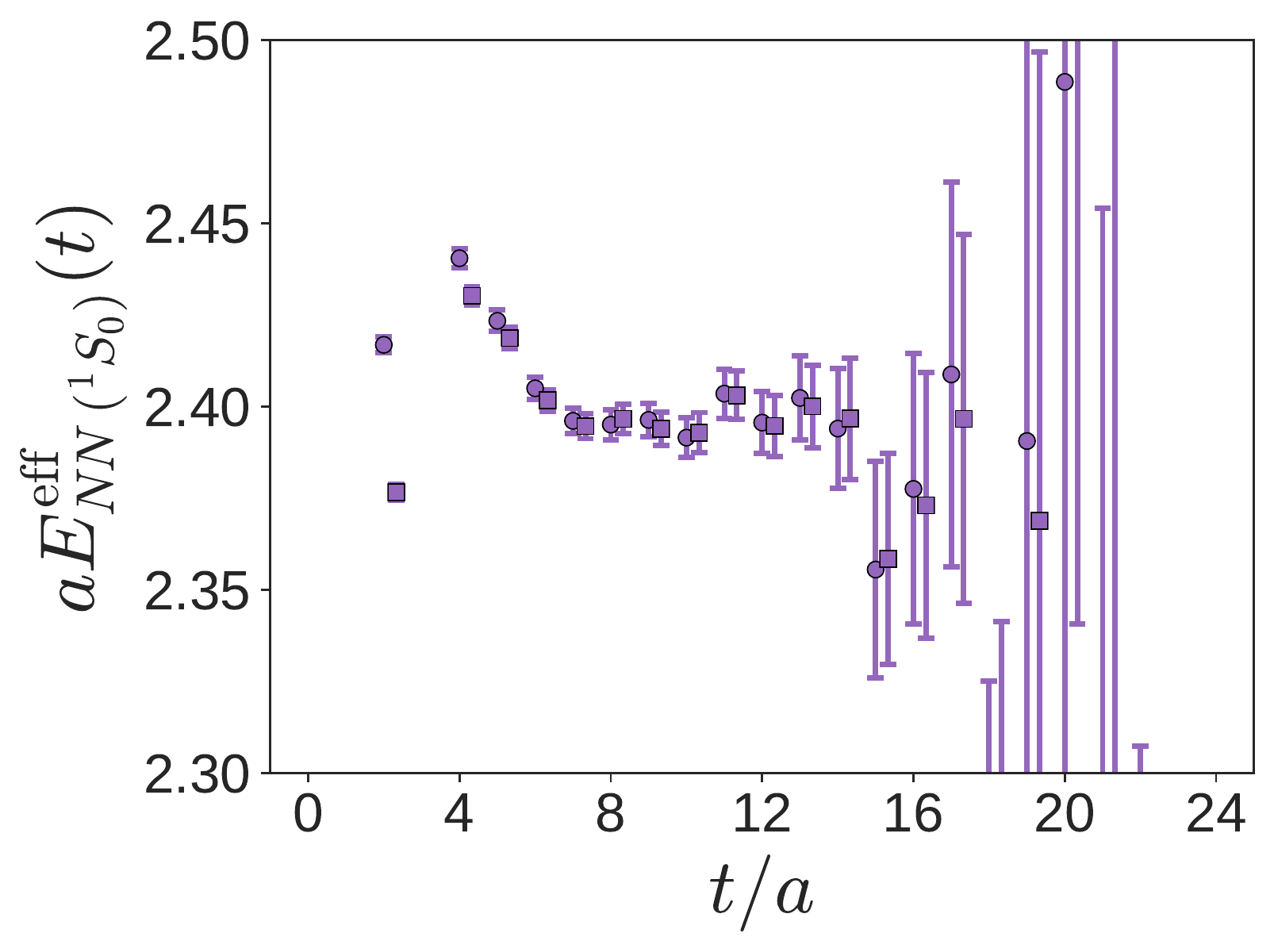}}
  \subfloat{\includegraphics[width=0.3\linewidth]{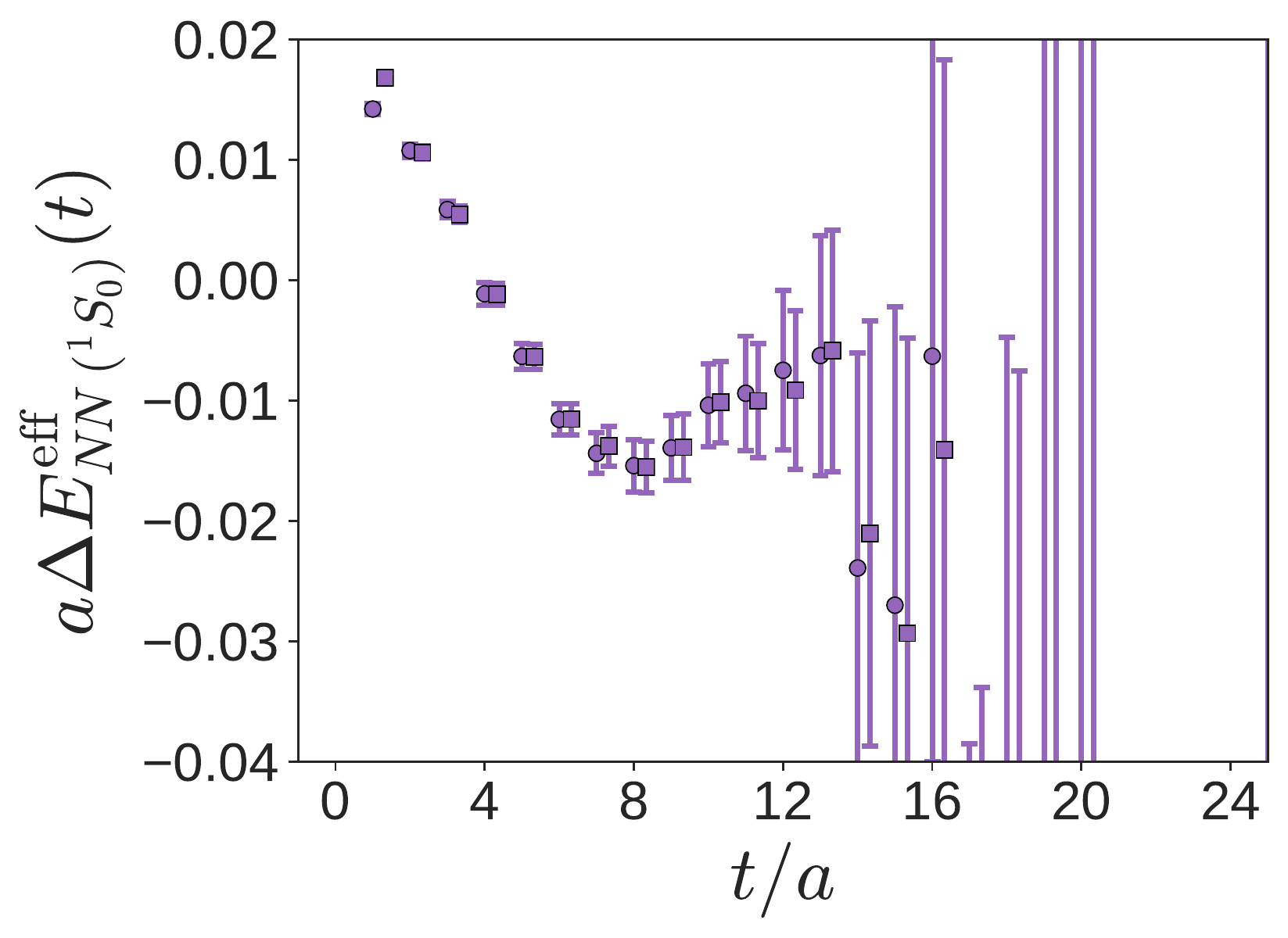}}
  \subfloat{\includegraphics[width=0.3\linewidth]{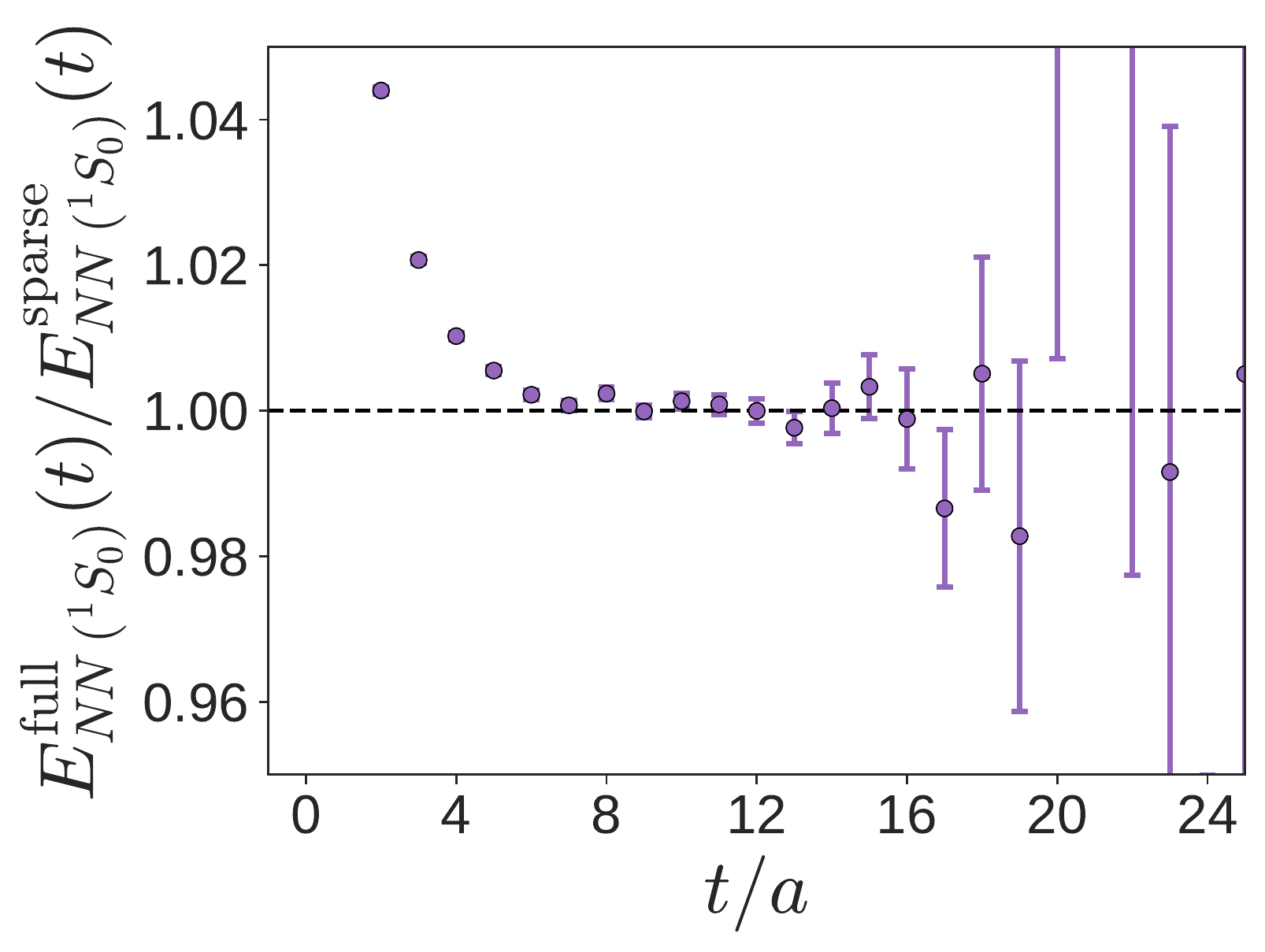}} \\
  \subfloat{\includegraphics[width=0.3\linewidth]{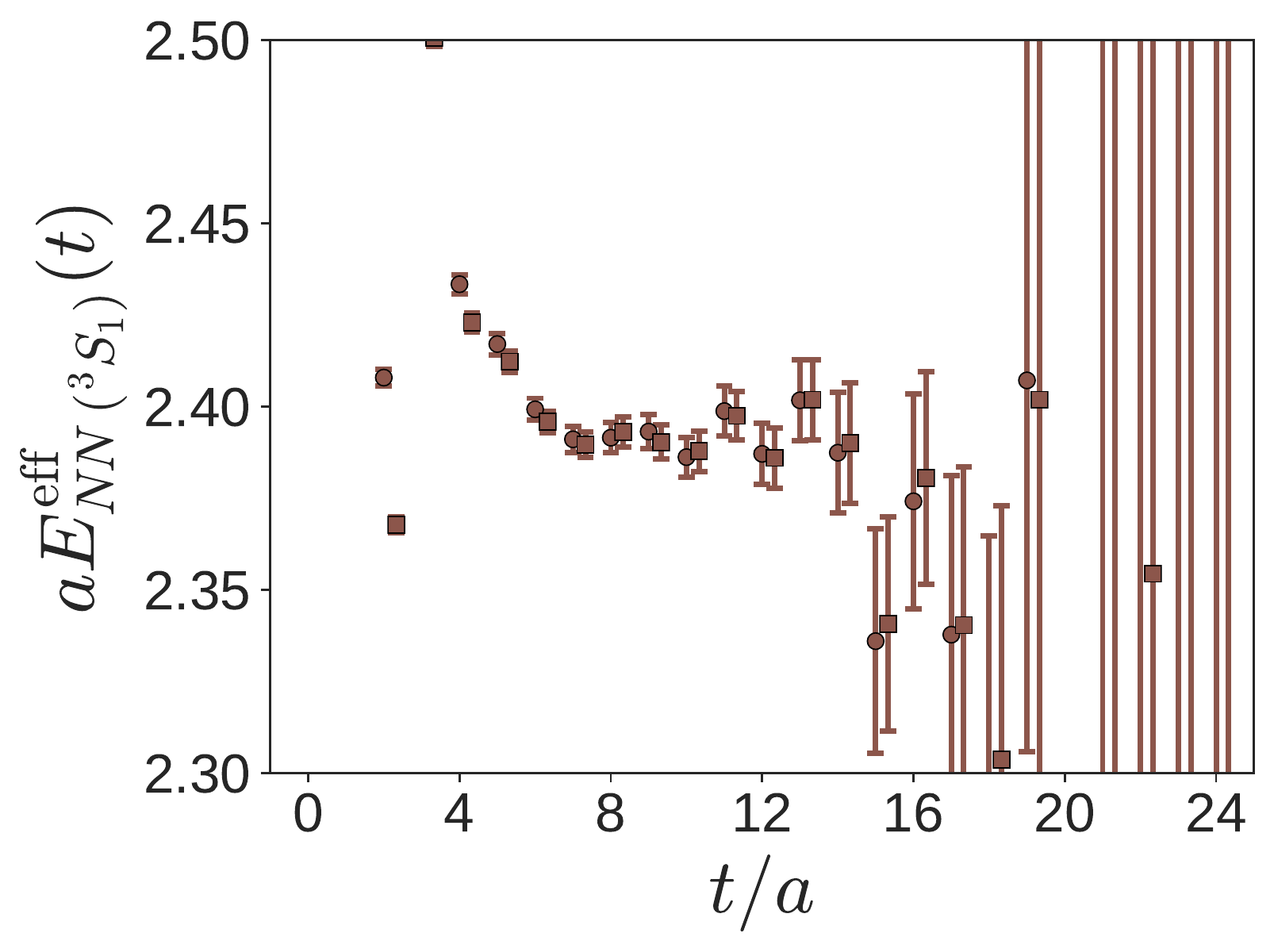}}
  \subfloat{\includegraphics[width=0.3\linewidth]{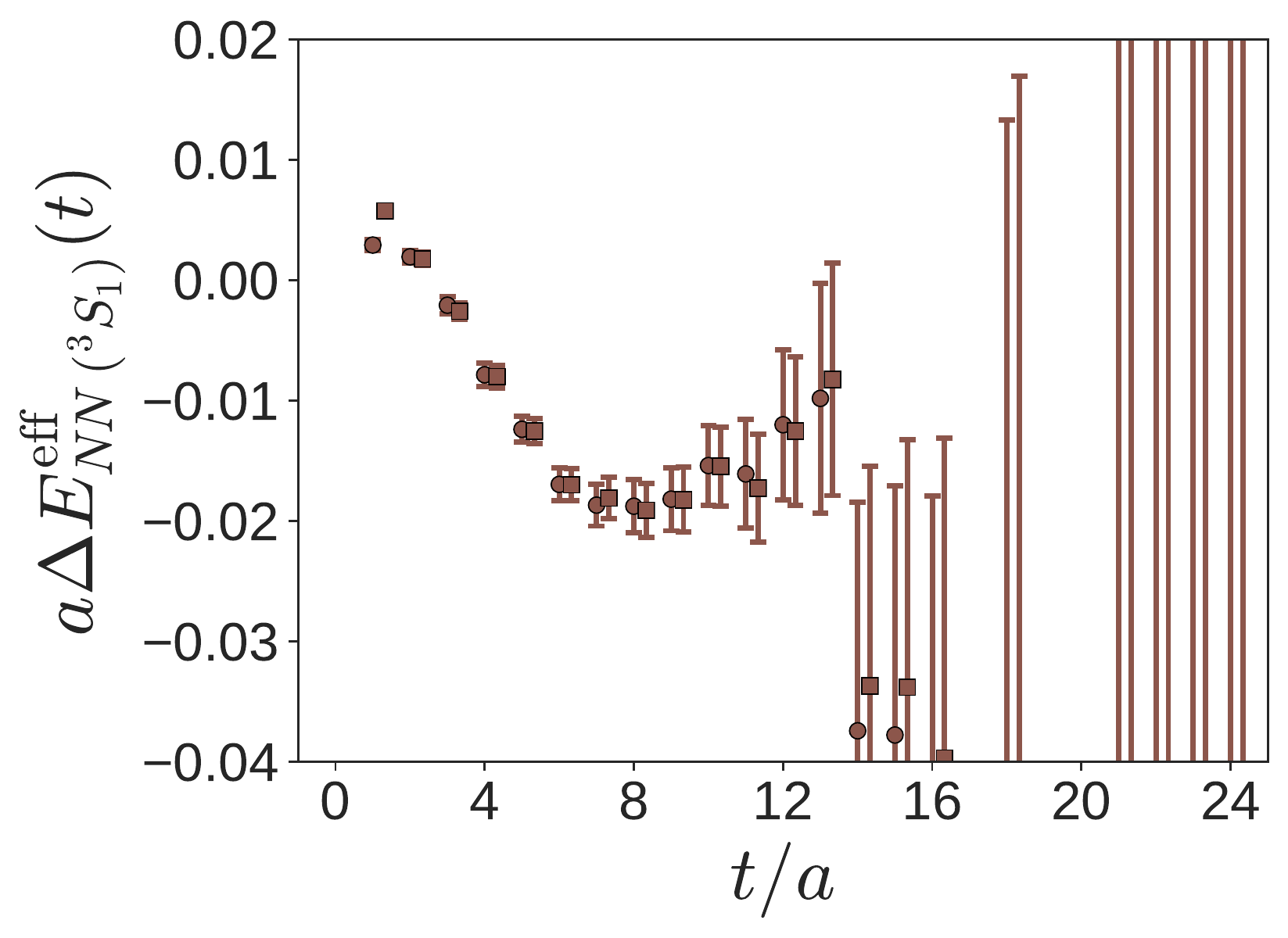}}
  \subfloat{\includegraphics[width=0.3\linewidth]{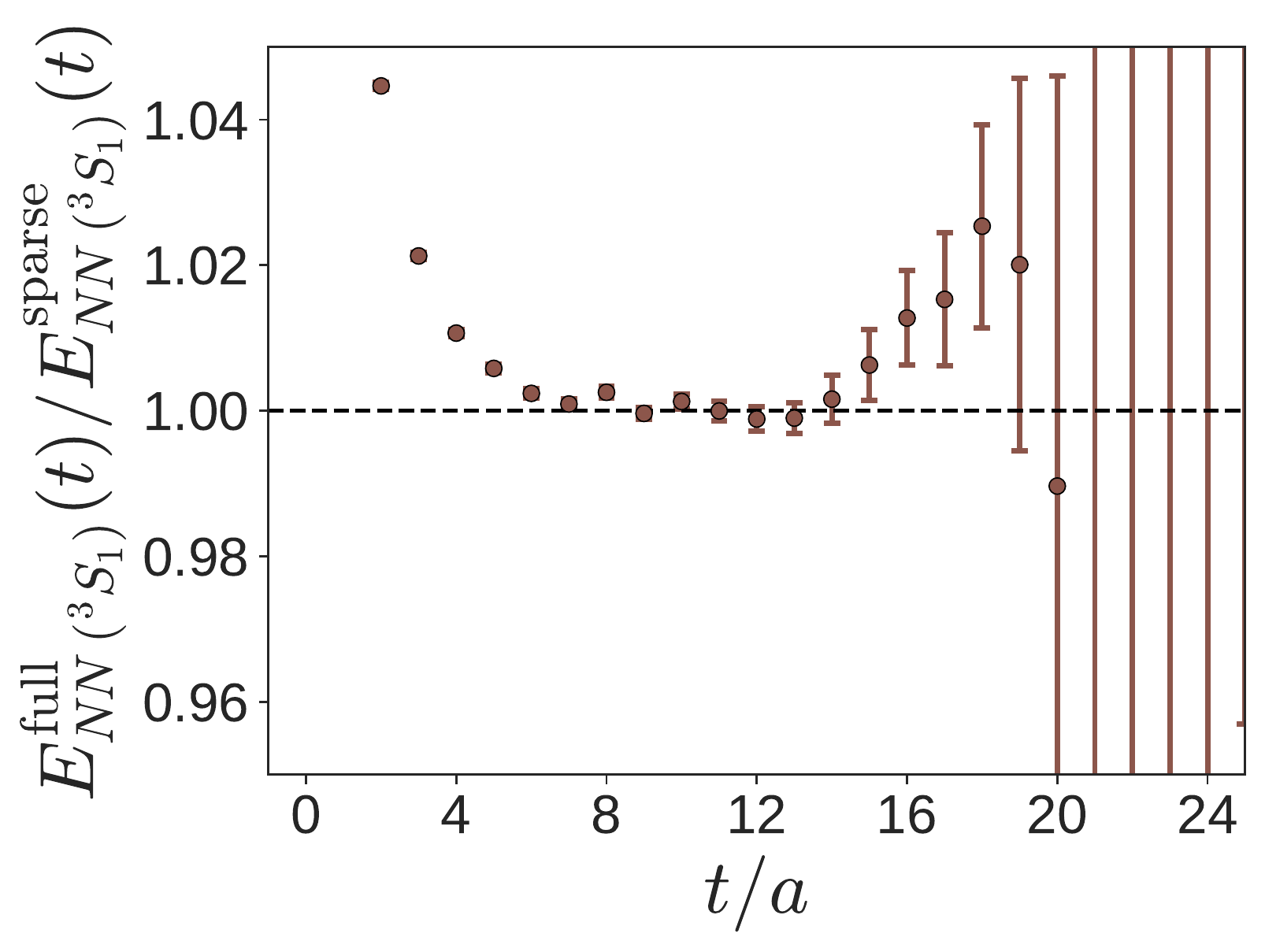}} \\
  \subfloat{\includegraphics[width=0.3\linewidth]{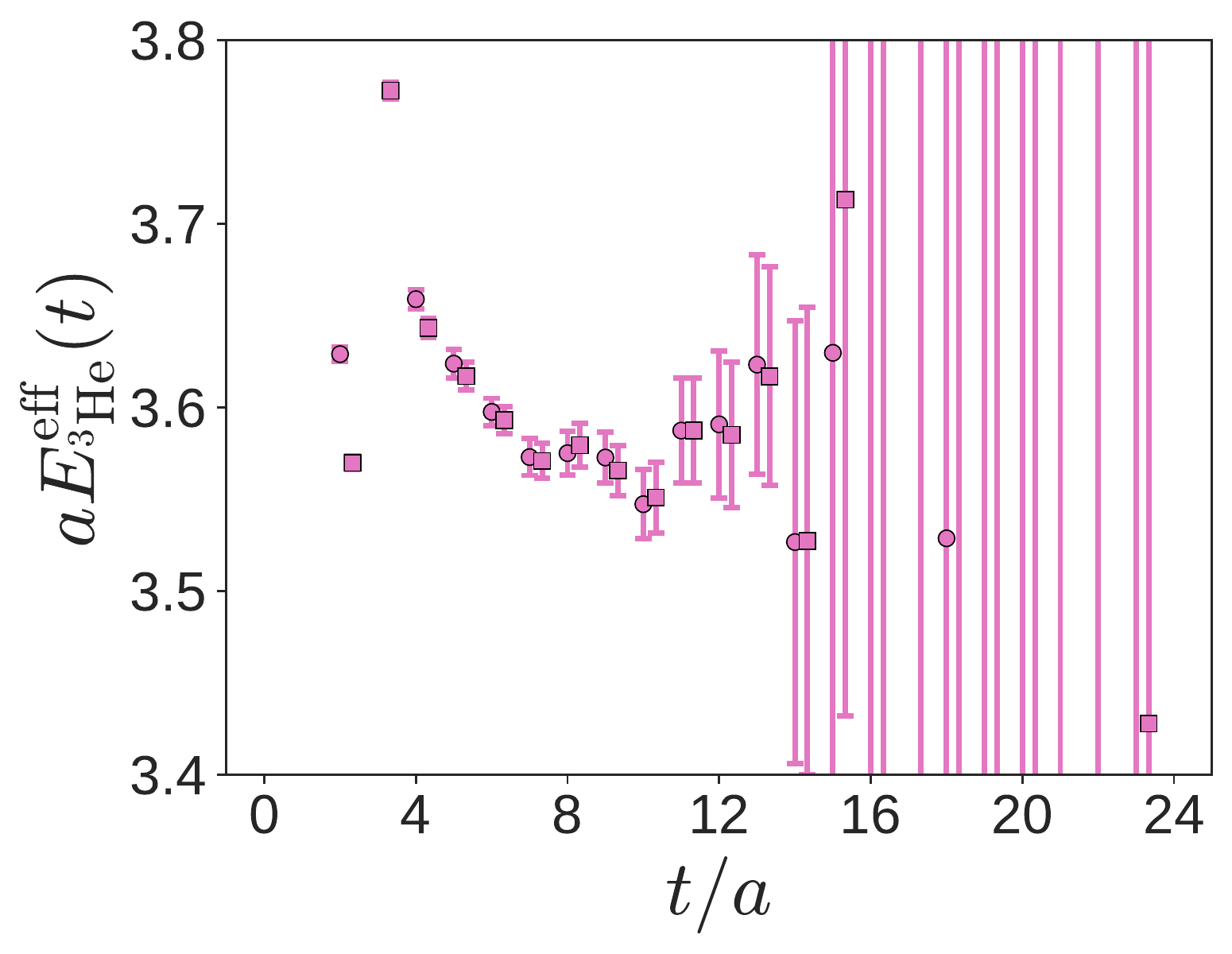}}
  \subfloat{\includegraphics[width=0.3\linewidth]{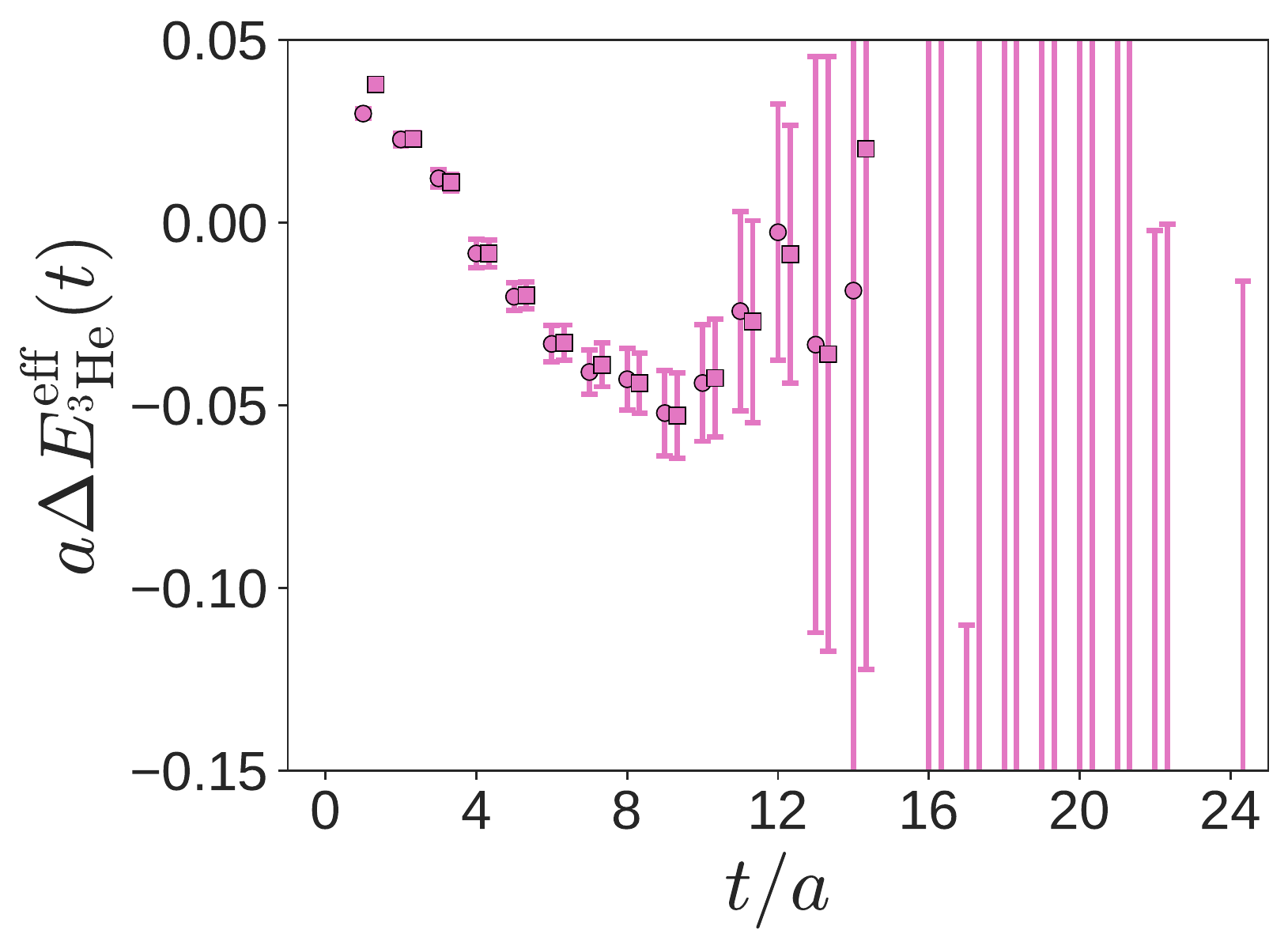}}
  \subfloat{\includegraphics[width=0.3\linewidth]{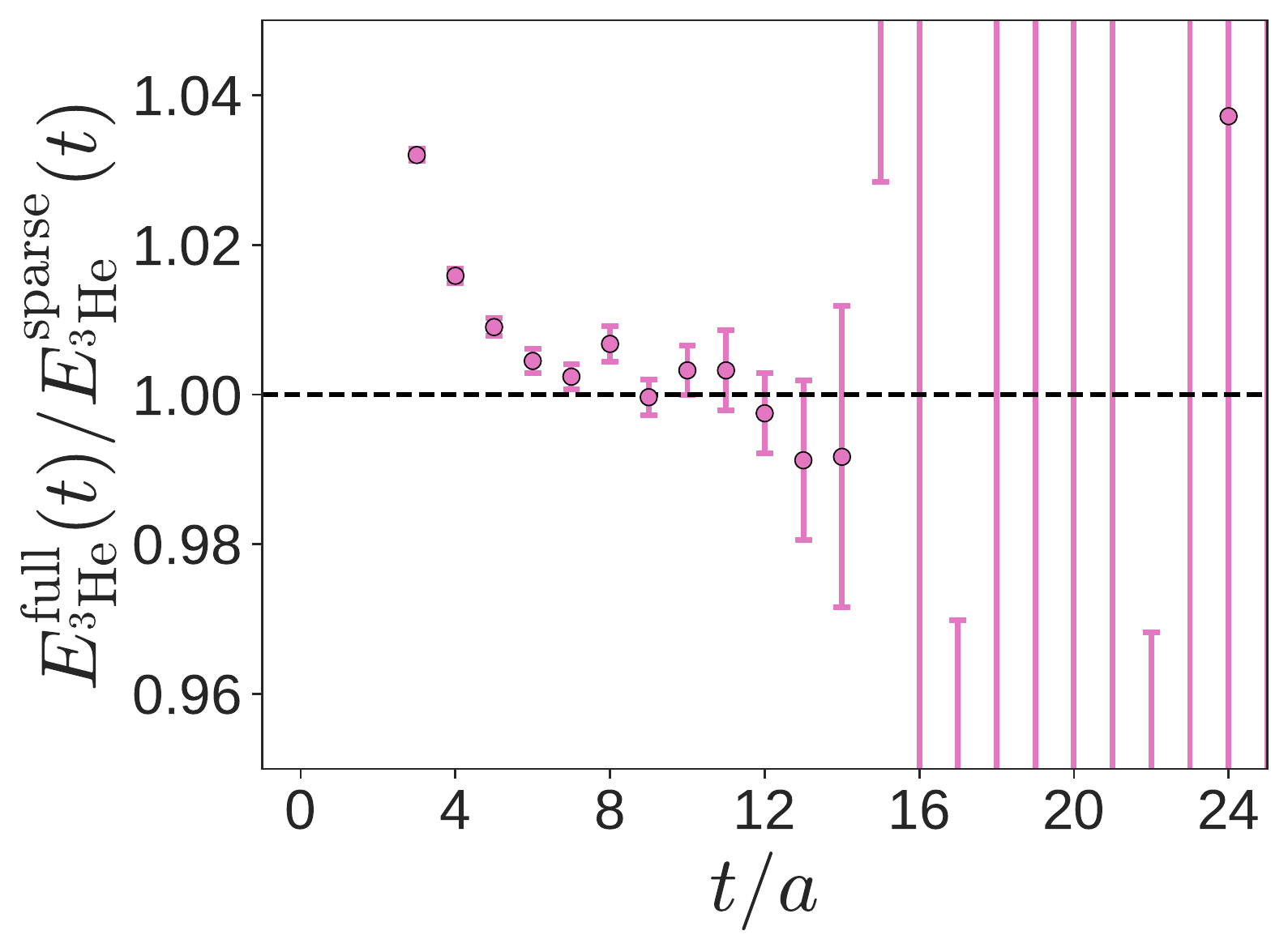}} \\
  \subfloat{\includegraphics[width=0.3\linewidth]{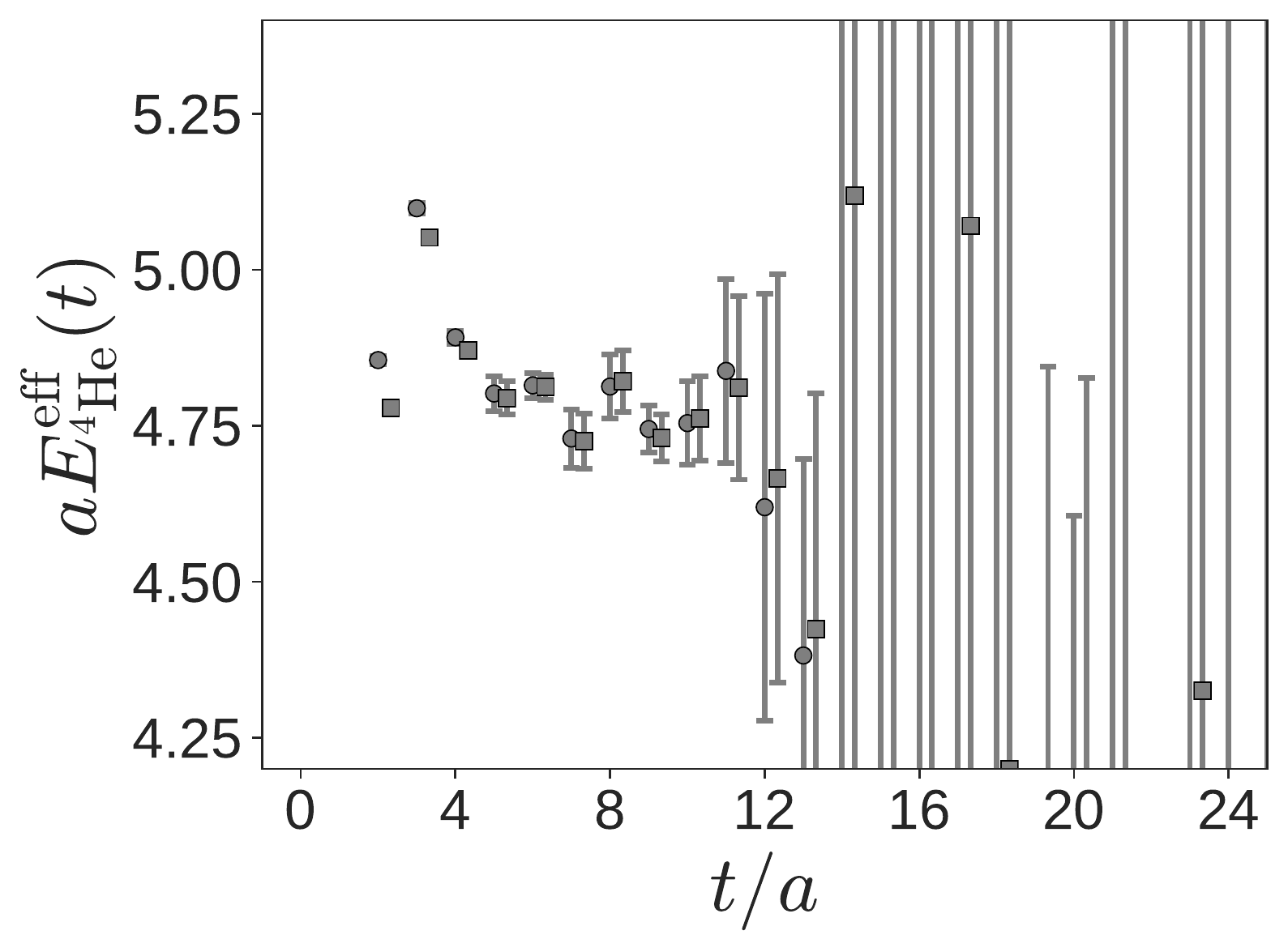}}
  \subfloat{\includegraphics[width=0.3\linewidth]{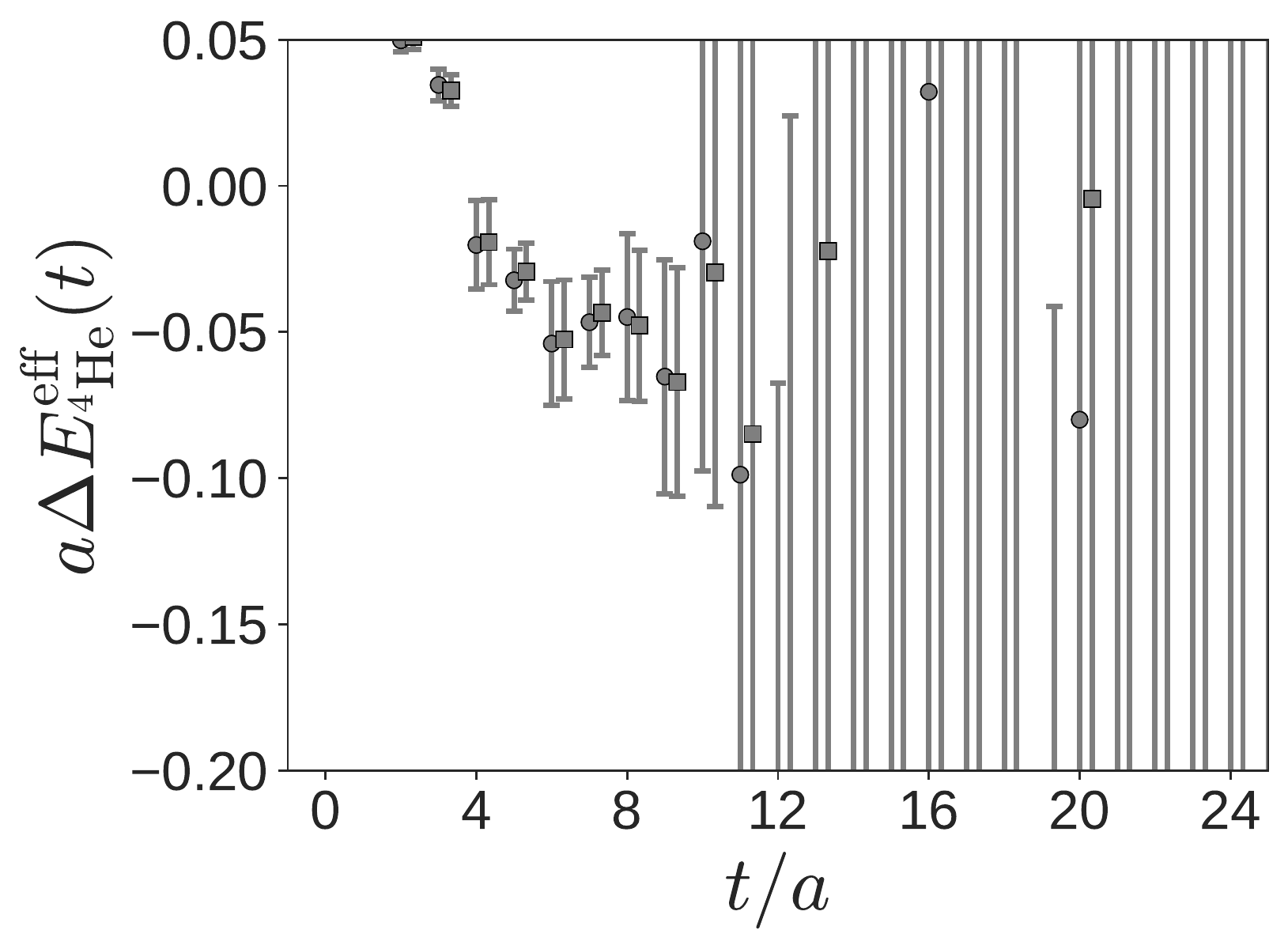}}
  \subfloat{\includegraphics[width=0.3\linewidth]{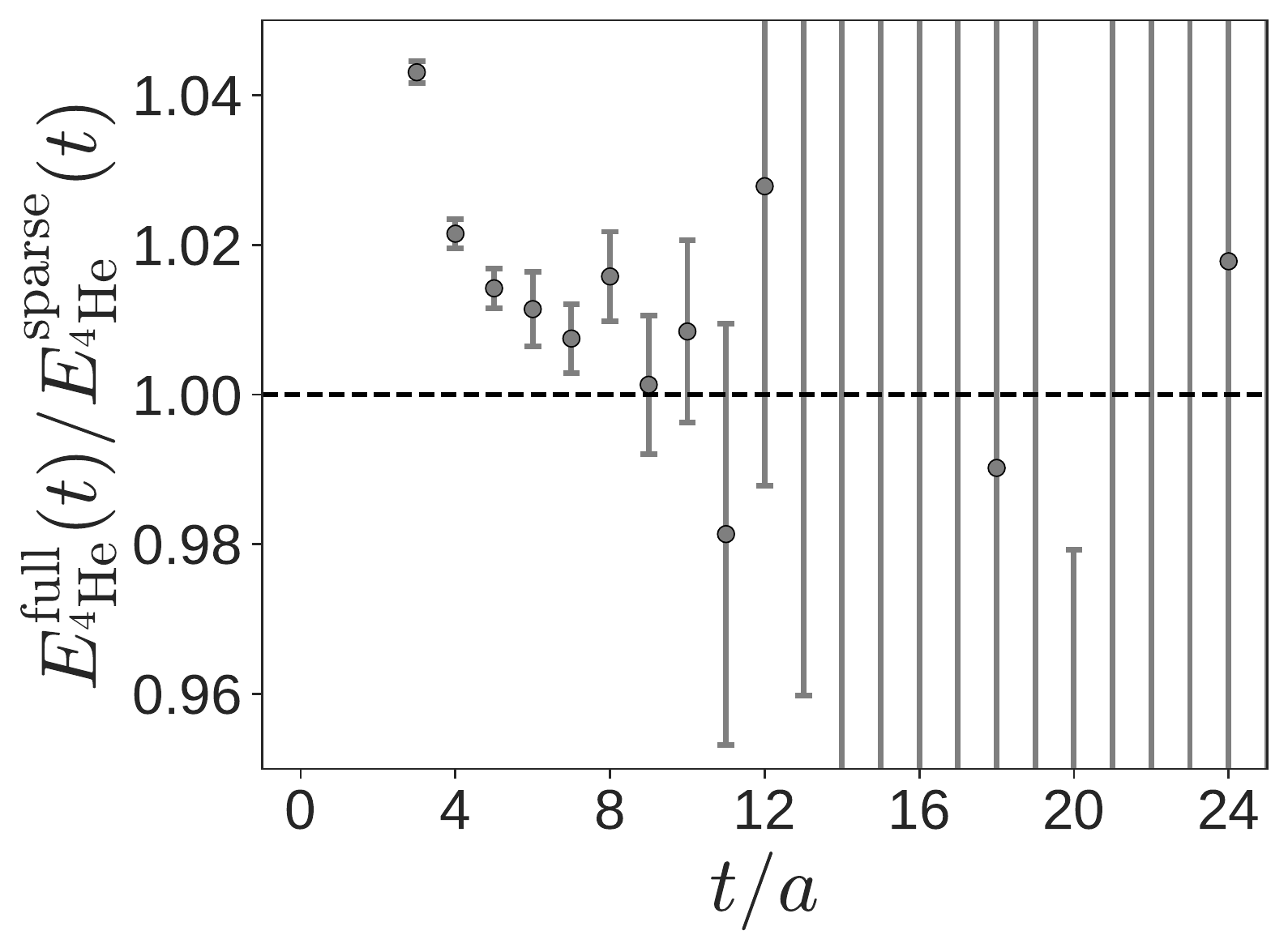}}
  \caption{Effective energies (left column), effective binding energies (middle column), and correlated ratios of full and sparsened effective energies (right column), for the diproton (\(NN\) (\(^{1}S_{0}\)), first row), the deuteron (\(NN\) (\(^{3}S_{1}\)), second row), \(^{3}{\rm He}\) (third row), and \(^{4}{\rm He}\) (fourth row). Circles denote results computed from full two-point correlation functions (Eq.~\eqref{eqn:full_corr_def}), whereas squares denote results computed from sparsened two-point correlation functions (Eq.~\eqref{eqn:sparse_corr_def}). The sparsened data has been shifted slightly along the time axis for clarity.}
\label{fig:nuclei_eff_mass}
\end{figure}

Figure \ref{fig:nuclei_eff_mass} depicts the effective masses, effective binding energies --- computed from Eqs.~\eqref{eqn:eff_energy} and \eqref{eqn:binding_energy_ratio} --- and correlated ratios of the full and sparsened effective energy signals, for the dinucleon, the deuteron, \(^{3}{\rm He}\), and \(^{4}{\rm He}\). Consistent with the single hadron case, the multi-hadron ground state plateaus agree within statistics between the full and sparsened data, and the correlators exhibit percent-scale deviations of the correlated ratios of full to sparse from unity in the early time, excited-state dominated regime. Likewise, in the summaries of fits to the ground state energies and binding energies detailed in Tables \ref{tab:nuclei_mass_fits} and \ref{tab:nuclei_binding_energy_fits}, respectively, there are again no observable discrepancies between fits to the full data and fits to the sparsened data, in terms of both the energies extracted and their statistical uncertainties. 

\begin{table}[!h]
\setlength{\tabcolsep}{6pt}
\begin{tabular}{c|c|ccc|ccc}
\hline
\hline
  \rule{0cm}{0.4cm}& & & \textbf{Full} & & & \textbf{Sparse} & \\
  \textbf{State} & $\bm{T_{\rm fit}}$ & $\bm{a E}$ & $\bm{\chi^{2}/}$\textbf{dof} & $\bm{\kappa \left( \Sigma \right)}$ & $\bm{a E}$ & $\bm{\chi^{2}/}$\textbf{dof} & $\bm{\kappa \left( \Sigma \right)}$ \\
\hline
  \rule{0cm}{0.4cm}$NN$ ($^{1}S_{0}$) & $[7,14]$ & $2.3961(25)$ & $0.41(52)$ & $8.84 \times 10^{13}$ & $2.3961(25)$ & $0.35(48)$ & $8.65 \times 10^{13}$ \\
  $NN$ ($^{3}S_{1}$) & $[7,14]$ & $2.3919(25)$ & $0.61(65)$ & $8.24 \times 10^{13}$ & $2.3918(25)$ & $0.53(60)$ & $8.32 \times 10^{13}$ \\
  $^{3}{\rm He}$ & $[7,12]$ & $3.5726(83)$ & $0.55(74)$ & $6.88 \times 10^{14}$ & $3.5726(84)$ & $0.55(72)$ & $6.84 \times 10^{14}$ \\
  $^{4}{\rm He}$ & $[7,11]$ & $4.769(29)$ & $0.37(57)$ & $7.85 \times 10^{15}$ & $4.766(27)$ & $0.57(72)$ & $6.98 \times 10^{15}$ \\
\hline
\hline
\end{tabular}
\caption{Summary of fits to extract the ground state energies of the dinucleon (\(N N\) (\(^{1}S_{0}\))), the deuteron (\(NN\) (\(^{3}S_{1}\))), \(^{3}{\rm He}\), and \(^{4}{\rm He}\). \(T_{\rm fit}\) denotes the range of Euclidean times included in the fit in lattice units, \( a E \) is the extracted ground state energy, \( \chi^{2} \)/dof is obtained by minimizing Eq.~\eqref{eqn:correlated_chi2pdof}, and \(\kappa(\Sigma)\) denotes the condition number of the covariance matrix (Eq.~\eqref{eqn:cov_matrix}). The first set of fit results (middle three columns) are from fits to full correlation functions (Eq.~\eqref{eqn:full_corr_def}), while the second set of fit results (rightmost three columns) are from fits to sparsened correlation functions (Eq.~\eqref{eqn:sparse_corr_def}). The statistical uncertainties of fitted quantities are computed using the jackknife resampling technique. These results are consistent with previous determinations of these quantities \cite{Beane:2012vq}.}
\label{tab:nuclei_mass_fits}
\end{table}

\begin{table}[!h]
\setlength{\tabcolsep}{6pt}
\begin{tabular}{c|c|ccc|ccc}
\hline
\hline
  \rule{0cm}{0.4cm}& & & \textbf{Full} & & & \textbf{Sparse} & \\
  \textbf{State} & $\bm{T_{\rm fit}}$ & $\bm{a \Delta E}$ & $\bm{\chi^{2}/}$\textbf{dof} & $\bm{\kappa \left( \Sigma \right)}$ & $\bm{a \Delta E}$ & $\bm{\chi^{2}/}$\textbf{dof} & $\bm{\kappa \left( \Sigma \right)}$ \\
\hline
  \rule{0cm}{0.4cm}$NN$ ($^{1}S_{0}$) & $[7,12]$ & $-0.0140(18)$ & $0.46(67)$ & $8.85 \times 10^{1}$ & $-0.0138(18)$ & $0.53(72)$ & $9.42 \times 10^{1}$ \\
  $NN$ ($^{3}S_{1}$) & $[7,12]$ & $-0.0180(17)$ & $0.26(50)$ & $8.78 \times 10^{1}$ & $-0.0180(17)$ & $0.29(53)$ & $9.40 \times 10^{1}$ \\
  $^{3}{\rm He}$ & $[7,11]$ & $-0.0434(72)$ & $0.37(67)$ & $6.08 \times 10^{1}$ & $-0.0431(75)$ & $0.44(74)$ & $6.90 \times 10^{1}$ \\
  $^{4}{\rm He}$ & $[6,10]$ & $-0.055(13)$ & $0.36(49)$ & $8.38 \times 10^{1}$ & $-0.054(13)$ & $0.55(67)$ & $1.01 \times 10^{2}$ \\
\hline
\hline
\end{tabular}
  \caption{Summary of fits to extract the binding energies, \( a \Delta E \), of the dinucleon (\(N N\) (\(^{1}S_{0}\))), the deuteron (\(NN\) (\(^{3}S_{1}\))), \(^{3}{\rm He}\), and \(^{4}{\rm He}\). The notation is otherwise the same as that of Table \ref{tab:nuclei_mass_fits}.}
\label{tab:nuclei_binding_energy_fits}
\end{table}

\section{Conclusions}
This work has introduced an algorithm for reducing the numerical resources required to compute multi-hadron correlation functions in lattice QCD simulations, based on sparsening. It has been demonstrated that a relatively simple prescription for sparsening --- uniformly blocking the lattice in the spatial directions, and taking the value of the propagator evaluated at the first site in each block to define sparsened propagators and correlation functions --- is sufficient to preserve the ground state energies and uncertainties extracted from a lattice QCD simulation. It has also been noted that this sparsening procedure alters the couplings to excited states observed at early Euclidean times for single- and multi-hadron correlation functions; however, a simple modification of the sparsified correlation functions can efficiently remove these modified excited state effects, if desired. Since sparsening differentially distorts the UV components of correlation functions, it is not surprising that it modifies the overlaps onto excited states at early Euclidean times.

The sparsening techniques that are presented here enable \( \mathcal{O}(10 - 100) \) fold speedups in the contraction stage of lattice calculations of nuclear physics. This factor will further increase as the continuum limit is approached, since it is possible to block more aggressively as the scale of the lattice cutoff grows in comparison to the scale of the systems being studied. Future work will explore the application of sparsening to more complicated observables, such as three-point functions describing the gluonic structure of light nuclei \cite{Winter:2017bfs}.

\begin{acknowledgements}
The authors wish to thank the members of the NPLQCD collaboration --- including S.R.~Beane, Z.~Davoudi, M.~Illa, K.~Orginos, and A.~Parre\~{n}o --- for helpful discussions in support of this work. The authors additionally thank R.~Edwards, B.~Jo\'{o}, and K.~Orginos for generating and allowing access to the ensembles used in this study. This research used resources of the Oak Ridge Leadership Computing Facility at the Oak Ridge National Laboratory, which is supported by the Office of Science of the U.S. Department of Energy under Contract number DE-AC05-00OR22725, as well as facilities of the USQCD Collaboration, which are funded by the Office of Science of the U.S. Department of Energy, and the PRACE Research Infrastructure resource Marconi at Cineca, Italy, supported under project 2016163877. The Chroma software library~\cite{Edwards:2004sx} was used in the data analysis. WD, DJM, PES, and MLW are supported in part by the U.S.~Department of Energy, Office of Science, Office of Nuclear Physics under grant Contract Number DE-SC0011090. WD is also supported within the framework of the TMD Topical Collaboration of the U.S.~Department of Energy, Office of Science, Office of Nuclear Physics, and  by the SciDAC4 award DE-SC0018121. AVP is supported by SciDAC4 award DE-SC0018121. MJS is supported by U.S. Department of Energy grant No. DE-FG02-00ER41132. PES is also supported by the National Science Foundation under CAREER Award 1841699. MLW is also supported by a MIT Pappalardo Fellowship. 

\end{acknowledgements}

\bibliographystyle{apsrev4-2}
\bibliography{sparsification}

\begin{thebibliography}{62}%
\makeatletter
\providecommand \@ifxundefined [1]{%
 \@ifx{#1\undefined}
}%
\providecommand \@ifnum [1]{%
 \ifnum #1\expandafter \@firstoftwo
 \else \expandafter \@secondoftwo
 \fi
}%
\providecommand \@ifx [1]{%
 \ifx #1\expandafter \@firstoftwo
 \else \expandafter \@secondoftwo
 \fi
}%
\providecommand \natexlab [1]{#1}%
\providecommand \enquote  [1]{``#1''}%
\providecommand \bibnamefont  [1]{#1}%
\providecommand \bibfnamefont [1]{#1}%
\providecommand \citenamefont [1]{#1}%
\providecommand \href@noop [0]{\@secondoftwo}%
\providecommand \href [0]{\begingroup \@sanitize@url \@href}%
\providecommand \@href[1]{\@@startlink{#1}\@@href}%
\providecommand \@@href[1]{\endgroup#1\@@endlink}%
\providecommand \@sanitize@url [0]{\catcode `\\12\catcode `\$12\catcode
  `\&12\catcode `\#12\catcode `\^12\catcode `\_12\catcode `\%12\relax}%
\providecommand \@@startlink[1]{}%
\providecommand \@@endlink[0]{}%
\providecommand \url  [0]{\begingroup\@sanitize@url \@url }%
\providecommand \@url [1]{\endgroup\@href {#1}{\urlprefix }}%
\providecommand \urlprefix  [0]{URL }%
\providecommand \Eprint [0]{\href }%
\providecommand \doibase [0]{https://doi.org/}%
\providecommand \selectlanguage [0]{\@gobble}%
\providecommand \bibinfo  [0]{\@secondoftwo}%
\providecommand \bibfield  [0]{\@secondoftwo}%
\providecommand \translation [1]{[#1]}%
\providecommand \BibitemOpen [0]{}%
\providecommand \bibitemStop [0]{}%
\providecommand \bibitemNoStop [0]{.\EOS\space}%
\providecommand \EOS [0]{\spacefactor3000\relax}%
\providecommand \BibitemShut  [1]{\csname bibitem#1\endcsname}%
\let\auto@bib@innerbib\@empty
\bibitem [{\citenamefont {Joó}\ \emph {et~al.}(2019)\citenamefont {Joó},
  \citenamefont {Jung}, \citenamefont {Christ}, \citenamefont {Detmold},
  \citenamefont {Edwards}, \citenamefont {Savage},\ and\ \citenamefont
  {Shanahan}}]{Joo:2019byq}%
  \BibitemOpen
  \bibfield  {author} {\bibinfo {author} {\bibfnamefont {B.}~\bibnamefont
  {Joó}}, \bibinfo {author} {\bibfnamefont {C.}~\bibnamefont {Jung}}, \bibinfo
  {author} {\bibfnamefont {N.~H.}\ \bibnamefont {Christ}}, \bibinfo {author}
  {\bibfnamefont {W.}~\bibnamefont {Detmold}}, \bibinfo {author} {\bibfnamefont
  {R.}~\bibnamefont {Edwards}}, \bibinfo {author} {\bibfnamefont
  {M.}~\bibnamefont {Savage}},\ and\ \bibinfo {author} {\bibfnamefont
  {P.}~\bibnamefont {Shanahan}} (\bibinfo {collaboration} {USQCD}),\
  }\href@noop {} {\  (\bibinfo {year} {2019})},\ \Eprint
  {https://arxiv.org/abs/1904.09725} {arXiv:1904.09725 [hep-lat]} \BibitemShut
  {NoStop}%
\bibitem [{\citenamefont {Aoki}\ \emph {et~al.}(2012)\citenamefont {Aoki},
  \citenamefont {Doi}, \citenamefont {Hatsuda}, \citenamefont {Ikeda},
  \citenamefont {Inoue}, \citenamefont {Ishii}, \citenamefont {Murano},
  \citenamefont {Nemura},\ and\ \citenamefont {Sasaki}}]{Aoki:2012tk}%
  \BibitemOpen
  \bibfield  {author} {\bibinfo {author} {\bibfnamefont {S.}~\bibnamefont
  {Aoki}}, \bibinfo {author} {\bibfnamefont {T.}~\bibnamefont {Doi}}, \bibinfo
  {author} {\bibfnamefont {T.}~\bibnamefont {Hatsuda}}, \bibinfo {author}
  {\bibfnamefont {Y.}~\bibnamefont {Ikeda}}, \bibinfo {author} {\bibfnamefont
  {T.}~\bibnamefont {Inoue}}, \bibinfo {author} {\bibfnamefont
  {N.}~\bibnamefont {Ishii}}, \bibinfo {author} {\bibfnamefont
  {K.}~\bibnamefont {Murano}}, \bibinfo {author} {\bibfnamefont
  {H.}~\bibnamefont {Nemura}},\ and\ \bibinfo {author} {\bibfnamefont
  {K.}~\bibnamefont {Sasaki}} (\bibinfo {collaboration} {HAL QCD}),\ }\href
  {https://doi.org/10.1093/ptep/pts010} {\bibfield  {journal} {\bibinfo
  {journal} {PTEP}\ }\textbf {\bibinfo {volume} {2012}},\ \bibinfo {pages}
  {01A105} (\bibinfo {year} {2012})},\ \Eprint
  {https://arxiv.org/abs/1206.5088} {arXiv:1206.5088 [hep-lat]} \BibitemShut
  {NoStop}%
\bibitem [{\citenamefont {Beane}\ \emph
  {et~al.}(2006{\natexlab{a}})\citenamefont {Beane}, \citenamefont {Bedaque},
  \citenamefont {Orginos},\ and\ \citenamefont {Savage}}]{Beane:2006mx}%
  \BibitemOpen
  \bibfield  {author} {\bibinfo {author} {\bibfnamefont {S.~R.}\ \bibnamefont
  {Beane}}, \bibinfo {author} {\bibfnamefont {P.~F.}\ \bibnamefont {Bedaque}},
  \bibinfo {author} {\bibfnamefont {K.}~\bibnamefont {Orginos}},\ and\ \bibinfo
  {author} {\bibfnamefont {M.~J.}\ \bibnamefont {Savage}},\ }\href
  {https://doi.org/10.1103/PhysRevLett.97.012001} {\bibfield  {journal}
  {\bibinfo  {journal} {Phys. Rev. Lett.}\ }\textbf {\bibinfo {volume} {97}},\
  \bibinfo {pages} {012001} (\bibinfo {year} {2006}{\natexlab{a}})},\ \Eprint
  {https://arxiv.org/abs/hep-lat/0602010} {arXiv:hep-lat/0602010 [hep-lat]}
  \BibitemShut {NoStop}%
\bibitem [{\citenamefont {Beane}\ \emph {et~al.}(2010)\citenamefont {Beane},
  \citenamefont {Detmold}, \citenamefont {Lin}, \citenamefont {Luu},
  \citenamefont {Orginos}, \citenamefont {Savage}, \citenamefont {Torok},\ and\
  \citenamefont {Walker-Loud}}]{Beane:2009py}%
  \BibitemOpen
  \bibfield  {author} {\bibinfo {author} {\bibfnamefont {S.~R.}\ \bibnamefont
  {Beane}}, \bibinfo {author} {\bibfnamefont {W.}~\bibnamefont {Detmold}},
  \bibinfo {author} {\bibfnamefont {H.-W.}\ \bibnamefont {Lin}}, \bibinfo
  {author} {\bibfnamefont {T.~C.}\ \bibnamefont {Luu}}, \bibinfo {author}
  {\bibfnamefont {K.}~\bibnamefont {Orginos}}, \bibinfo {author} {\bibfnamefont
  {M.~J.}\ \bibnamefont {Savage}}, \bibinfo {author} {\bibfnamefont
  {A.}~\bibnamefont {Torok}},\ and\ \bibinfo {author} {\bibfnamefont
  {A.}~\bibnamefont {Walker-Loud}} (\bibinfo {collaboration} {NPLQCD}),\ }\href
  {https://doi.org/10.1103/PhysRevD.81.054505} {\bibfield  {journal} {\bibinfo
  {journal} {Phys. Rev.}\ }\textbf {\bibinfo {volume} {D81}},\ \bibinfo {pages}
  {054505} (\bibinfo {year} {2010})},\ \Eprint
  {https://arxiv.org/abs/0912.4243} {arXiv:0912.4243 [hep-lat]} \BibitemShut
  {NoStop}%
\bibitem [{\citenamefont {Beane}\ \emph {et~al.}(2009)\citenamefont {Beane},
  \citenamefont {Detmold}, \citenamefont {Luu}, \citenamefont {Orginos},
  \citenamefont {Parre\~{n}o}, \citenamefont {Savage}, \citenamefont {Torok},\
  and\ \citenamefont {Walker-Loud}}]{Beane:2009gs}%
  \BibitemOpen
  \bibfield  {author} {\bibinfo {author} {\bibfnamefont {S.~R.}\ \bibnamefont
  {Beane}}, \bibinfo {author} {\bibfnamefont {W.}~\bibnamefont {Detmold}},
  \bibinfo {author} {\bibfnamefont {T.~C.}\ \bibnamefont {Luu}}, \bibinfo
  {author} {\bibfnamefont {K.}~\bibnamefont {Orginos}}, \bibinfo {author}
  {\bibfnamefont {A.}~\bibnamefont {Parre\~{n}o}}, \bibinfo {author}
  {\bibfnamefont {M.~J.}\ \bibnamefont {Savage}}, \bibinfo {author}
  {\bibfnamefont {A.}~\bibnamefont {Torok}},\ and\ \bibinfo {author}
  {\bibfnamefont {A.}~\bibnamefont {Walker-Loud}},\ }\href
  {https://doi.org/10.1103/PhysRevD.80.074501} {\bibfield  {journal} {\bibinfo
  {journal} {Phys. Rev.}\ }\textbf {\bibinfo {volume} {D80}},\ \bibinfo {pages}
  {074501} (\bibinfo {year} {2009})},\ \Eprint
  {https://arxiv.org/abs/0905.0466} {arXiv:0905.0466 [hep-lat]} \BibitemShut
  {NoStop}%
\bibitem [{\citenamefont {Beane}\ \emph
  {et~al.}(2011{\natexlab{a}})\citenamefont {Beane} \emph
  {et~al.}}]{Beane:2010hg}%
  \BibitemOpen
  \bibfield  {author} {\bibinfo {author} {\bibfnamefont {S.~R.}\ \bibnamefont
  {Beane}} \emph {et~al.} (\bibinfo {collaboration} {NPLQCD}),\ }\href
  {https://doi.org/10.1103/PhysRevLett.106.162001} {\bibfield  {journal}
  {\bibinfo  {journal} {Phys. Rev. Lett.}\ }\textbf {\bibinfo {volume} {106}},\
  \bibinfo {pages} {162001} (\bibinfo {year} {2011}{\natexlab{a}})},\ \Eprint
  {https://arxiv.org/abs/1012.3812} {arXiv:1012.3812 [hep-lat]} \BibitemShut
  {NoStop}%
\bibitem [{\citenamefont {Beane}\ \emph
  {et~al.}(2011{\natexlab{b}})\citenamefont {Beane}, \citenamefont {Detmold},
  \citenamefont {Orginos},\ and\ \citenamefont {Savage}}]{Beane:2010em}%
  \BibitemOpen
  \bibfield  {author} {\bibinfo {author} {\bibfnamefont {S.~R.}\ \bibnamefont
  {Beane}}, \bibinfo {author} {\bibfnamefont {W.}~\bibnamefont {Detmold}},
  \bibinfo {author} {\bibfnamefont {K.}~\bibnamefont {Orginos}},\ and\ \bibinfo
  {author} {\bibfnamefont {M.~J.}\ \bibnamefont {Savage}},\ }\href
  {https://doi.org/10.1016/j.ppnp.2010.08.002} {\bibfield  {journal} {\bibinfo
  {journal} {Prog. Part. Nucl. Phys.}\ }\textbf {\bibinfo {volume} {66}},\
  \bibinfo {pages} {1} (\bibinfo {year} {2011}{\natexlab{b}})},\ \Eprint
  {https://arxiv.org/abs/1004.2935} {arXiv:1004.2935 [hep-lat]} \BibitemShut
  {NoStop}%
\bibitem [{\citenamefont {Beane}\ \emph
  {et~al.}(2012{\natexlab{a}})\citenamefont {Beane}, \citenamefont {Chang},
  \citenamefont {Detmold}, \citenamefont {Lin}, \citenamefont {Luu},
  \citenamefont {Orginos}, \citenamefont {Parreno}, \citenamefont {Savage},
  \citenamefont {Torok},\ and\ \citenamefont {Walker-Loud}}]{Beane:2011iw}%
  \BibitemOpen
  \bibfield  {author} {\bibinfo {author} {\bibfnamefont {S.~R.}\ \bibnamefont
  {Beane}}, \bibinfo {author} {\bibfnamefont {E.}~\bibnamefont {Chang}},
  \bibinfo {author} {\bibfnamefont {W.}~\bibnamefont {Detmold}}, \bibinfo
  {author} {\bibfnamefont {H.~W.}\ \bibnamefont {Lin}}, \bibinfo {author}
  {\bibfnamefont {T.~C.}\ \bibnamefont {Luu}}, \bibinfo {author} {\bibfnamefont
  {K.}~\bibnamefont {Orginos}}, \bibinfo {author} {\bibfnamefont
  {A.}~\bibnamefont {Parreno}}, \bibinfo {author} {\bibfnamefont {M.~J.}\
  \bibnamefont {Savage}}, \bibinfo {author} {\bibfnamefont {A.}~\bibnamefont
  {Torok}},\ and\ \bibinfo {author} {\bibfnamefont {A.}~\bibnamefont
  {Walker-Loud}} (\bibinfo {collaboration} {NPLQCD}),\ }\href
  {https://doi.org/10.1103/PhysRevD.85.054511} {\bibfield  {journal} {\bibinfo
  {journal} {Phys. Rev.}\ }\textbf {\bibinfo {volume} {D85}},\ \bibinfo {pages}
  {054511} (\bibinfo {year} {2012}{\natexlab{a}})},\ \Eprint
  {https://arxiv.org/abs/1109.2889} {arXiv:1109.2889 [hep-lat]} \BibitemShut
  {NoStop}%
\bibitem [{\citenamefont {Beane}\ \emph
  {et~al.}(2013{\natexlab{a}})\citenamefont {Beane}, \citenamefont {Chang},
  \citenamefont {Cohen}, \citenamefont {Detmold}, \citenamefont {Lin},
  \citenamefont {Luu}, \citenamefont {Orginos}, \citenamefont {Parre\~{n}o},
  \citenamefont {Savage},\ and\ \citenamefont {Walker-Loud}}]{Beane:2012vq}%
  \BibitemOpen
  \bibfield  {author} {\bibinfo {author} {\bibfnamefont {S.~R.}\ \bibnamefont
  {Beane}}, \bibinfo {author} {\bibfnamefont {E.}~\bibnamefont {Chang}},
  \bibinfo {author} {\bibfnamefont {S.~D.}\ \bibnamefont {Cohen}}, \bibinfo
  {author} {\bibfnamefont {W.}~\bibnamefont {Detmold}}, \bibinfo {author}
  {\bibfnamefont {H.~W.}\ \bibnamefont {Lin}}, \bibinfo {author} {\bibfnamefont
  {T.~C.}\ \bibnamefont {Luu}}, \bibinfo {author} {\bibfnamefont
  {K.}~\bibnamefont {Orginos}}, \bibinfo {author} {\bibfnamefont
  {A.}~\bibnamefont {Parre\~{n}o}}, \bibinfo {author} {\bibfnamefont {M.~J.}\
  \bibnamefont {Savage}},\ and\ \bibinfo {author} {\bibfnamefont
  {A.}~\bibnamefont {Walker-Loud}} (\bibinfo {collaboration} {NPLQCD}),\ }\href
  {https://doi.org/10.1103/PhysRevD.87.034506} {\bibfield  {journal} {\bibinfo
  {journal} {Phys. Rev.}\ }\textbf {\bibinfo {volume} {D87}},\ \bibinfo {pages}
  {034506} (\bibinfo {year} {2013}{\natexlab{a}})},\ \Eprint
  {https://arxiv.org/abs/1206.5219} {arXiv:1206.5219 [hep-lat]} \BibitemShut
  {NoStop}%
\bibitem [{\citenamefont {Beane}\ \emph
  {et~al.}(2012{\natexlab{b}})\citenamefont {Beane}, \citenamefont {Chang},
  \citenamefont {Cohen}, \citenamefont {Detmold}, \citenamefont {Lin},
  \citenamefont {Luu}, \citenamefont {Orginos}, \citenamefont {Parreno},
  \citenamefont {Savage},\ and\ \citenamefont {Walker-Loud}}]{Beane:2012ey}%
  \BibitemOpen
  \bibfield  {author} {\bibinfo {author} {\bibfnamefont {S.~R.}\ \bibnamefont
  {Beane}}, \bibinfo {author} {\bibfnamefont {E.}~\bibnamefont {Chang}},
  \bibinfo {author} {\bibfnamefont {S.~D.}\ \bibnamefont {Cohen}}, \bibinfo
  {author} {\bibfnamefont {W.}~\bibnamefont {Detmold}}, \bibinfo {author}
  {\bibfnamefont {H.~W.}\ \bibnamefont {Lin}}, \bibinfo {author} {\bibfnamefont
  {T.~C.}\ \bibnamefont {Luu}}, \bibinfo {author} {\bibfnamefont
  {K.}~\bibnamefont {Orginos}}, \bibinfo {author} {\bibfnamefont
  {A.}~\bibnamefont {Parreno}}, \bibinfo {author} {\bibfnamefont {M.~J.}\
  \bibnamefont {Savage}},\ and\ \bibinfo {author} {\bibfnamefont
  {A.}~\bibnamefont {Walker-Loud}},\ }\href
  {https://doi.org/10.1103/PhysRevLett.109.172001} {\bibfield  {journal}
  {\bibinfo  {journal} {Phys. Rev. Lett.}\ }\textbf {\bibinfo {volume} {109}},\
  \bibinfo {pages} {172001} (\bibinfo {year} {2012}{\natexlab{b}})},\ \Eprint
  {https://arxiv.org/abs/1204.3606} {arXiv:1204.3606 [hep-lat]} \BibitemShut
  {NoStop}%
\bibitem [{\citenamefont {Beane}\ \emph
  {et~al.}(2013{\natexlab{b}})\citenamefont {Beane} \emph
  {et~al.}}]{Beane:2013br}%
  \BibitemOpen
  \bibfield  {author} {\bibinfo {author} {\bibfnamefont {S.~R.}\ \bibnamefont
  {Beane}} \emph {et~al.} (\bibinfo {collaboration} {NPLQCD}),\ }\href
  {https://doi.org/10.1103/PhysRevC.88.024003} {\bibfield  {journal} {\bibinfo
  {journal} {Phys. Rev.}\ }\textbf {\bibinfo {volume} {C88}},\ \bibinfo {pages}
  {024003} (\bibinfo {year} {2013}{\natexlab{b}})},\ \Eprint
  {https://arxiv.org/abs/1301.5790} {arXiv:1301.5790 [hep-lat]} \BibitemShut
  {NoStop}%
\bibitem [{\citenamefont {Beane}\ \emph {et~al.}(2014)\citenamefont {Beane},
  \citenamefont {Chang}, \citenamefont {Cohen}, \citenamefont {Detmold},
  \citenamefont {Lin}, \citenamefont {Orginos}, \citenamefont {Parreno},
  \citenamefont {Savage},\ and\ \citenamefont {Tiburzi}}]{Beane:2014ora}%
  \BibitemOpen
  \bibfield  {author} {\bibinfo {author} {\bibfnamefont {S.~R.}\ \bibnamefont
  {Beane}}, \bibinfo {author} {\bibfnamefont {E.}~\bibnamefont {Chang}},
  \bibinfo {author} {\bibfnamefont {S.}~\bibnamefont {Cohen}}, \bibinfo
  {author} {\bibfnamefont {W.}~\bibnamefont {Detmold}}, \bibinfo {author}
  {\bibfnamefont {H.~W.}\ \bibnamefont {Lin}}, \bibinfo {author} {\bibfnamefont
  {K.}~\bibnamefont {Orginos}}, \bibinfo {author} {\bibfnamefont
  {A.}~\bibnamefont {Parreno}}, \bibinfo {author} {\bibfnamefont {M.~J.}\
  \bibnamefont {Savage}},\ and\ \bibinfo {author} {\bibfnamefont {B.~C.}\
  \bibnamefont {Tiburzi}},\ }\href
  {https://doi.org/10.1103/PhysRevLett.113.252001} {\bibfield  {journal}
  {\bibinfo  {journal} {Phys. Rev. Lett.}\ }\textbf {\bibinfo {volume} {113}},\
  \bibinfo {pages} {252001} (\bibinfo {year} {2014})},\ \Eprint
  {https://arxiv.org/abs/1409.3556} {arXiv:1409.3556 [hep-lat]} \BibitemShut
  {NoStop}%
\bibitem [{\citenamefont {Berkowitz}\ \emph {et~al.}(2017)\citenamefont
  {Berkowitz}, \citenamefont {Kurth}, \citenamefont {Nicholson}, \citenamefont
  {Joo}, \citenamefont {Rinaldi}, \citenamefont {Strother}, \citenamefont
  {Vranas},\ and\ \citenamefont {Walker-Loud}}]{Berkowitz:2015eaa}%
  \BibitemOpen
  \bibfield  {author} {\bibinfo {author} {\bibfnamefont {E.}~\bibnamefont
  {Berkowitz}}, \bibinfo {author} {\bibfnamefont {T.}~\bibnamefont {Kurth}},
  \bibinfo {author} {\bibfnamefont {A.}~\bibnamefont {Nicholson}}, \bibinfo
  {author} {\bibfnamefont {B.}~\bibnamefont {Joo}}, \bibinfo {author}
  {\bibfnamefont {E.}~\bibnamefont {Rinaldi}}, \bibinfo {author} {\bibfnamefont
  {M.}~\bibnamefont {Strother}}, \bibinfo {author} {\bibfnamefont {P.~M.}\
  \bibnamefont {Vranas}},\ and\ \bibinfo {author} {\bibfnamefont
  {A.}~\bibnamefont {Walker-Loud}},\ }\href
  {https://doi.org/10.1016/j.physletb.2016.12.024} {\bibfield  {journal}
  {\bibinfo  {journal} {Phys. Lett.}\ }\textbf {\bibinfo {volume} {B765}},\
  \bibinfo {pages} {285} (\bibinfo {year} {2017})},\ \Eprint
  {https://arxiv.org/abs/1508.00886} {arXiv:1508.00886 [hep-lat]} \BibitemShut
  {NoStop}%
\bibitem [{\citenamefont {Beane}\ \emph {et~al.}(2015)\citenamefont {Beane},
  \citenamefont {Chang}, \citenamefont {Detmold}, \citenamefont {Orginos},
  \citenamefont {Parreño}, \citenamefont {Savage},\ and\ \citenamefont
  {Tiburzi}}]{Beane:2015yha}%
  \BibitemOpen
  \bibfield  {author} {\bibinfo {author} {\bibfnamefont {S.~R.}\ \bibnamefont
  {Beane}}, \bibinfo {author} {\bibfnamefont {E.}~\bibnamefont {Chang}},
  \bibinfo {author} {\bibfnamefont {W.}~\bibnamefont {Detmold}}, \bibinfo
  {author} {\bibfnamefont {K.}~\bibnamefont {Orginos}}, \bibinfo {author}
  {\bibfnamefont {A.}~\bibnamefont {Parreño}}, \bibinfo {author}
  {\bibfnamefont {M.~J.}\ \bibnamefont {Savage}},\ and\ \bibinfo {author}
  {\bibfnamefont {B.~C.}\ \bibnamefont {Tiburzi}} (\bibinfo {collaboration}
  {NPLQCD}),\ }\href {https://doi.org/10.1103/PhysRevLett.115.132001}
  {\bibfield  {journal} {\bibinfo  {journal} {Phys. Rev. Lett.}\ }\textbf
  {\bibinfo {volume} {115}},\ \bibinfo {pages} {132001} (\bibinfo {year}
  {2015})},\ \Eprint {https://arxiv.org/abs/1505.02422} {arXiv:1505.02422
  [hep-lat]} \BibitemShut {NoStop}%
\bibitem [{\citenamefont {Berkowitz}\ \emph
  {et~al.}(2018{\natexlab{a}})\citenamefont {Berkowitz}, \citenamefont
  {Nicholson}, \citenamefont {Chang}, \citenamefont {Rinaldi}, \citenamefont
  {Clark}, \citenamefont {Joó}, \citenamefont {Kurth}, \citenamefont
  {Vranas},\ and\ \citenamefont {Walker-Loud}}]{Berkowitz:2017smo}%
  \BibitemOpen
  \bibfield  {author} {\bibinfo {author} {\bibfnamefont {E.}~\bibnamefont
  {Berkowitz}}, \bibinfo {author} {\bibfnamefont {A.}~\bibnamefont
  {Nicholson}}, \bibinfo {author} {\bibfnamefont {C.~C.}\ \bibnamefont
  {Chang}}, \bibinfo {author} {\bibfnamefont {E.}~\bibnamefont {Rinaldi}},
  \bibinfo {author} {\bibfnamefont {M.~A.}\ \bibnamefont {Clark}}, \bibinfo
  {author} {\bibfnamefont {B.}~\bibnamefont {Joó}}, \bibinfo {author}
  {\bibfnamefont {T.}~\bibnamefont {Kurth}}, \bibinfo {author} {\bibfnamefont
  {P.}~\bibnamefont {Vranas}},\ and\ \bibinfo {author} {\bibfnamefont
  {A.}~\bibnamefont {Walker-Loud}},\ }\bibfield  {booktitle} {\emph {\bibinfo
  {booktitle} {{Proceedings, 35th International Symposium on Lattice Field
  Theory (Lattice 2017): Granada, Spain, June 18-24, 2017}}},\ }\href
  {https://doi.org/10.1051/epjconf/201817505029} {\bibfield  {journal}
  {\bibinfo  {journal} {EPJ Web Conf.}\ }\textbf {\bibinfo {volume} {175}},\
  \bibinfo {pages} {05029} (\bibinfo {year} {2018}{\natexlab{a}})},\ \Eprint
  {https://arxiv.org/abs/1710.05642} {arXiv:1710.05642 [hep-lat]} \BibitemShut
  {NoStop}%
\bibitem [{\citenamefont {Berkowitz}\ \emph
  {et~al.}(2018{\natexlab{b}})\citenamefont {Berkowitz} \emph
  {et~al.}}]{Berkowitz:2019yrf}%
  \BibitemOpen
  \bibfield  {author} {\bibinfo {author} {\bibfnamefont {E.}~\bibnamefont
  {Berkowitz}} \emph {et~al.},\ }\bibfield  {booktitle} {\emph {\bibinfo
  {booktitle} {{Proceedings, 36th International Symposium on Lattice Field
  Theory (Lattice 2018): East Lansing, MI, United States, July 22-28, 2018}}},\
  }\href {https://doi.org/10.22323/1.334.0003} {\bibfield  {journal} {\bibinfo
  {journal} {PoS}\ }\textbf {\bibinfo {volume} {LATTICE2018}},\ \bibinfo
  {pages} {003} (\bibinfo {year} {2018}{\natexlab{b}})},\ \Eprint
  {https://arxiv.org/abs/1902.09416} {arXiv:1902.09416 [hep-lat]} \BibitemShut
  {NoStop}%
\bibitem [{\citenamefont {Chang}\ \emph {et~al.}(2015)\citenamefont {Chang},
  \citenamefont {Detmold}, \citenamefont {Orginos}, \citenamefont
  {Parre\~{n}o}, \citenamefont {Savage}, \citenamefont {Tiburzi},\ and\
  \citenamefont {Beane}}]{Chang:2015qxa}%
  \BibitemOpen
  \bibfield  {author} {\bibinfo {author} {\bibfnamefont {E.}~\bibnamefont
  {Chang}}, \bibinfo {author} {\bibfnamefont {W.}~\bibnamefont {Detmold}},
  \bibinfo {author} {\bibfnamefont {K.}~\bibnamefont {Orginos}}, \bibinfo
  {author} {\bibfnamefont {A.}~\bibnamefont {Parre\~{n}o}}, \bibinfo {author}
  {\bibfnamefont {M.~J.}\ \bibnamefont {Savage}}, \bibinfo {author}
  {\bibfnamefont {B.~C.}\ \bibnamefont {Tiburzi}},\ and\ \bibinfo {author}
  {\bibfnamefont {S.~R.}\ \bibnamefont {Beane}} (\bibinfo {collaboration}
  {NPLQCD}),\ }\href {https://doi.org/10.1103/PhysRevD.92.114502} {\bibfield
  {journal} {\bibinfo  {journal} {Phys. Rev.}\ }\textbf {\bibinfo {volume}
  {D92}},\ \bibinfo {pages} {114502} (\bibinfo {year} {2015})},\ \Eprint
  {https://arxiv.org/abs/1506.05518} {arXiv:1506.05518 [hep-lat]} \BibitemShut
  {NoStop}%
\bibitem [{\citenamefont {Chang}\ \emph {et~al.}(2018)\citenamefont {Chang},
  \citenamefont {Davoudi}, \citenamefont {Detmold}, \citenamefont {Gambhir},
  \citenamefont {Orginos}, \citenamefont {Savage}, \citenamefont {Shanahan},
  \citenamefont {Wagman},\ and\ \citenamefont {Winter}}]{Chang:2017eiq}%
  \BibitemOpen
  \bibfield  {author} {\bibinfo {author} {\bibfnamefont {E.}~\bibnamefont
  {Chang}}, \bibinfo {author} {\bibfnamefont {Z.}~\bibnamefont {Davoudi}},
  \bibinfo {author} {\bibfnamefont {W.}~\bibnamefont {Detmold}}, \bibinfo
  {author} {\bibfnamefont {A.~S.}\ \bibnamefont {Gambhir}}, \bibinfo {author}
  {\bibfnamefont {K.}~\bibnamefont {Orginos}}, \bibinfo {author} {\bibfnamefont
  {M.~J.}\ \bibnamefont {Savage}}, \bibinfo {author} {\bibfnamefont {P.~E.}\
  \bibnamefont {Shanahan}}, \bibinfo {author} {\bibfnamefont {M.~L.}\
  \bibnamefont {Wagman}},\ and\ \bibinfo {author} {\bibfnamefont
  {F.}~\bibnamefont {Winter}} (\bibinfo {collaboration} {NPLQCD}),\ }\href
  {https://doi.org/10.1103/PhysRevLett.120.152002} {\bibfield  {journal}
  {\bibinfo  {journal} {Phys. Rev. Lett.}\ }\textbf {\bibinfo {volume} {120}},\
  \bibinfo {pages} {152002} (\bibinfo {year} {2018})},\ \Eprint
  {https://arxiv.org/abs/1712.03221} {arXiv:1712.03221 [hep-lat]} \BibitemShut
  {NoStop}%
\bibitem [{\citenamefont {Doi}\ and\ \citenamefont
  {Endres}(2013)}]{Doi:2012xd}%
  \BibitemOpen
  \bibfield  {author} {\bibinfo {author} {\bibfnamefont {T.}~\bibnamefont
  {Doi}}\ and\ \bibinfo {author} {\bibfnamefont {M.~G.}\ \bibnamefont
  {Endres}},\ }\href {https://doi.org/10.1016/j.cpc.2012.09.004} {\bibfield
  {journal} {\bibinfo  {journal} {Comput. Phys. Commun.}\ }\textbf {\bibinfo
  {volume} {184}},\ \bibinfo {pages} {117} (\bibinfo {year} {2013})},\ \Eprint
  {https://arxiv.org/abs/1205.0585} {arXiv:1205.0585 [hep-lat]} \BibitemShut
  {NoStop}%
\bibitem [{\citenamefont {Francis}\ \emph {et~al.}(2014)\citenamefont
  {Francis}, \citenamefont {Miao}, \citenamefont {Rae},\ and\ \citenamefont
  {Wittig}}]{Francis:2013lva}%
  \BibitemOpen
  \bibfield  {author} {\bibinfo {author} {\bibfnamefont {A.}~\bibnamefont
  {Francis}}, \bibinfo {author} {\bibfnamefont {C.}~\bibnamefont {Miao}},
  \bibinfo {author} {\bibfnamefont {T.~D.}\ \bibnamefont {Rae}},\ and\ \bibinfo
  {author} {\bibfnamefont {H.}~\bibnamefont {Wittig}},\ }\bibfield  {booktitle}
  {\emph {\bibinfo {booktitle} {{Proceedings, 31st International Symposium on
  Lattice Field Theory (Lattice 2013): Mainz, Germany, July 29-August 3,
  2013}}},\ }\href {https://doi.org/10.22323/1.187.0440} {\bibfield  {journal}
  {\bibinfo  {journal} {PoS}\ }\textbf {\bibinfo {volume} {LATTICE2013}},\
  \bibinfo {pages} {440} (\bibinfo {year} {2014})},\ \Eprint
  {https://arxiv.org/abs/1311.3933} {arXiv:1311.3933 [hep-lat]} \BibitemShut
  {NoStop}%
\bibitem [{\citenamefont {Francis}\ \emph {et~al.}(2019)\citenamefont
  {Francis}, \citenamefont {Green}, \citenamefont {Junnarkar}, \citenamefont
  {Miao}, \citenamefont {Rae},\ and\ \citenamefont {Wittig}}]{Francis:2018qch}%
  \BibitemOpen
  \bibfield  {author} {\bibinfo {author} {\bibfnamefont {A.}~\bibnamefont
  {Francis}}, \bibinfo {author} {\bibfnamefont {J.~R.}\ \bibnamefont {Green}},
  \bibinfo {author} {\bibfnamefont {P.~M.}\ \bibnamefont {Junnarkar}}, \bibinfo
  {author} {\bibfnamefont {C.}~\bibnamefont {Miao}}, \bibinfo {author}
  {\bibfnamefont {T.~D.}\ \bibnamefont {Rae}},\ and\ \bibinfo {author}
  {\bibfnamefont {H.}~\bibnamefont {Wittig}},\ }\href
  {https://doi.org/10.1103/PhysRevD.99.074505} {\bibfield  {journal} {\bibinfo
  {journal} {Phys. Rev.}\ }\textbf {\bibinfo {volume} {D99}},\ \bibinfo {pages}
  {074505} (\bibinfo {year} {2019})},\ \Eprint
  {https://arxiv.org/abs/1805.03966} {arXiv:1805.03966 [hep-lat]} \BibitemShut
  {NoStop}%
\bibitem [{\citenamefont {Ishii}\ \emph {et~al.}(2012)\citenamefont {Ishii},
  \citenamefont {Aoki}, \citenamefont {Doi}, \citenamefont {Hatsuda},
  \citenamefont {Ikeda}, \citenamefont {Inoue}, \citenamefont {Murano},
  \citenamefont {Nemura},\ and\ \citenamefont {Sasaki}}]{HALQCD:2012aa}%
  \BibitemOpen
  \bibfield  {author} {\bibinfo {author} {\bibfnamefont {N.}~\bibnamefont
  {Ishii}}, \bibinfo {author} {\bibfnamefont {S.}~\bibnamefont {Aoki}},
  \bibinfo {author} {\bibfnamefont {T.}~\bibnamefont {Doi}}, \bibinfo {author}
  {\bibfnamefont {T.}~\bibnamefont {Hatsuda}}, \bibinfo {author} {\bibfnamefont
  {Y.}~\bibnamefont {Ikeda}}, \bibinfo {author} {\bibfnamefont
  {T.}~\bibnamefont {Inoue}}, \bibinfo {author} {\bibfnamefont
  {K.}~\bibnamefont {Murano}}, \bibinfo {author} {\bibfnamefont
  {H.}~\bibnamefont {Nemura}},\ and\ \bibinfo {author} {\bibfnamefont
  {K.}~\bibnamefont {Sasaki}} (\bibinfo {collaboration} {HAL QCD}),\ }\href
  {https://doi.org/10.1016/j.physletb.2012.04.076} {\bibfield  {journal}
  {\bibinfo  {journal} {Phys. Lett.}\ }\textbf {\bibinfo {volume} {B712}},\
  \bibinfo {pages} {437} (\bibinfo {year} {2012})},\ \Eprint
  {https://arxiv.org/abs/1203.3642} {arXiv:1203.3642 [hep-lat]} \BibitemShut
  {NoStop}%
\bibitem [{\citenamefont {Hanlon}\ \emph {et~al.}(2018)\citenamefont {Hanlon},
  \citenamefont {Francis}, \citenamefont {Green}, \citenamefont {Junnarkar},\
  and\ \citenamefont {Wittig}}]{Hanlon:2018yfv}%
  \BibitemOpen
  \bibfield  {author} {\bibinfo {author} {\bibfnamefont {A.}~\bibnamefont
  {Hanlon}}, \bibinfo {author} {\bibfnamefont {A.}~\bibnamefont {Francis}},
  \bibinfo {author} {\bibfnamefont {J.}~\bibnamefont {Green}}, \bibinfo
  {author} {\bibfnamefont {P.}~\bibnamefont {Junnarkar}},\ and\ \bibinfo
  {author} {\bibfnamefont {H.}~\bibnamefont {Wittig}},\ }\bibfield  {booktitle}
  {\emph {\bibinfo {booktitle} {{Proceedings, 36th International Symposium on
  Lattice Field Theory (Lattice 2018): East Lansing, MI, United States, July
  22-28, 2018}}},\ }\href {https://doi.org/10.22323/1.334.0081} {\bibfield
  {journal} {\bibinfo  {journal} {PoS}\ }\textbf {\bibinfo {volume}
  {LATTICE2018}},\ \bibinfo {pages} {081} (\bibinfo {year} {2018})},\ \Eprint
  {https://arxiv.org/abs/1810.13282} {arXiv:1810.13282 [hep-lat]} \BibitemShut
  {NoStop}%
\bibitem [{\citenamefont {Iritani}\ \emph
  {et~al.}(2019{\natexlab{a}})\citenamefont {Iritani}, \citenamefont {Aoki},
  \citenamefont {Doi}, \citenamefont {Gongyo}, \citenamefont {Hatsuda},
  \citenamefont {Ikeda}, \citenamefont {Inoue}, \citenamefont {Ishii},
  \citenamefont {Nemura},\ and\ \citenamefont {Sasaki}}]{Iritani:2018zbt}%
  \BibitemOpen
  \bibfield  {author} {\bibinfo {author} {\bibfnamefont {T.}~\bibnamefont
  {Iritani}}, \bibinfo {author} {\bibfnamefont {S.}~\bibnamefont {Aoki}},
  \bibinfo {author} {\bibfnamefont {T.}~\bibnamefont {Doi}}, \bibinfo {author}
  {\bibfnamefont {S.}~\bibnamefont {Gongyo}}, \bibinfo {author} {\bibfnamefont
  {T.}~\bibnamefont {Hatsuda}}, \bibinfo {author} {\bibfnamefont
  {Y.}~\bibnamefont {Ikeda}}, \bibinfo {author} {\bibfnamefont
  {T.}~\bibnamefont {Inoue}}, \bibinfo {author} {\bibfnamefont
  {N.}~\bibnamefont {Ishii}}, \bibinfo {author} {\bibfnamefont
  {H.}~\bibnamefont {Nemura}},\ and\ \bibinfo {author} {\bibfnamefont
  {K.}~\bibnamefont {Sasaki}} (\bibinfo {collaboration} {HAL QCD}),\ }\href
  {https://doi.org/10.1103/PhysRevD.99.014514} {\bibfield  {journal} {\bibinfo
  {journal} {Phys. Rev.}\ }\textbf {\bibinfo {volume} {D99}},\ \bibinfo {pages}
  {014514} (\bibinfo {year} {2019}{\natexlab{a}})},\ \Eprint
  {https://arxiv.org/abs/1805.02365} {arXiv:1805.02365 [hep-lat]} \BibitemShut
  {NoStop}%
\bibitem [{\citenamefont {Iritani}\ \emph
  {et~al.}(2019{\natexlab{b}})\citenamefont {Iritani} \emph
  {et~al.}}]{Iritani:2018sra}%
  \BibitemOpen
  \bibfield  {author} {\bibinfo {author} {\bibfnamefont {T.}~\bibnamefont
  {Iritani}} \emph {et~al.} (\bibinfo {collaboration} {HAL QCD}),\ }\href
  {https://doi.org/10.1016/j.physletb.2019.03.050} {\bibfield  {journal}
  {\bibinfo  {journal} {Phys. Lett.}\ }\textbf {\bibinfo {volume} {B792}},\
  \bibinfo {pages} {284} (\bibinfo {year} {2019}{\natexlab{b}})},\ \Eprint
  {https://arxiv.org/abs/1810.03416} {arXiv:1810.03416 [hep-lat]} \BibitemShut
  {NoStop}%
\bibitem [{\citenamefont {Ishii}\ \emph {et~al.}(2007)\citenamefont {Ishii},
  \citenamefont {Aoki},\ and\ \citenamefont {Hatsuda}}]{Ishii:2006ec}%
  \BibitemOpen
  \bibfield  {author} {\bibinfo {author} {\bibfnamefont {N.}~\bibnamefont
  {Ishii}}, \bibinfo {author} {\bibfnamefont {S.}~\bibnamefont {Aoki}},\ and\
  \bibinfo {author} {\bibfnamefont {T.}~\bibnamefont {Hatsuda}},\ }\href
  {https://doi.org/10.1103/PhysRevLett.99.022001} {\bibfield  {journal}
  {\bibinfo  {journal} {Phys. Rev. Lett.}\ }\textbf {\bibinfo {volume} {99}},\
  \bibinfo {pages} {022001} (\bibinfo {year} {2007})},\ \Eprint
  {https://arxiv.org/abs/nucl-th/0611096} {arXiv:nucl-th/0611096 [nucl-th]}
  \BibitemShut {NoStop}%
\bibitem [{\citenamefont {Nemura}\ \emph {et~al.}(2017)\citenamefont {Nemura}
  \emph {et~al.}}]{Nemura:2017bbw}%
  \BibitemOpen
  \bibfield  {author} {\bibinfo {author} {\bibfnamefont {H.}~\bibnamefont
  {Nemura}} \emph {et~al.},\ }\bibfield  {booktitle} {\emph {\bibinfo
  {booktitle} {{Proceedings, 34th International Symposium on Lattice Field
  Theory (Lattice 2016): Southampton, UK, July 24-30, 2016}}},\ }\href
  {https://doi.org/10.22323/1.256.0101} {\bibfield  {journal} {\bibinfo
  {journal} {PoS}\ }\textbf {\bibinfo {volume} {LATTICE2016}},\ \bibinfo
  {pages} {101} (\bibinfo {year} {2017})},\ \Eprint
  {https://arxiv.org/abs/1702.00734} {arXiv:1702.00734 [hep-lat]} \BibitemShut
  {NoStop}%
\bibitem [{\citenamefont {Orginos}\ \emph {et~al.}(2015)\citenamefont
  {Orginos}, \citenamefont {Parreno}, \citenamefont {Savage}, \citenamefont
  {Beane}, \citenamefont {Chang},\ and\ \citenamefont
  {Detmold}}]{Orginos:2015aya}%
  \BibitemOpen
  \bibfield  {author} {\bibinfo {author} {\bibfnamefont {K.}~\bibnamefont
  {Orginos}}, \bibinfo {author} {\bibfnamefont {A.}~\bibnamefont {Parreno}},
  \bibinfo {author} {\bibfnamefont {M.~J.}\ \bibnamefont {Savage}}, \bibinfo
  {author} {\bibfnamefont {S.~R.}\ \bibnamefont {Beane}}, \bibinfo {author}
  {\bibfnamefont {E.}~\bibnamefont {Chang}},\ and\ \bibinfo {author}
  {\bibfnamefont {W.}~\bibnamefont {Detmold}},\ }\href
  {https://doi.org/10.1103/PhysRevD.92.114512} {\bibfield  {journal} {\bibinfo
  {journal} {Phys. Rev.}\ }\textbf {\bibinfo {volume} {D92}},\ \bibinfo {pages}
  {114512} (\bibinfo {year} {2015})},\ \Eprint
  {https://arxiv.org/abs/1508.07583} {arXiv:1508.07583 [hep-lat]} \BibitemShut
  {NoStop}%
\bibitem [{\citenamefont {Savage}\ \emph {et~al.}(2017)\citenamefont {Savage},
  \citenamefont {Shanahan}, \citenamefont {Tiburzi}, \citenamefont {Wagman},
  \citenamefont {Winter}, \citenamefont {Beane}, \citenamefont {Chang},
  \citenamefont {Davoudi}, \citenamefont {Detmold},\ and\ \citenamefont
  {Orginos}}]{Savage:2016kon}%
  \BibitemOpen
  \bibfield  {author} {\bibinfo {author} {\bibfnamefont {M.~J.}\ \bibnamefont
  {Savage}}, \bibinfo {author} {\bibfnamefont {P.~E.}\ \bibnamefont
  {Shanahan}}, \bibinfo {author} {\bibfnamefont {B.~C.}\ \bibnamefont
  {Tiburzi}}, \bibinfo {author} {\bibfnamefont {M.~L.}\ \bibnamefont {Wagman}},
  \bibinfo {author} {\bibfnamefont {F.}~\bibnamefont {Winter}}, \bibinfo
  {author} {\bibfnamefont {S.~R.}\ \bibnamefont {Beane}}, \bibinfo {author}
  {\bibfnamefont {E.}~\bibnamefont {Chang}}, \bibinfo {author} {\bibfnamefont
  {Z.}~\bibnamefont {Davoudi}}, \bibinfo {author} {\bibfnamefont
  {W.}~\bibnamefont {Detmold}},\ and\ \bibinfo {author} {\bibfnamefont
  {K.}~\bibnamefont {Orginos}},\ }\href
  {https://doi.org/10.1103/PhysRevLett.119.062002} {\bibfield  {journal}
  {\bibinfo  {journal} {Phys. Rev. Lett.}\ }\textbf {\bibinfo {volume} {119}},\
  \bibinfo {pages} {062002} (\bibinfo {year} {2017})},\ \Eprint
  {https://arxiv.org/abs/1610.04545} {arXiv:1610.04545 [hep-lat]} \BibitemShut
  {NoStop}%
\bibitem [{\citenamefont {Shanahan}\ \emph {et~al.}(2017)\citenamefont
  {Shanahan}, \citenamefont {Tiburzi}, \citenamefont {Wagman}, \citenamefont
  {Winter}, \citenamefont {Chang}, \citenamefont {Davoudi}, \citenamefont
  {Detmold}, \citenamefont {Orginos},\ and\ \citenamefont
  {Savage}}]{Shanahan:2017bgi}%
  \BibitemOpen
  \bibfield  {author} {\bibinfo {author} {\bibfnamefont {P.~E.}\ \bibnamefont
  {Shanahan}}, \bibinfo {author} {\bibfnamefont {B.~C.}\ \bibnamefont
  {Tiburzi}}, \bibinfo {author} {\bibfnamefont {M.~L.}\ \bibnamefont {Wagman}},
  \bibinfo {author} {\bibfnamefont {F.}~\bibnamefont {Winter}}, \bibinfo
  {author} {\bibfnamefont {E.}~\bibnamefont {Chang}}, \bibinfo {author}
  {\bibfnamefont {Z.}~\bibnamefont {Davoudi}}, \bibinfo {author} {\bibfnamefont
  {W.}~\bibnamefont {Detmold}}, \bibinfo {author} {\bibfnamefont
  {K.}~\bibnamefont {Orginos}},\ and\ \bibinfo {author} {\bibfnamefont {M.~J.}\
  \bibnamefont {Savage}},\ }\href
  {https://doi.org/10.1103/PhysRevLett.119.062003} {\bibfield  {journal}
  {\bibinfo  {journal} {Phys. Rev. Lett.}\ }\textbf {\bibinfo {volume} {119}},\
  \bibinfo {pages} {062003} (\bibinfo {year} {2017})},\ \Eprint
  {https://arxiv.org/abs/1701.03456} {arXiv:1701.03456 [hep-lat]} \BibitemShut
  {NoStop}%
\bibitem [{\citenamefont {Tiburzi}\ \emph {et~al.}(2017)\citenamefont
  {Tiburzi}, \citenamefont {Wagman}, \citenamefont {Winter}, \citenamefont
  {Chang}, \citenamefont {Davoudi}, \citenamefont {Detmold}, \citenamefont
  {Orginos}, \citenamefont {Savage},\ and\ \citenamefont
  {Shanahan}}]{Tiburzi:2017iux}%
  \BibitemOpen
  \bibfield  {author} {\bibinfo {author} {\bibfnamefont {B.~C.}\ \bibnamefont
  {Tiburzi}}, \bibinfo {author} {\bibfnamefont {M.~L.}\ \bibnamefont {Wagman}},
  \bibinfo {author} {\bibfnamefont {F.}~\bibnamefont {Winter}}, \bibinfo
  {author} {\bibfnamefont {E.}~\bibnamefont {Chang}}, \bibinfo {author}
  {\bibfnamefont {Z.}~\bibnamefont {Davoudi}}, \bibinfo {author} {\bibfnamefont
  {W.}~\bibnamefont {Detmold}}, \bibinfo {author} {\bibfnamefont
  {K.}~\bibnamefont {Orginos}}, \bibinfo {author} {\bibfnamefont {M.~J.}\
  \bibnamefont {Savage}},\ and\ \bibinfo {author} {\bibfnamefont {P.~E.}\
  \bibnamefont {Shanahan}},\ }\href
  {https://doi.org/10.1103/PhysRevD.96.054505} {\bibfield  {journal} {\bibinfo
  {journal} {Phys. Rev.}\ }\textbf {\bibinfo {volume} {D96}},\ \bibinfo {pages}
  {054505} (\bibinfo {year} {2017})},\ \Eprint
  {https://arxiv.org/abs/1702.02929} {arXiv:1702.02929 [hep-lat]} \BibitemShut
  {NoStop}%
\bibitem [{\citenamefont {Wagman}\ \emph {et~al.}(2017)\citenamefont {Wagman},
  \citenamefont {Winter}, \citenamefont {Chang}, \citenamefont {Davoudi},
  \citenamefont {Detmold}, \citenamefont {Orginos}, \citenamefont {Savage},\
  and\ \citenamefont {Shanahan}}]{Wagman:2017tmp}%
  \BibitemOpen
  \bibfield  {author} {\bibinfo {author} {\bibfnamefont {M.~L.}\ \bibnamefont
  {Wagman}}, \bibinfo {author} {\bibfnamefont {F.}~\bibnamefont {Winter}},
  \bibinfo {author} {\bibfnamefont {E.}~\bibnamefont {Chang}}, \bibinfo
  {author} {\bibfnamefont {Z.}~\bibnamefont {Davoudi}}, \bibinfo {author}
  {\bibfnamefont {W.}~\bibnamefont {Detmold}}, \bibinfo {author} {\bibfnamefont
  {K.}~\bibnamefont {Orginos}}, \bibinfo {author} {\bibfnamefont {M.~J.}\
  \bibnamefont {Savage}},\ and\ \bibinfo {author} {\bibfnamefont {P.~E.}\
  \bibnamefont {Shanahan}},\ }\href
  {https://doi.org/10.1103/PhysRevD.96.114510} {\bibfield  {journal} {\bibinfo
  {journal} {Phys. Rev.}\ }\textbf {\bibinfo {volume} {D96}},\ \bibinfo {pages}
  {114510} (\bibinfo {year} {2017})},\ \Eprint
  {https://arxiv.org/abs/1706.06550} {arXiv:1706.06550 [hep-lat]} \BibitemShut
  {NoStop}%
\bibitem [{\citenamefont {Winter}\ \emph {et~al.}(2017)\citenamefont {Winter},
  \citenamefont {Detmold}, \citenamefont {Gambhir}, \citenamefont {Orginos},
  \citenamefont {Savage}, \citenamefont {Shanahan},\ and\ \citenamefont
  {Wagman}}]{Winter:2017bfs}%
  \BibitemOpen
  \bibfield  {author} {\bibinfo {author} {\bibfnamefont {F.}~\bibnamefont
  {Winter}}, \bibinfo {author} {\bibfnamefont {W.}~\bibnamefont {Detmold}},
  \bibinfo {author} {\bibfnamefont {A.~S.}\ \bibnamefont {Gambhir}}, \bibinfo
  {author} {\bibfnamefont {K.}~\bibnamefont {Orginos}}, \bibinfo {author}
  {\bibfnamefont {M.~J.}\ \bibnamefont {Savage}}, \bibinfo {author}
  {\bibfnamefont {P.~E.}\ \bibnamefont {Shanahan}},\ and\ \bibinfo {author}
  {\bibfnamefont {M.~L.}\ \bibnamefont {Wagman}},\ }\href
  {https://doi.org/10.1103/PhysRevD.96.094512} {\bibfield  {journal} {\bibinfo
  {journal} {Phys. Rev.}\ }\textbf {\bibinfo {volume} {D96}},\ \bibinfo {pages}
  {094512} (\bibinfo {year} {2017})},\ \Eprint
  {https://arxiv.org/abs/1709.00395} {arXiv:1709.00395 [hep-lat]} \BibitemShut
  {NoStop}%
\bibitem [{\citenamefont {Yamazaki}\ \emph {et~al.}(2010)\citenamefont
  {Yamazaki}, \citenamefont {Kuramashi},\ and\ \citenamefont
  {Ukawa}}]{Yamazaki:2009ua}%
  \BibitemOpen
  \bibfield  {author} {\bibinfo {author} {\bibfnamefont {T.}~\bibnamefont
  {Yamazaki}}, \bibinfo {author} {\bibfnamefont {Y.}~\bibnamefont
  {Kuramashi}},\ and\ \bibinfo {author} {\bibfnamefont {A.}~\bibnamefont
  {Ukawa}} (\bibinfo {collaboration} {PACS-CS}),\ }\href
  {https://doi.org/10.1103/PhysRevD.81.111504} {\bibfield  {journal} {\bibinfo
  {journal} {Phys. Rev.}\ }\textbf {\bibinfo {volume} {D81}},\ \bibinfo {pages}
  {111504} (\bibinfo {year} {2010})},\ \Eprint
  {https://arxiv.org/abs/0912.1383} {arXiv:0912.1383 [hep-lat]} \BibitemShut
  {NoStop}%
\bibitem [{\citenamefont {Yamazaki}\ \emph {et~al.}(2011)\citenamefont
  {Yamazaki}, \citenamefont {Kuramashi},\ and\ \citenamefont
  {Ukawa}}]{Yamazaki:2011nd}%
  \BibitemOpen
  \bibfield  {author} {\bibinfo {author} {\bibfnamefont {T.}~\bibnamefont
  {Yamazaki}}, \bibinfo {author} {\bibfnamefont {Y.}~\bibnamefont
  {Kuramashi}},\ and\ \bibinfo {author} {\bibfnamefont {A.}~\bibnamefont
  {Ukawa}} (\bibinfo {collaboration} {PACS-CS}),\ }\href
  {https://doi.org/10.1103/PhysRevD.84.054506} {\bibfield  {journal} {\bibinfo
  {journal} {Phys. Rev.}\ }\textbf {\bibinfo {volume} {D84}},\ \bibinfo {pages}
  {054506} (\bibinfo {year} {2011})},\ \Eprint
  {https://arxiv.org/abs/1105.1418} {arXiv:1105.1418 [hep-lat]} \BibitemShut
  {NoStop}%
\bibitem [{\citenamefont {Yamazaki}\ \emph {et~al.}(2012)\citenamefont
  {Yamazaki}, \citenamefont {Ishikawa}, \citenamefont {Kuramashi},\ and\
  \citenamefont {Ukawa}}]{Yamazaki:2012hi}%
  \BibitemOpen
  \bibfield  {author} {\bibinfo {author} {\bibfnamefont {T.}~\bibnamefont
  {Yamazaki}}, \bibinfo {author} {\bibfnamefont {K.-i.}\ \bibnamefont
  {Ishikawa}}, \bibinfo {author} {\bibfnamefont {Y.}~\bibnamefont
  {Kuramashi}},\ and\ \bibinfo {author} {\bibfnamefont {A.}~\bibnamefont
  {Ukawa}},\ }\href {https://doi.org/10.1103/PhysRevD.86.074514} {\bibfield
  {journal} {\bibinfo  {journal} {Phys. Rev.}\ }\textbf {\bibinfo {volume}
  {D86}},\ \bibinfo {pages} {074514} (\bibinfo {year} {2012})},\ \Eprint
  {https://arxiv.org/abs/1207.4277} {arXiv:1207.4277 [hep-lat]} \BibitemShut
  {NoStop}%
\bibitem [{\citenamefont {Yamazaki}\ \emph {et~al.}(2015)\citenamefont
  {Yamazaki}, \citenamefont {Ishikawa}, \citenamefont {Kuramashi},\ and\
  \citenamefont {Ukawa}}]{Yamazaki:2015asa}%
  \BibitemOpen
  \bibfield  {author} {\bibinfo {author} {\bibfnamefont {T.}~\bibnamefont
  {Yamazaki}}, \bibinfo {author} {\bibfnamefont {K.-i.}\ \bibnamefont
  {Ishikawa}}, \bibinfo {author} {\bibfnamefont {Y.}~\bibnamefont
  {Kuramashi}},\ and\ \bibinfo {author} {\bibfnamefont {A.}~\bibnamefont
  {Ukawa}},\ }\href {https://doi.org/10.1103/PhysRevD.92.014501} {\bibfield
  {journal} {\bibinfo  {journal} {Phys. Rev.}\ }\textbf {\bibinfo {volume}
  {D92}},\ \bibinfo {pages} {014501} (\bibinfo {year} {2015})},\ \Eprint
  {https://arxiv.org/abs/1502.04182} {arXiv:1502.04182 [hep-lat]} \BibitemShut
  {NoStop}%
\bibitem [{\citenamefont {Yamazaki}(2016)}]{Yamazaki:2015vjn}%
  \BibitemOpen
  \bibfield  {author} {\bibinfo {author} {\bibfnamefont {T.}~\bibnamefont
  {Yamazaki}} (\bibinfo {collaboration} {PACS}),\ }\bibfield  {booktitle}
  {\emph {\bibinfo {booktitle} {{Proceedings, 33rd International Symposium on
  Lattice Field Theory (Lattice 2015): Kobe, Japan, July 14-18, 2015}}},\
  }\href {https://doi.org/10.22323/1.251.0081} {\bibfield  {journal} {\bibinfo
  {journal} {PoS}\ }\textbf {\bibinfo {volume} {LATTICE2015}},\ \bibinfo
  {pages} {081} (\bibinfo {year} {2016})},\ \Eprint
  {https://arxiv.org/abs/1511.09179} {arXiv:1511.09179 [hep-lat]} \BibitemShut
  {NoStop}%
\bibitem [{\citenamefont {Yamazaki}\ \emph {et~al.}(2018)\citenamefont
  {Yamazaki}, \citenamefont {Ishikawa},\ and\ \citenamefont
  {Kuramashi}}]{Yamazaki:2017jfh}%
  \BibitemOpen
  \bibfield  {author} {\bibinfo {author} {\bibfnamefont {T.}~\bibnamefont
  {Yamazaki}}, \bibinfo {author} {\bibfnamefont {K.-i.}\ \bibnamefont
  {Ishikawa}},\ and\ \bibinfo {author} {\bibfnamefont {Y.}~\bibnamefont
  {Kuramashi}} (\bibinfo {collaboration} {PACS}),\ }\bibfield  {booktitle}
  {\emph {\bibinfo {booktitle} {{Proceedings, 35th International Symposium on
  Lattice Field Theory (Lattice 2017): Granada, Spain, June 18-24, 2017}}},\
  }\href {https://doi.org/10.1051/epjconf/201817505019} {\bibfield  {journal}
  {\bibinfo  {journal} {EPJ Web Conf.}\ }\textbf {\bibinfo {volume} {175}},\
  \bibinfo {pages} {05019} (\bibinfo {year} {2018})},\ \Eprint
  {https://arxiv.org/abs/1710.08066} {arXiv:1710.08066 [hep-lat]} \BibitemShut
  {NoStop}%
\bibitem [{\citenamefont {Duane}\ \emph {et~al.}(1987)\citenamefont {Duane},
  \citenamefont {Kennedy}, \citenamefont {Pendleton},\ and\ \citenamefont
  {Roweth}}]{Duane:1987de}%
  \BibitemOpen
  \bibfield  {author} {\bibinfo {author} {\bibfnamefont {S.}~\bibnamefont
  {Duane}}, \bibinfo {author} {\bibfnamefont {A.~D.}\ \bibnamefont {Kennedy}},
  \bibinfo {author} {\bibfnamefont {B.~J.}\ \bibnamefont {Pendleton}},\ and\
  \bibinfo {author} {\bibfnamefont {D.}~\bibnamefont {Roweth}},\ }\href
  {https://doi.org/10.1016/0370-2693(87)91197-X} {\bibfield  {journal}
  {\bibinfo  {journal} {Phys. Lett.}\ }\textbf {\bibinfo {volume} {B195}},\
  \bibinfo {pages} {216} (\bibinfo {year} {1987})}\BibitemShut {NoStop}%
\bibitem [{\citenamefont {Babich}\ \emph {et~al.}(2010)\citenamefont {Babich},
  \citenamefont {Brannick}, \citenamefont {Brower}, \citenamefont {Clark},
  \citenamefont {Manteuffel}, \citenamefont {McCormick}, \citenamefont
  {Osborn},\ and\ \citenamefont {Rebbi}}]{Babich:2010qb}%
  \BibitemOpen
  \bibfield  {author} {\bibinfo {author} {\bibfnamefont {R.}~\bibnamefont
  {Babich}}, \bibinfo {author} {\bibfnamefont {J.}~\bibnamefont {Brannick}},
  \bibinfo {author} {\bibfnamefont {R.~C.}\ \bibnamefont {Brower}}, \bibinfo
  {author} {\bibfnamefont {M.~A.}\ \bibnamefont {Clark}}, \bibinfo {author}
  {\bibfnamefont {T.~A.}\ \bibnamefont {Manteuffel}}, \bibinfo {author}
  {\bibfnamefont {S.~F.}\ \bibnamefont {McCormick}}, \bibinfo {author}
  {\bibfnamefont {J.~C.}\ \bibnamefont {Osborn}},\ and\ \bibinfo {author}
  {\bibfnamefont {C.}~\bibnamefont {Rebbi}},\ }\href
  {https://doi.org/10.1103/PhysRevLett.105.201602} {\bibfield  {journal}
  {\bibinfo  {journal} {Phys. Rev. Lett.}\ }\textbf {\bibinfo {volume} {105}},\
  \bibinfo {pages} {201602} (\bibinfo {year} {2010})},\ \Eprint
  {https://arxiv.org/abs/1005.3043} {arXiv:1005.3043 [hep-lat]} \BibitemShut
  {NoStop}%
\bibitem [{\citenamefont {Osborn}\ \emph {et~al.}(2010)\citenamefont {Osborn},
  \citenamefont {Babich}, \citenamefont {Brannick}, \citenamefont {Brower},
  \citenamefont {Clark}, \citenamefont {Cohen},\ and\ \citenamefont
  {Rebbi}}]{Osborn:2010mb}%
  \BibitemOpen
  \bibfield  {author} {\bibinfo {author} {\bibfnamefont {J.~C.}\ \bibnamefont
  {Osborn}}, \bibinfo {author} {\bibfnamefont {R.}~\bibnamefont {Babich}},
  \bibinfo {author} {\bibfnamefont {J.}~\bibnamefont {Brannick}}, \bibinfo
  {author} {\bibfnamefont {R.~C.}\ \bibnamefont {Brower}}, \bibinfo {author}
  {\bibfnamefont {M.~A.}\ \bibnamefont {Clark}}, \bibinfo {author}
  {\bibfnamefont {S.~D.}\ \bibnamefont {Cohen}},\ and\ \bibinfo {author}
  {\bibfnamefont {C.}~\bibnamefont {Rebbi}},\ }\bibfield  {booktitle} {\emph
  {\bibinfo {booktitle} {{Proceedings, 28th International Symposium on Lattice
  field theory (Lattice 2010): Villasimius, Italy, June 14-19, 2010}}},\ }\href
  {https://doi.org/10.22323/1.105.0037} {\bibfield  {journal} {\bibinfo
  {journal} {PoS}\ }\textbf {\bibinfo {volume} {LATTICE2010}},\ \bibinfo
  {pages} {037} (\bibinfo {year} {2010})},\ \Eprint
  {https://arxiv.org/abs/1011.2775} {arXiv:1011.2775 [hep-lat]} \BibitemShut
  {NoStop}%
\bibitem [{\citenamefont {Boyle}(2014)}]{Boyle:2014rwa}%
  \BibitemOpen
  \bibfield  {author} {\bibinfo {author} {\bibfnamefont {P.~A.}\ \bibnamefont
  {Boyle}},\ }\href@noop {} {\  (\bibinfo {year} {2014})},\ \Eprint
  {https://arxiv.org/abs/1402.2585} {arXiv:1402.2585 [hep-lat]} \BibitemShut
  {NoStop}%
\bibitem [{\citenamefont {Endres}\ \emph {et~al.}(2015)\citenamefont {Endres},
  \citenamefont {Brower}, \citenamefont {Detmold}, \citenamefont {Orginos},\
  and\ \citenamefont {Pochinsky}}]{PhysRevD.92.114516}%
  \BibitemOpen
  \bibfield  {author} {\bibinfo {author} {\bibfnamefont {M.~G.}\ \bibnamefont
  {Endres}}, \bibinfo {author} {\bibfnamefont {R.~C.}\ \bibnamefont {Brower}},
  \bibinfo {author} {\bibfnamefont {W.}~\bibnamefont {Detmold}}, \bibinfo
  {author} {\bibfnamefont {K.}~\bibnamefont {Orginos}},\ and\ \bibinfo {author}
  {\bibfnamefont {A.~V.}\ \bibnamefont {Pochinsky}},\ }\href
  {https://doi.org/10.1103/PhysRevD.92.114516} {\bibfield  {journal} {\bibinfo
  {journal} {Phys. Rev. D}\ }\textbf {\bibinfo {volume} {92}},\ \bibinfo
  {pages} {114516} (\bibinfo {year} {2015})}\BibitemShut {NoStop}%
\bibitem [{\citenamefont {Detmold}\ and\ \citenamefont
  {Endres}(2016)}]{Detmold:2016rnh}%
  \BibitemOpen
  \bibfield  {author} {\bibinfo {author} {\bibfnamefont {W.}~\bibnamefont
  {Detmold}}\ and\ \bibinfo {author} {\bibfnamefont {M.~G.}\ \bibnamefont
  {Endres}},\ }\href {https://doi.org/10.1103/PhysRevD.94.114502} {\bibfield
  {journal} {\bibinfo  {journal} {Phys. Rev.}\ }\textbf {\bibinfo {volume}
  {D94}},\ \bibinfo {pages} {114502} (\bibinfo {year} {2016})},\ \Eprint
  {https://arxiv.org/abs/1605.09650} {arXiv:1605.09650 [hep-lat]} \BibitemShut
  {NoStop}%
\bibitem [{\citenamefont {Clark}\ \emph {et~al.}(2016)\citenamefont {Clark},
  \citenamefont {Joó}, \citenamefont {Strelchenko}, \citenamefont {Cheng},
  \citenamefont {Gambhir},\ and\ \citenamefont {Brower}}]{Clark:2016rdz}%
  \BibitemOpen
  \bibfield  {author} {\bibinfo {author} {\bibfnamefont {M.~A.}\ \bibnamefont
  {Clark}}, \bibinfo {author} {\bibfnamefont {B.}~\bibnamefont {Joó}},
  \bibinfo {author} {\bibfnamefont {A.}~\bibnamefont {Strelchenko}}, \bibinfo
  {author} {\bibfnamefont {M.}~\bibnamefont {Cheng}}, \bibinfo {author}
  {\bibfnamefont {A.}~\bibnamefont {Gambhir}},\ and\ \bibinfo {author}
  {\bibfnamefont {R.}~\bibnamefont {Brower}},\ }\href@noop {} {\  (\bibinfo
  {year} {2016})},\ \Eprint {https://arxiv.org/abs/1612.07873}
  {arXiv:1612.07873 [hep-lat]} \BibitemShut {NoStop}%
\bibitem [{\citenamefont {Yamaguchi}\ and\ \citenamefont
  {Boyle}(2016)}]{Yamaguchi:2016kop}%
  \BibitemOpen
  \bibfield  {author} {\bibinfo {author} {\bibfnamefont {A.}~\bibnamefont
  {Yamaguchi}}\ and\ \bibinfo {author} {\bibfnamefont {P.}~\bibnamefont
  {Boyle}},\ }\bibfield  {booktitle} {\emph {\bibinfo {booktitle}
  {{Proceedings, 34th International Symposium on Lattice Field Theory (Lattice
  2016): Southampton, UK, July 24-30, 2016}}},\ }\href
  {https://doi.org/10.22323/1.256.0374} {\bibfield  {journal} {\bibinfo
  {journal} {PoS}\ }\textbf {\bibinfo {volume} {LATTICE2016}},\ \bibinfo
  {pages} {374} (\bibinfo {year} {2016})},\ \Eprint
  {https://arxiv.org/abs/1611.06944} {arXiv:1611.06944 [hep-lat]} \BibitemShut
  {NoStop}%
\bibitem [{\citenamefont {Bacchio}\ \emph {et~al.}(2018)\citenamefont
  {Bacchio}, \citenamefont {Alexandrou},\ and\ \citenamefont
  {Finkerath}}]{Bacchio:2017pcp}%
  \BibitemOpen
  \bibfield  {author} {\bibinfo {author} {\bibfnamefont {S.}~\bibnamefont
  {Bacchio}}, \bibinfo {author} {\bibfnamefont {C.}~\bibnamefont
  {Alexandrou}},\ and\ \bibinfo {author} {\bibfnamefont {J.}~\bibnamefont
  {Finkerath}},\ }\bibfield  {booktitle} {\emph {\bibinfo {booktitle}
  {{Proceedings, 35th International Symposium on Lattice Field Theory (Lattice
  2017): Granada, Spain, June 18-24, 2017}}},\ }\href
  {https://doi.org/10.1051/epjconf/201817502002} {\bibfield  {journal}
  {\bibinfo  {journal} {EPJ Web Conf.}\ }\textbf {\bibinfo {volume} {175}},\
  \bibinfo {pages} {02002} (\bibinfo {year} {2018})},\ \Eprint
  {https://arxiv.org/abs/1710.06198} {arXiv:1710.06198 [hep-lat]} \BibitemShut
  {NoStop}%
\bibitem [{\citenamefont {Brower}\ \emph {et~al.}(2018)\citenamefont {Brower},
  \citenamefont {Clark}, \citenamefont {Strelchenko},\ and\ \citenamefont
  {Weinberg}}]{Brower:2018ymy}%
  \BibitemOpen
  \bibfield  {author} {\bibinfo {author} {\bibfnamefont {R.~C.}\ \bibnamefont
  {Brower}}, \bibinfo {author} {\bibfnamefont {M.~A.}\ \bibnamefont {Clark}},
  \bibinfo {author} {\bibfnamefont {A.}~\bibnamefont {Strelchenko}},\ and\
  \bibinfo {author} {\bibfnamefont {E.}~\bibnamefont {Weinberg}},\ }\href
  {https://doi.org/10.1103/PhysRevD.97.114513} {\bibfield  {journal} {\bibinfo
  {journal} {Phys. Rev.}\ }\textbf {\bibinfo {volume} {D97}},\ \bibinfo {pages}
  {114513} (\bibinfo {year} {2018})},\ \Eprint
  {https://arxiv.org/abs/1801.07823} {arXiv:1801.07823 [hep-lat]} \BibitemShut
  {NoStop}%
\bibitem [{\citenamefont {Richtmann}\ \emph {et~al.}(2019)\citenamefont
  {Richtmann}, \citenamefont {Boyle},\ and\ \citenamefont
  {Wettig}}]{Richtmann:2019eyj}%
  \BibitemOpen
  \bibfield  {author} {\bibinfo {author} {\bibfnamefont {D.}~\bibnamefont
  {Richtmann}}, \bibinfo {author} {\bibfnamefont {P.~A.}\ \bibnamefont
  {Boyle}},\ and\ \bibinfo {author} {\bibfnamefont {T.}~\bibnamefont
  {Wettig}},\ }in\ \href@noop {} {\emph {\bibinfo {booktitle} {{36th
  International Symposium on Lattice Field Theory (Lattice 2018) East Lansing,
  MI, United States, July 22-28, 2018}}}}\ (\bibinfo {year} {2019})\ \Eprint
  {https://arxiv.org/abs/1904.08678} {arXiv:1904.08678 [hep-lat]} \BibitemShut
  {NoStop}%
\bibitem [{\citenamefont {Clark}\ \emph {et~al.}(2018)\citenamefont {Clark},
  \citenamefont {Jung},\ and\ \citenamefont {Lehner}}]{Clark:2017wom}%
  \BibitemOpen
  \bibfield  {author} {\bibinfo {author} {\bibfnamefont {M.~A.}\ \bibnamefont
  {Clark}}, \bibinfo {author} {\bibfnamefont {C.}~\bibnamefont {Jung}},\ and\
  \bibinfo {author} {\bibfnamefont {C.}~\bibnamefont {Lehner}},\ }\bibfield
  {booktitle} {\emph {\bibinfo {booktitle} {{Proceedings, 35th International
  Symposium on Lattice Field Theory (Lattice 2017): Granada, Spain, June 18-24,
  2017}}},\ }\href {https://doi.org/10.1051/epjconf/201817514023} {\bibfield
  {journal} {\bibinfo  {journal} {EPJ Web Conf.}\ }\textbf {\bibinfo {volume}
  {175}},\ \bibinfo {pages} {14023} (\bibinfo {year} {2018})},\ \Eprint
  {https://arxiv.org/abs/1710.06884} {arXiv:1710.06884 [hep-lat]} \BibitemShut
  {NoStop}%
\bibitem [{\citenamefont {Basak}\ \emph {et~al.}(2005)\citenamefont {Basak},
  \citenamefont {Edwards}, \citenamefont {Fleming}, \citenamefont {Heller},
  \citenamefont {Morningstar}, \citenamefont {Richards}, \citenamefont {Sato},\
  and\ \citenamefont {Wallace}}]{Basak:2005ir}%
  \BibitemOpen
  \bibfield  {author} {\bibinfo {author} {\bibfnamefont {S.}~\bibnamefont
  {Basak}}, \bibinfo {author} {\bibfnamefont {R.}~\bibnamefont {Edwards}},
  \bibinfo {author} {\bibfnamefont {G.~T.}\ \bibnamefont {Fleming}}, \bibinfo
  {author} {\bibfnamefont {U.~M.}\ \bibnamefont {Heller}}, \bibinfo {author}
  {\bibfnamefont {C.}~\bibnamefont {Morningstar}}, \bibinfo {author}
  {\bibfnamefont {D.}~\bibnamefont {Richards}}, \bibinfo {author}
  {\bibfnamefont {I.}~\bibnamefont {Sato}},\ and\ \bibinfo {author}
  {\bibfnamefont {S.~J.}\ \bibnamefont {Wallace}} (\bibinfo {collaboration}
  {Lattice Hadron Physics (LHPC)}),\ }\href
  {https://doi.org/10.1103/PhysRevD.72.074501} {\bibfield  {journal} {\bibinfo
  {journal} {Phys. Rev.}\ }\textbf {\bibinfo {volume} {D72}},\ \bibinfo {pages}
  {074501} (\bibinfo {year} {2005})},\ \Eprint
  {https://arxiv.org/abs/hep-lat/0508018} {arXiv:hep-lat/0508018 [hep-lat]}
  \BibitemShut {NoStop}%
\bibitem [{\citenamefont {Beane}\ \emph
  {et~al.}(2006{\natexlab{b}})\citenamefont {Beane}, \citenamefont {Bedaque},
  \citenamefont {Orginos},\ and\ \citenamefont {Savage}}]{Beane:2005rj}%
  \BibitemOpen
  \bibfield  {author} {\bibinfo {author} {\bibfnamefont {S.~R.}\ \bibnamefont
  {Beane}}, \bibinfo {author} {\bibfnamefont {P.~F.}\ \bibnamefont {Bedaque}},
  \bibinfo {author} {\bibfnamefont {K.}~\bibnamefont {Orginos}},\ and\ \bibinfo
  {author} {\bibfnamefont {M.~J.}\ \bibnamefont {Savage}} (\bibinfo
  {collaboration} {NPLQCD}),\ }\href
  {https://doi.org/10.1103/PhysRevD.73.054503} {\bibfield  {journal} {\bibinfo
  {journal} {Phys. Rev.}\ }\textbf {\bibinfo {volume} {D73}},\ \bibinfo {pages}
  {054503} (\bibinfo {year} {2006}{\natexlab{b}})},\ \Eprint
  {https://arxiv.org/abs/hep-lat/0506013} {arXiv:hep-lat/0506013 [hep-lat]}
  \BibitemShut {NoStop}%
\bibitem [{\citenamefont {Beane}\ \emph {et~al.}(2008)\citenamefont {Beane},
  \citenamefont {Orginos},\ and\ \citenamefont {Savage}}]{Beane:2008dv}%
  \BibitemOpen
  \bibfield  {author} {\bibinfo {author} {\bibfnamefont {S.~R.}\ \bibnamefont
  {Beane}}, \bibinfo {author} {\bibfnamefont {K.}~\bibnamefont {Orginos}},\
  and\ \bibinfo {author} {\bibfnamefont {M.~J.}\ \bibnamefont {Savage}},\
  }\href {https://doi.org/10.1142/S0218301308010404} {\bibfield  {journal}
  {\bibinfo  {journal} {Int. J. Mod. Phys.}\ }\textbf {\bibinfo {volume}
  {E17}},\ \bibinfo {pages} {1157} (\bibinfo {year} {2008})},\ \Eprint
  {https://arxiv.org/abs/0805.4629} {arXiv:0805.4629 [hep-lat]} \BibitemShut
  {NoStop}%
\bibitem [{\citenamefont {Detmold}\ and\ \citenamefont
  {Orginos}(2013)}]{Detmold:2012eu}%
  \BibitemOpen
  \bibfield  {author} {\bibinfo {author} {\bibfnamefont {W.}~\bibnamefont
  {Detmold}}\ and\ \bibinfo {author} {\bibfnamefont {K.}~\bibnamefont
  {Orginos}},\ }\href {https://doi.org/10.1103/PhysRevD.87.114512} {\bibfield
  {journal} {\bibinfo  {journal} {Phys. Rev.}\ }\textbf {\bibinfo {volume}
  {D87}},\ \bibinfo {pages} {114512} (\bibinfo {year} {2013})},\ \Eprint
  {https://arxiv.org/abs/1207.1452} {arXiv:1207.1452 [hep-lat]} \BibitemShut
  {NoStop}%
\bibitem [{\citenamefont {Sheikholeslami}\ and\ \citenamefont
  {Wohlert}(1985)}]{Sheikholeslami:1985ij}%
  \BibitemOpen
  \bibfield  {author} {\bibinfo {author} {\bibfnamefont {B.}~\bibnamefont
  {Sheikholeslami}}\ and\ \bibinfo {author} {\bibfnamefont {R.}~\bibnamefont
  {Wohlert}},\ }\href {https://doi.org/10.1016/0550-3213(85)90002-1} {\bibfield
   {journal} {\bibinfo  {journal} {Nucl. Phys.}\ }\textbf {\bibinfo {volume}
  {B259}},\ \bibinfo {pages} {572} (\bibinfo {year} {1985})}\BibitemShut
  {NoStop}%
\bibitem [{\citenamefont {L{\"u}scher}\ and\ \citenamefont
  {Weisz}(1985)}]{Luscher:1984xn}%
  \BibitemOpen
  \bibfield  {author} {\bibinfo {author} {\bibfnamefont {M.}~\bibnamefont
  {L{\"u}scher}}\ and\ \bibinfo {author} {\bibfnamefont {P.}~\bibnamefont
  {Weisz}},\ }\href {https://doi.org/10.1007/BF01206178} {\bibfield  {journal}
  {\bibinfo  {journal} {Commun. Math. Phys.}\ }\textbf {\bibinfo {volume}
  {97}},\ \bibinfo {pages} {59} (\bibinfo {year} {1985})},\ \bibinfo {note}
  {[Erratum: Commun. Math. Phys.98,433(1985)]}\BibitemShut {NoStop}%
\bibitem [{\citenamefont {Albanese}\ \emph {et~al.}(1987)\citenamefont
  {Albanese} \emph {et~al.}}]{Albanese:1987ds}%
  \BibitemOpen
  \bibfield  {author} {\bibinfo {author} {\bibfnamefont {M.}~\bibnamefont
  {Albanese}} \emph {et~al.} (\bibinfo {collaboration} {APE}),\ }\href
  {https://doi.org/10.1016/0370-2693(87)91160-9} {\bibfield  {journal}
  {\bibinfo  {journal} {Phys. Lett.}\ }\textbf {\bibinfo {volume} {B192}},\
  \bibinfo {pages} {163} (\bibinfo {year} {1987})}\BibitemShut {NoStop}%
\bibitem [{\citenamefont {Wagman}\ and\ \citenamefont
  {Savage}(2017)}]{Wagman:2016bam}%
  \BibitemOpen
  \bibfield  {author} {\bibinfo {author} {\bibfnamefont {M.~L.}\ \bibnamefont
  {Wagman}}\ and\ \bibinfo {author} {\bibfnamefont {M.~J.}\ \bibnamefont
  {Savage}},\ }\href {https://doi.org/10.1103/PhysRevD.96.114508} {\bibfield
  {journal} {\bibinfo  {journal} {Phys. Rev.}\ }\textbf {\bibinfo {volume}
  {D96}},\ \bibinfo {pages} {114508} (\bibinfo {year} {2017})},\ \Eprint
  {https://arxiv.org/abs/1611.07643} {arXiv:1611.07643 [hep-lat]} \BibitemShut
  {NoStop}%
\bibitem [{\citenamefont {Detmold}\ \emph {et~al.}(2019)\citenamefont
  {Detmold}, \citenamefont {Murphy}, \citenamefont {Pochinsky}, \citenamefont
  {Savage}, \citenamefont {Shanahan},\ and\ \citenamefont
  {Wagman}}]{Murphy:2019lp}%
  \BibitemOpen
  \bibfield  {author} {\bibinfo {author} {\bibfnamefont {W.}~\bibnamefont
  {Detmold}}, \bibinfo {author} {\bibfnamefont {D.~J.}\ \bibnamefont {Murphy}},
  \bibinfo {author} {\bibfnamefont {A.~V.}\ \bibnamefont {Pochinsky}}, \bibinfo
  {author} {\bibfnamefont {M.~J.}\ \bibnamefont {Savage}}, \bibinfo {author}
  {\bibfnamefont {P.~E.}\ \bibnamefont {Shanahan}},\ and\ \bibinfo {author}
  {\bibfnamefont {M.~L.}\ \bibnamefont {Wagman}},\ }\bibfield  {booktitle}
  {\emph {\bibinfo {booktitle} {{Proceedings, 37th International Symposium on
  Lattice Field Theory (Lattice 2019): Wuhan, China, June 16-22, 2019}}},\
  }\href@noop {} {\bibfield  {journal} {\bibinfo  {journal} {PoS}\ }\textbf
  {\bibinfo {volume} {LATTICE2019}},\ \bibinfo {pages} {104} (\bibinfo {year}
  {2019})}\BibitemShut {NoStop}%
\bibitem [{\citenamefont {Shintani}\ \emph {et~al.}(2015)\citenamefont
  {Shintani}, \citenamefont {Arthur}, \citenamefont {Blum}, \citenamefont
  {Izubuchi}, \citenamefont {Jung},\ and\ \citenamefont
  {Lehner}}]{Shintani:2014vja}%
  \BibitemOpen
  \bibfield  {author} {\bibinfo {author} {\bibfnamefont {E.}~\bibnamefont
  {Shintani}}, \bibinfo {author} {\bibfnamefont {R.}~\bibnamefont {Arthur}},
  \bibinfo {author} {\bibfnamefont {T.}~\bibnamefont {Blum}}, \bibinfo {author}
  {\bibfnamefont {T.}~\bibnamefont {Izubuchi}}, \bibinfo {author}
  {\bibfnamefont {C.}~\bibnamefont {Jung}},\ and\ \bibinfo {author}
  {\bibfnamefont {C.}~\bibnamefont {Lehner}},\ }\href
  {https://doi.org/10.1103/PhysRevD.91.114511} {\bibfield  {journal} {\bibinfo
  {journal} {Phys. Rev.}\ }\textbf {\bibinfo {volume} {D91}},\ \bibinfo {pages}
  {114511} (\bibinfo {year} {2015})},\ \Eprint
  {https://arxiv.org/abs/1402.0244} {arXiv:1402.0244 [hep-lat]} \BibitemShut
  {NoStop}%
\bibitem [{\citenamefont {Edwards}\ and\ \citenamefont
  {Jo\'{o}}(2005)}]{Edwards:2004sx}%
  \BibitemOpen
  \bibfield  {author} {\bibinfo {author} {\bibfnamefont {R.~G.}\ \bibnamefont
  {Edwards}}\ and\ \bibinfo {author} {\bibfnamefont {B.}~\bibnamefont
  {Jo\'{o}}} (\bibinfo {collaboration} {SciDAC, LHPC, UKQCD}),\ }\bibfield
  {booktitle} {\emph {\bibinfo {booktitle} {{Lattice field theory. Proceedings,
  22nd International Symposium, Lattice 2004, Batavia, USA, June 21-26,
  2004}}},\ }\href {https://doi.org/10.1016/j.nuclphysbps.2004.11.254}
  {\bibfield  {journal} {\bibinfo  {journal} {Nucl. Phys. Proc. Suppl.}\
  }\textbf {\bibinfo {volume} {140}},\ \bibinfo {pages} {832} (\bibinfo {year}
  {2005})},\ \bibinfo {note} {[,832(2004)]},\ \Eprint
  {https://arxiv.org/abs/hep-lat/0409003} {arXiv:hep-lat/0409003 [hep-lat]}
  \BibitemShut {NoStop}%
\end{thebibliography}%

\end{document}